\documentclass[epj,nopacs]{svjour}
\usepackage[T1]{fontenc}
\usepackage[italian,english]{babel}
\usepackage{hyperref}
\usepackage{ifpdf}
\usepackage{subfigure}
\usepackage{amssymb}
\usepackage{amsfonts}
\usepackage{epsf}
\usepackage{rotating}
\usepackage{graphicx}
\usepackage{amsmath}
\usepackage{fancyhdr}
\usepackage{lineno}
\usepackage{babel}
\usepackage{graphics}
\usepackage{pstricks}
\usepackage{color}
\usepackage{multirow}
\usepackage{nicefrac}
\usepackage{bm}
\usepackage{slashed}
\usepackage{float}
\usepackage{array}
\usepackage{boldline}
\usepackage{mathrsfs}
\usepackage{xcolor}

\usepackage{hyperref}
\hypersetup{colorlinks=true,linkcolor=blue,citecolor=red,urlcolor=lightseagreen}

\definecolor{lightseagreen}{rgb}{0.13, 0.7, 0.67}

\usepackage[numbers,sort&compress]{natbib}
\usepackage{tikz}
\usepackage{tikz-feynman}
\tikzfeynmanset{compat=1.0.0}

\DeclareFontFamily{OT1}{pzc}{}
\DeclareFontShape{OT1}{pzc}{m}{it}{<-> s * [1.2] pzcmi7t}{}
\DeclareMathAlphabet{\mathpzc}{OT1}{pzc}{m}{it}

\newcommand{\e}{\mathpzc{e}}
\newcommand{\Q}{\bm{Q}}
\newcommand{\LL}{\bm{L}}
\newcommand{\cl}{\mathpzc{l}}
\newcommand{\uq}{\mathpzc{u}}
\newcommand{\dq}{\mathpzc{d}}

\newcommand*\xbar[1]{%
	\hbox{%
		\vbox{%
			\hrule height 0.65pt 
			\kern0.4ex
			\hbox{%
				\kern-0.05em
				\ensuremath{#1}%
				\kern0.0em
			}%
		}%
	}%
}

\def\ep{$e$-$p$ }
\def\pbi{$\text{pb}^{-1}$ }
\def\fbi{$\text{fb}^{-1}$ }

\begin{document}
	\title{Zeros of Amplitude in the Associated Production of Photon and Leptoquark at $\bm e$-$\bm p$ Collider}
	\author{Priyotosh Bandyopadhyay\thanks{\email{\href{bpriyo@phy.iith.ac.in}{bpriyo@phy.iith.ac.in}}} \and Saunak Dutta\thanks{\email{\href{ph17resch11002@iith.ac.in}{ph17resch11002@iith.ac.in}}} \and Anirban Karan\thanks{\email{\href{kanirban@iith.ac.in}{kanirban@iith.ac.in}}}}                
	\institute{Indian Institute of Technology Hyderabad, Kandi, Sangareddy-502285, Telangana, India }
   \date{}
	\authorrunning{{}}
	\titlerunning{{}}
	
	\abstract{Though various extensions of the Standard Model with higher gauge group predict the existence of leptoquarks, none of them has been observed yet at any of the colliders. In this paper, we study the prospect of several past and future \ep colliders like HERA, LHeC and FCC-he to detect them through \textit{radiation amplitude zero}. We find that the leptoquarks showing zeros in the tree-level single-photon amplitudes at \ep collider lie within the complementary set of those exhibiting zeros at $e$-$\gamma$ collider. We present a PYTHIA-based analysis for HERA, LHeC and FCC-he (run II) to detect the leptoquarks with masses 70 GeV, 900 GeV and 1.5 TeV (2.0 TeV)  respectively through radiation amplitude zero.}
	\maketitle

\flushbottom
\setcounter{tocdepth}{1}
\setcounter{tocdepth}{2}
\setcounter{tocdepth}{3}
\tableofcontents

\section{Introduction}
\label{sec:intro}
The phenomenon of \textit{radiation amplitude zero} (RAZ) was discovered long ago by
Brown, Mikaelian, Sahdev, and Samuel \cite{Mikaelian:1979nr,Brown:1979ux} in the context of probing electromagnetic properties of $W$-boson. Soon after, it was noticed that in case of any gauge theory tree-level amplitude for any four-particle process involving one or more massless gauge fields in external legs  gets factorized into two parts comprehending several internal symmetry (charge) dependence and spin (polarization) dependence separately  governed by a \textit{spatial generalized Jacobi identity} \cite{Goebel:1980es,Zhu:1980sz}. Hinging on charge and four momenta of the four external particles, the first factor for single photon\footnote{The tree-level factorization holds for single gluon amplitude also, but the angular distribution does not show any zero due to colour summation \cite{Deshpande:1994vf}.} tree-level amplitude vanishes sometimes at certain regions of phase space and thus RAZ occurs. Later, RAZ was realized as a relativistic and quantum mechanical generalization of the classical event displaying no dipole radiation in the collision of non-relativistic particles with same charge-to-mass ratio and $g$-factor \cite{Brodsky:1982sh,Brown:1982xx}. Thus, the kinematic condition for single photon amplitude at tree-level to vanish is that the ratio $(Q_i/p_i\cdot k)$ must be same for all external particles (labelled by $i$) where their charges and four momenta are given by $Q_i$ and $p_i^\mu$ respectively with $k^\mu$ being the four momentum of the photon \cite{Brodsky:1982sh,Brown:1982xx}. The criteria for these zeros to lie inside the kinematically allowed region of phase space are also well-studied \cite{Passarino:1982hh,Samuel:1983eg,Samuel:1984ru}. Although various contaminations like initial and final state photon radiations, QED and QCD loop contributions, uncertainties in partonic centre of momentum reconstruction, etc., smear the effect of RAZ \cite{Stroughair:1984pj,Baur:1988qt,Cortes:1985yi,Valenzuela:1985dp,Laursen:1982iv,Laursen:1983kw,Smith:1989xz,Ohnemus:1992jn,Baur:1989gk}, noticeable dip in the angular distribution still remains \cite{Baur:1993ir,Baur:1994sa,Capdevilla:2019zbx}.

On the other hand, leptoquarks are electromagnetically charged, colour triplet proposed bosons that emerge naturally in several extensions of the Standard Model (SM) unifying matter fields \cite{Pati:1973uk,Pati:1974yy,Georgi:1974my,Georgi:1974sy,Dimopoulos:1979es,Farhi:1980xs,Schrempp:1984nj,Wudka:1985ef,Nilles:1983ge,Haber:1984rc}. The peculiarity of these particles is that they carry both non-zero lepton and  baryon numbers simultaneously and therefore are able to transform a quark into a lepton or vice-versa \cite{Hewett:1997ce,Dorsner:2016wpm}. Much effort has been devoted through last few decades for the direct detection of these particles at different colliders \cite{Behrend:1986jz,Bartel:1987de,Kim:1989qz,Decamp:1991uy,Adriani:1993gk,Abbiendi:2003iv,Abreu:1998fw,Aaron:2011zz,Collaboration:2011qaa,Abramowicz:2012tg,Abramowicz:2019uti,Alitti:1991dn,Acosta:2005ge,Abulencia:2005ua,Aaltonen:2007rb,Abazov:2008np,Abazov:2010wq,Abazov:2011qj,Aaboud:2019bye,Aaboud:2019jcc,Sirunyan:2018btu,Sirunyan:2018kzh,Sirunyan:2018nkj,Sirunyan:2018ryt,Sirunyan:2018vhk,Buchmuller:1986iq,Buchmuller:1986zs,Hewett:1987yg,Belyaev:2005ew,Bandyopadhyay:2018syt,Bhaskar:2020kdr,Chandak:2019iwj,Mandal:2015vfa,Alves:2018krf,Dorsner:2019vgp,Mandal:2018qpg,Padhan:2019dcp,Fuentes-Martin:2019ign,Baker:2019sli,Bhattacharyya:1994ig,Hewett:1987bh,Plehn:1997az,Kramer:1997hh,Cuypers:1995ax,Eboli:1993qx,Nadeau:1993zv,Atag:1994hk,Atag:1994np,Dobado:1987pj,Ilyin:1995jv,Gunion:1987ge,Bandyopadhyay:2020klr,Bandyopadhyay:2020wfv,Angelescu:2018tyl}, but no conclusive evidence advocating their existence has been discovered so far. In this paper, we have scrutinized the possibility of detecting all kinds of leptoquarks (if there is any), that can be produced at any electron-proton collider, through the phenomenon of RAZ. Though hadronic colliders have the splendour of performing collisions at very high energy, the main disadvantage there is the emergence of large SM background which makes the searches for beyond Standard Model (BSM) scenarios unfathomable. In contrast, electron-hadron and electron-photon colliders produce extremely limited number of SM processes as background making them preferable for any BSM hunt. The drawback of lower centre on momentum energy for these kind of machines will be tackled in various proposed colliders that are going to be built in near future with TeV scale energy.

 Based on the centre of momentum (CM) energies for various past and future \ep colliders \cite{19436,Klein:2008di,Agostini:2020fmq,Bordry:2018gri} like HERA, LHeC and FCC-he (run II), we analyse the production channels of leptoquarks for masses 70 GeV, 900 GeV and 1.5 TeV (2.0 TeV) respectively associated with a photon.  Though current experimental bounds seem to rule out leptoquarks with mass below 1.5 TeV, most of those analyses are performed surmising the coupling of leptoquarks to one generation of quark and lepton only. While considering their coupling to all the generations, the constraints become more relaxed and it turns out that the above mentioned masses are still allowed depending on specific couplings to different quarks and leptons. Now, if a leptoquark is produced, it will eventually decay to a quark and a lepton; therefore we look for mono-jet plus mono-lepton events associated with single photon at the detector for our analysis. In order to keep SM background null, we leave the electron events and search for the muon events only. In a PYTHIA-based simulation, we reconstruct the leptoquark from the invariant mass of the muon-jet pair, then we boost the whole system back to the CM frame and finally study the angular distribution of the  process with respect to the angle between radiated photon and electron beam. Using the charge separation of final state jets, model backgrounds have also been estimated. Unlike $e$-$\gamma$ collider \cite{Eboli:1993qx,Bandyopadhyay:2020klr,Cuypers:1995ax,Nadeau:1993zv}, the zeros of single-photon tree-level amplitude in this case do not depend on masses of leptoquarks or energies of collisions, rather they are controlled solely by electromagnetic charges of leptoquarks. However, it is very important to notice that the leptoquarks showing RAZ at \ep collider lie in the complementary set of those displaying the zero in single-photon tree-level amplitude at $e$-$\gamma$ collider. It is also noteworthy that while distinguishing signatures of different leptoquarks,  $e$-$\gamma$ and \ep colliders have great advantage over $pp$ or $p\bar p$ colliders. Although the angular distribution of pair production in hadronic colliders can categorize the leptoquarks according to their spins, it fails to separate various $SU(2)_L$ multiplets with same spin. In this context, \ep and $e$-$\gamma$ colliders together do this job through zeros of single photon tree level amplitude.

This paper is organized in the following way. The forthcoming section (Sect. \ref{Sec:Th}) illustrates the theoretical description of RAZ for various scalar and vector leptoquarks. Bounds on masses, couplings and branching fractions of leptoquarks from different experiments and specification of benchmark points for our simulation have been summarized in Sect. \ref{Sec:BP}. The succeeding section (Sect. \ref{Sec:collider}) deals with the set up needed for this collider simulation. In Sect. \ref{Sec:hera}, we reconsider the aspects of HERA for the search of low mass leptoquarks. The prospects of LHeC and FCC-he in detecting the heavy leptoquarks have been demonstrated in the next two sections (Sects. \ref{Sec:LHeC} and \ref{Sec:FCC}). We finally sum up and conclude in Sect \ref{Sec:concl}.

\section{Theoretical Framework}
\label{Sec:Th}

\begin{figure}[h!]
	\begin{center}
	\begin{tikzpicture}
	\begin{feynman}
	\vertex (a2);
	\vertex [left=0.8cm of a2] (a1){\(\bar q\)} ;
	\vertex [right=0.8cm of a2] (a3){\(\gamma\)};
	\vertex [below=1.8cm of a2] (b2);
	\vertex [left=0.8cm of b2] (b1){\(e\)};
	\vertex [right=0.8cm of b2] (b3){\(\phi\)};
	\diagram{(a1)--[anti fermion](a2)--[boson](a3),
		(a2)--[anti fermion, edge label'=\(q\)](b2),
		(b2)--[scalar](b3),
		(b1)--[fermion](b2)};
	\end{feynman}
	\end{tikzpicture}
		\hfill
		\begin{tikzpicture}
		\begin{feynman}
		\vertex (a2);
		\vertex [above left=1cm of a2] (z1){\(\bar q\)} ;
		\vertex [below left=1cm of a2] (b1){\(e\)};
		\vertex [right=0.8cm of a2] (a3);
		\vertex [above right=1cm of a3] (z4){\(\gamma\)};
		\vertex [below right=1cm of a3] (b4){{\(\phi\)}};
		\diagram{(z1)--[anti fermion](a2)--[scalar,edge label=\(\phi\)](a3),
			(a2)--[anti fermion](b1),
			(z4)--[boson](a3)--[scalar](b4)};
		\end{feynman}
		\end{tikzpicture}
	\hfill
	\begin{tikzpicture}
	\begin{feynman}
	\vertex (a2);
	\vertex [left=0.8cm of a2] (a1){\(\bar q\)} ;
	\vertex [right=0.8cm of a2] (a3){\(\phi\)};
	\vertex [below=1.8cm of a2] (b2);
	\vertex [left=0.8cm of b2] (b1){\(e\)};
	\vertex [right=0.8cm of b2] (b3){\(\gamma\)};
	\diagram{(a1)--[anti fermion](a2)--[scalar](a3),
		(a2)--[anti fermion, edge label'=\(e\)](b2),
		(b2)--[boson](b3),
		(b1)--[fermion](b2)};
	\end{feynman}
	\end{tikzpicture}
	\caption{Feynman diagrams for $e\,\bar q \to \gamma\,\phi$}
	\label{fig:eqbpg}
\end{center}
\end{figure}
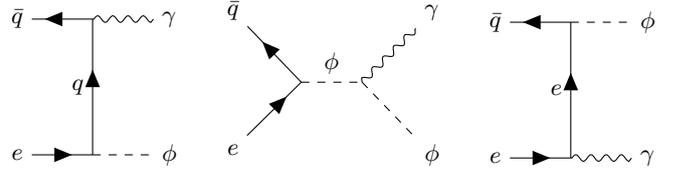
In this section, we discuss the theoretical aspects of observing RAZ at \ep collider for the production of any leptoquark $\phi$ associated with a photon. The parton level interactions responsible for this process are $e\,\bar q \,(\text{or } q)\to \gamma\,\phi$. We present here the analytic calculations for $e\,\bar q \to \gamma\,\phi$ only. It is straight forward to find the expressions for the other parton level process by repeating the same procedure. 

There are three Feynman diagrams contributing to the process $e\,\bar q \to \gamma\,\phi$ as shown in Fig. \ref{fig:eqbpg}. Combining the matrix elements for all the diagrams, the total amplitudes for production of scalar and vector leptoquarks along with a photon from the collision of electrons and anti-quarks are respectively given by:  
\begin{align}
\label{eq:AmpS}
&\mathcal{M}^S=-i\e\,\Big(\frac{Q_\phi}{p_\phi. k}+\frac{1}{p_e.k}\Big)\,\epsilon^{\gamma*}_\mu\,\bar v(p_{\bar q})\,(Y_L^{eq}P_L+Y_R^{eq}P_R)\nonumber\\
&\hspace{1.3cm}\times\bigg[p_\phi^\mu-\Big(\frac{p_\phi.k}{p_{\bar q}.k}\Big)\big(p_{\bar q}^\mu-\frac{1}{2}\gamma^\mu\slashed k\big)\bigg]u(p_e)\,,\\
\label{eq:AmpV}
&\mathcal{M}^V=-i\e\,\Big(\frac{Q_\phi}{p_\phi. k}+\frac{1}{p_e.k}\Big)\,\epsilon_\mu^{\gamma*}\epsilon_\nu^{\phi*}\,\bar v(p_{\bar q})\,\Big[\frac{1}{2}\gamma^\nu\gamma^\mu\slashed k\nonumber\\
&\hspace{1.3cm}+\gamma^\nu p_e^\mu+\Big(\frac{p_e.k}{p_{\bar q}.k}\Big)\big(\gamma^\mu p_e^\nu+\frac{1}{2}\gamma^\mu \gamma^\nu \slashed p_\phi\big)\Big]\nonumber\\
&\hspace{1.3cm}\times(Y_L^{eq} P_L+Y_R^{eq} P_R)\, u(p_e)\,, 
\end{align}
where $p_e^\mu$, $p_{\bar q}^\mu$, $p_\phi^\mu$, and $k^\mu$ are the four momenta of electron, anti-quark, leptoquark and photon respectively, $Y_{L,R}^{eq}$ are the  couplings of leptoquark with different generations of left-handed and right-handed leptons and quarks, $-\e$ and $Q_{\phi}\e$ denote the charges of electron and leptoquark  respectively, $\epsilon_\mu^{\gamma,\phi}$ signify the polarizations of the photon and the vector leptoquark and $P_{L,R}\equiv(1\mp\gamma^5)/2$. While deriving amplitudes in Eqs. \eqref{eq:AmpS} and \eqref{eq:AmpV}, we have presumed the electron and anti-quark to be massless. However, at the time of simulation, we have taken the masses of the SM fermions accordingly.

The factorization of the single-photon tree-level amplitude for the above-mentioned four-particle process, as described in Refs. \cite{Goebel:1980es,Zhu:1980sz,Deshpande:1994vf}, is quite apparent from Eqs. \eqref{eq:AmpS} and \eqref{eq:AmpV}. The charge-dependent factor $(\frac{Q_\phi}{p_\phi. k}+\frac{1}{p_e.k})$ in these two equations vanishes at particular points of phase space for some leptoquarks \cite{Brown:1982xx,Brodsky:1982sh} which in turn causes RAZ. Therefore, if $\theta^*$ be the angle between photon and electron in CM frame at which RAZ happens, then:
\begin{equation}
\label{eq:RAZ}
\cos\theta^*=1+\frac{2}{Q_{\phi}}\,.
\end{equation}
 This implies that the necessary condition for RAZ occurring inside the physical region of an \ep collider is given by:
 \begin{equation}
 \label{eq:cond}
 Q_\phi<-1\,.
 \end{equation}
 In Table \ref{tab:LQ}, we epitomize all the leptoquarks that can be produced at \ep collider and indicate which of them will show RAZ. Refs.\cite{Hewett:1997ce,Dorsner:2016wpm,Buchmuller:1986zs,Hewett:1987yg,Belyaev:2005ew,Davidson:1993qk} contain comprehensive list of all the leptoquarks and we follow similar notations.  Ref.\cite{Davies:1990sc} also performed similar study with generic exotic scalars.
 The subscript 1, 2 and 3 in the name of leptoquarks signify singlet, doublet and triplet leptoquarks under $SU(2)_L$ gauge group; additionally, the presence of Lorentz index $\mu$ indicates the corresponding leptoquark to be a vector particle. In Table \ref{tab:LQ}, we have also mentioned the weak hypercharge $(\mathtt Y_\phi)$, the third component of weak isospin $(T_3)$ and electromagnetic charge $(Q_{\phi})$ of different leptoquarks.
 The gauge representations of all the leptoquark multiplets have been explicitly mentioned in the subsequent subsections of Section \ref{Sec:hera} where they are introduced and discussed. However, while discussing a particular component of a doublet or triplet leptoquark, we explicitly mention its electromagnetic charge in the superscript to distinguish it from the other excitations of the same multiplet.

 It is noteworthy from Table \ref{tab:LQ} that $S_3^{\nicefrac{-2}{3}}$, $\widetilde R_2^{\nicefrac{-1}{3}}$, $\widetilde V_{2\mu}^{\nicefrac{-2}{3}}$ and $U_{3\mu}^{\nicefrac{-1}{3}}$ will never be produced at \ep collider since the gauge-structure of the Lagrangian prohibits them to interact with any charged lepton. Another important remark to make at this point is that the zeros of single-photon tree-level amplitudes for production of leptoquark associated with a quark at $e$-$\gamma$ collider occur only if $-1<Q_\phi<0$ \cite{Bandyopadhyay:2020klr} which lies beyond the parameter-space for $Q_\phi$ specified by Eq. \eqref{eq:cond}. Hence, the leptoquark whose angular distribution for the production channel does not show any zero at $e$-$\gamma$ collider will definitely exhibit RAZ here.

\begin{table}[h!]
	\begin{tabular*}{0.49\textwidth}{@{\hspace*{4mm}\extracolsep{\fill}}cccccc}
		\hline
		\\[-2mm]
		$\phi$ & $\mathtt Y_\phi$ & $T_3$ & $Q_{\phi}$  &  Production Channel & $\cos\theta^*$ \\[1mm]
		\hline
		\\[-1mm]
		\multicolumn{5}{l}{\textbf{Scalar Leptoquarks}}\\[1mm]
		$S_1$  & $\nicefrac{2}{3}$ & 0 & $\nicefrac{1}{3}$   & $e^-\, u\to \gamma\, \Big({S_1^{\nicefrac{+1}{3}}}\Big)^c$ & ---\\[3mm]
		
		$\widetilde S_1$ & $\nicefrac{8}{3}$ & 0 & $\nicefrac{4}{3}$  & $e^-\, d \to \gamma\, \Big({\widetilde S_1^{\nicefrac{+4}{3}}}\Big)^c$ & $\nicefrac{-1}{2}$ \\[3mm]
		
			$R_2$ & $\nicefrac{7}{3}$ & $\nicefrac{1}{2}$ & $\nicefrac{5}{3}$ & $e^-\,\bar u \to \gamma\,  \Big({R_2^{\nicefrac{+5}{3}}}\Big)^c$ & $\nicefrac{-1}{5}$ \\
			&  & $\nicefrac{-1}{2}$ &$\nicefrac{2}{3}$   &  $e^-\,\bar d\to \gamma\, \Big({R_2^{\nicefrac{+2}{3}}}\Big)^c$ & ---\\[3mm]
			
			$\widetilde{R}_2$ & \nicefrac{1}{3} & $\nicefrac{1}{2}$ &\nicefrac{2}{3} & $e^-\,\bar d\to \gamma\,  \Big({\widetilde R_2^{\nicefrac{+2}{3}}}\Big)^c$ & --- \\
			&  & $\nicefrac{-1}{2}$ & $\nicefrac{-1}{3}$   &  --- & ---\\[3mm]

	$\vec{S}_3$ & $\nicefrac{2}{3}$ & 1 & $\nicefrac{4}{3}$    & $e^-\, d \to \gamma\,  \Big({S_3^{\nicefrac{+4}{3}}}\Big)^c$ & $\nicefrac{-1}{2}$ \\
		&   & 0 & $\nicefrac{1}{3}$  &  $e^-\, u\to \gamma\, \Big({S_3^{\nicefrac{+1}{3}}}\Big)^c$ & ---\\
		&  & $-1$ &$\nicefrac{-2}{3}$   & --- & ---\\[3mm]

		\multicolumn{5}{l}{\textbf{Vector Leptoquarks}}\\[1mm]
		
		$U_{1\mu}$ & $\nicefrac{4}{3}$ & 0 & $\nicefrac{2}{3}$  & $e^-\,\bar d\to \gamma\,\Big({ U_{1\mu}^{\nicefrac{+2}{3}}}\Big)^c$ & ---\\[3mm]
		
		$\widetilde U_{1\mu}$ & $\nicefrac{10}{3}$ & 0 & $\nicefrac{5}{3}$ & $e^-\,\bar u\to \gamma \,\Big({ \widetilde U_{1\mu}^{\nicefrac{+5}{3}}}\Big)^c$ & $\nicefrac{-1}{5}$ \\[3mm]
		
		$V_{2\mu}$& $\nicefrac{5}{3}$ & $\nicefrac{1}{2}$ & $\nicefrac{4}{3}$ & $e^-\, d\to\gamma \, \Big({ V_{2\mu}^{\nicefrac{+4}{3}}}\Big)^c$ & $\nicefrac{-1}{2}$ \\
		& & $\nicefrac{-1}{2}$ &$\nicefrac{1}{3}$  & $e^-\, u\to\gamma\, \Big({V_{2\mu}^{\nicefrac{+1}{3}}}\Big)^c$ & ---\\[3mm]
		
		$\widetilde V_{2\mu}$ & $\nicefrac{-1}{3}$ & $\nicefrac{1}{2}$ & $\nicefrac{1}{3}$  & $e^-\, u \to \gamma\, \Big({\widetilde V_{2\mu}^{\nicefrac{+1}{3}}}\Big)^c$ &---  \\
		&  & $\nicefrac{-1}{2}$ &$\nicefrac{-2}{3}$ &--- & ---\\[3mm]
		
		$\vec{U}_{3\mu}$& $\nicefrac{4}{3}$ & 1 & $\nicefrac{5}{3}$ & $e^-\, \bar u \to \gamma\,  \Big({U_{3\mu}^{\nicefrac{+5}{3}}}\Big)^c$ & $\nicefrac{-1}{5}$ \\
		&  & 0 & $\nicefrac{2}{3}$ & $ e^-\, \bar d\to \gamma\,\Big({ U_{3\mu}^{\nicefrac{+2}{3}}}\Big)^c$ & ---\\
		&  & $-1$ &$\nicefrac{-1}{3}$  & --- & ---\\[2mm]
		\hline
	\end{tabular*}
	\caption{Specification of different leptoquarks produced at \ep collider. We follow the notations introduced in Refs.\cite{Hewett:1997ce,Dorsner:2016wpm,Buchmuller:1986zs,Hewett:1987yg,Belyaev:2005ew,Davidson:1993qk} where a comprehensive studies of all leptoquark models have been carried out.}
	\label{tab:LQ}
\end{table}

The angular distribution in CM frame with respect to the angle $(\theta)$ between electron and photon for the channel $e\,\bar q_{\alpha} \to \gamma\,\phi_\alpha$ ($\alpha$ is the colour-index) is given by:
\begin{equation}
\label{eq:ang_dist}
\frac{d\sigma}{d\cos\theta}=\frac{s-M_\phi^2}{32\pi s^2}\,\Big(\frac{1}{4}\sum_{\text{spin}}\big|\mathcal M^{S,V}\big|^2\Big)\,,
\end{equation}
where, $s=(p_e+p_{\bar q})^2$, $M_\phi$ represents the mass of leptoquark and $\displaystyle\Big(\sum_{\text{spin}}\big|\mathcal M^{S,V}\big|^2\Big)$ indicates the sum over polarizations of all final state particles for the absolute square of matrix element which can be expressed for scalar and vector leptoquark cases as: 
 \begin{align}
\label{eq:modMsqSLQ}
\sum_{\text{spin}}\big|\mathcal{M}^S\big|^2=e^2\,\big[&(Y_L^{eq})^2+(Y_R^{eq})^2\big]\,\bigg[1+\frac{(s+M_\phi^2)^2}{(s-M_\phi^2)^2}\bigg] \nonumber\\&\times\mathrm{cosec}^2\,\theta\,\big[Q_\phi\cos\theta-(2+Q_\phi)\big]^2\,,
\end{align}
\begin{align}
\label{eq:modMsqVLQ}
\sum_{\text{spin}}\big|\mathcal{M}^V\big|^2=2e^2\,&\big[(Y_L^{eq})^2+(Y_R^{eq})^2\big]\,\bigg[\cos^2\theta+\frac{(s+M_\phi^2)^2}{(s-M_\phi^2)^2}\bigg] \nonumber\\&\times\mathrm{cosec}^2\,\theta\,\big[Q_\phi\cos\theta-(2+Q_\phi)\big]^2\,.
\end{align}


However, one should not expect Eq. \eqref{eq:ang_dist} to hold explicitly for associate production of a leptoquark with a photon in any \ep collider since the parton distribution function of proton would modify the angular distribution accordingly. Moreover, the uncertainty in the parton distribution function may also lead to slight shift in the position of RAZ.

\section{Experimental Bounds \& Benchmark Points}
\label{Sec:BP}

\begin{figure*}
	\centering
\includegraphics[scale=0.2]{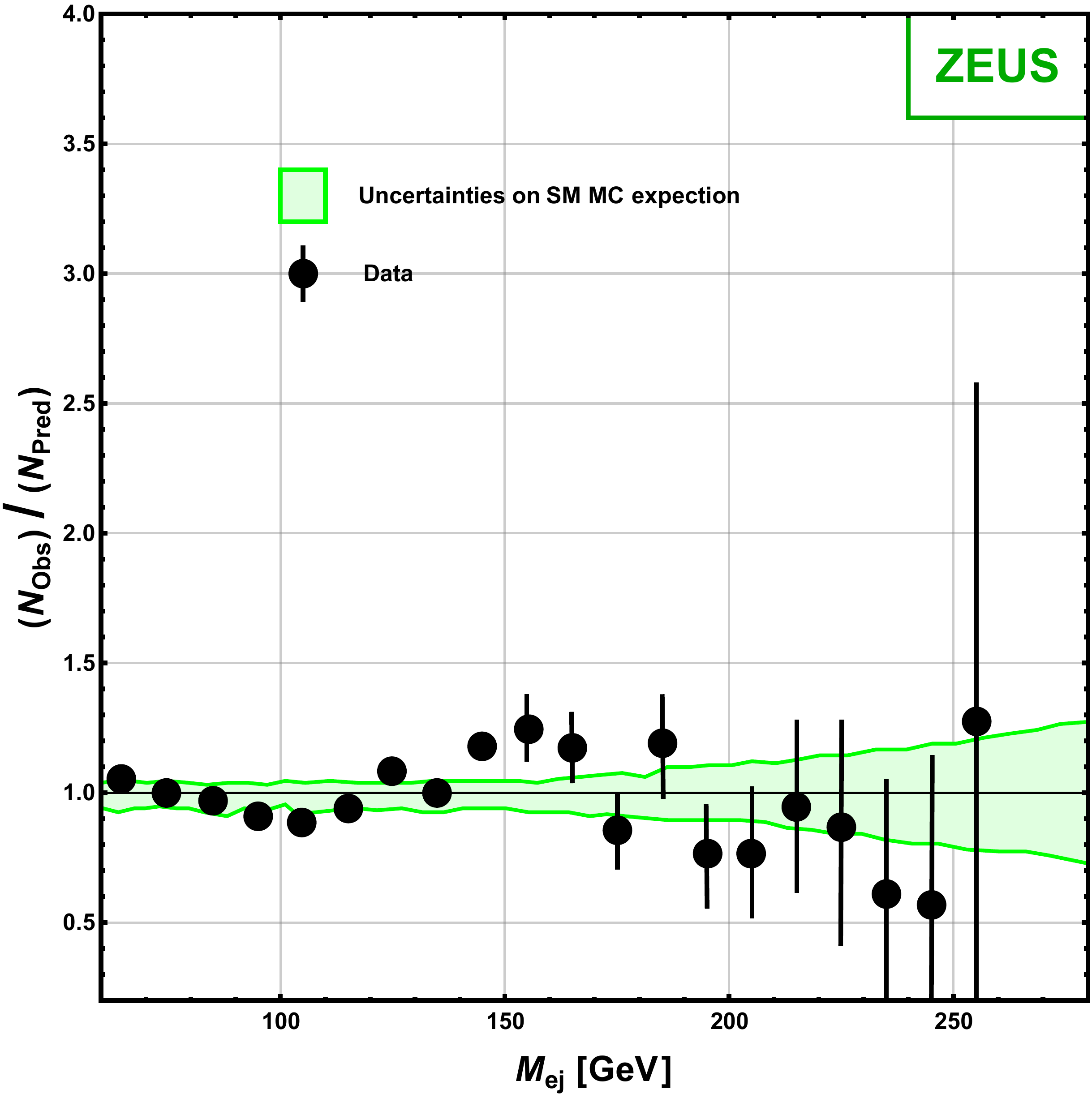}
\hfil
\includegraphics[scale=0.2]{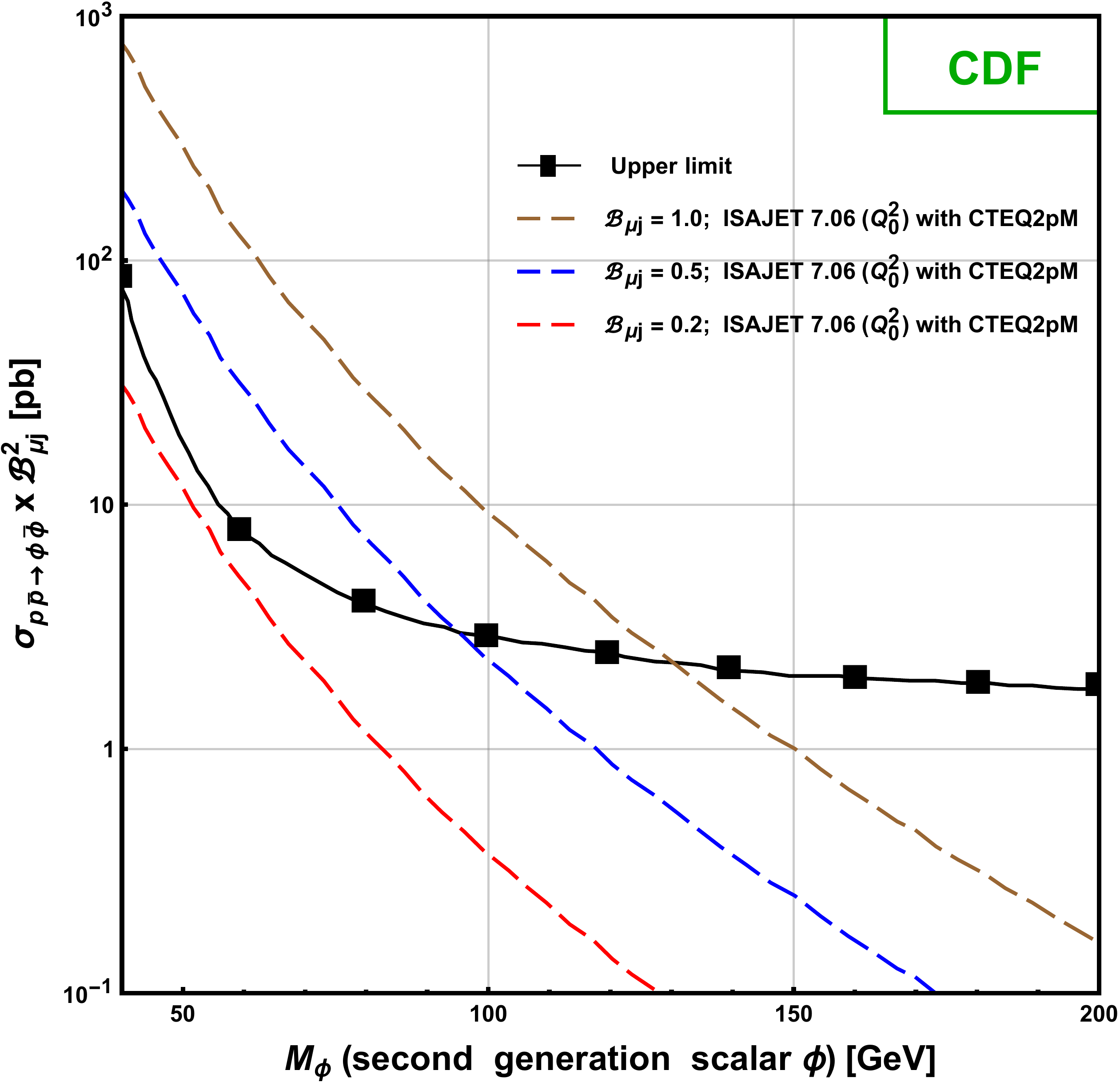}
\hfil
\includegraphics[scale=0.2]{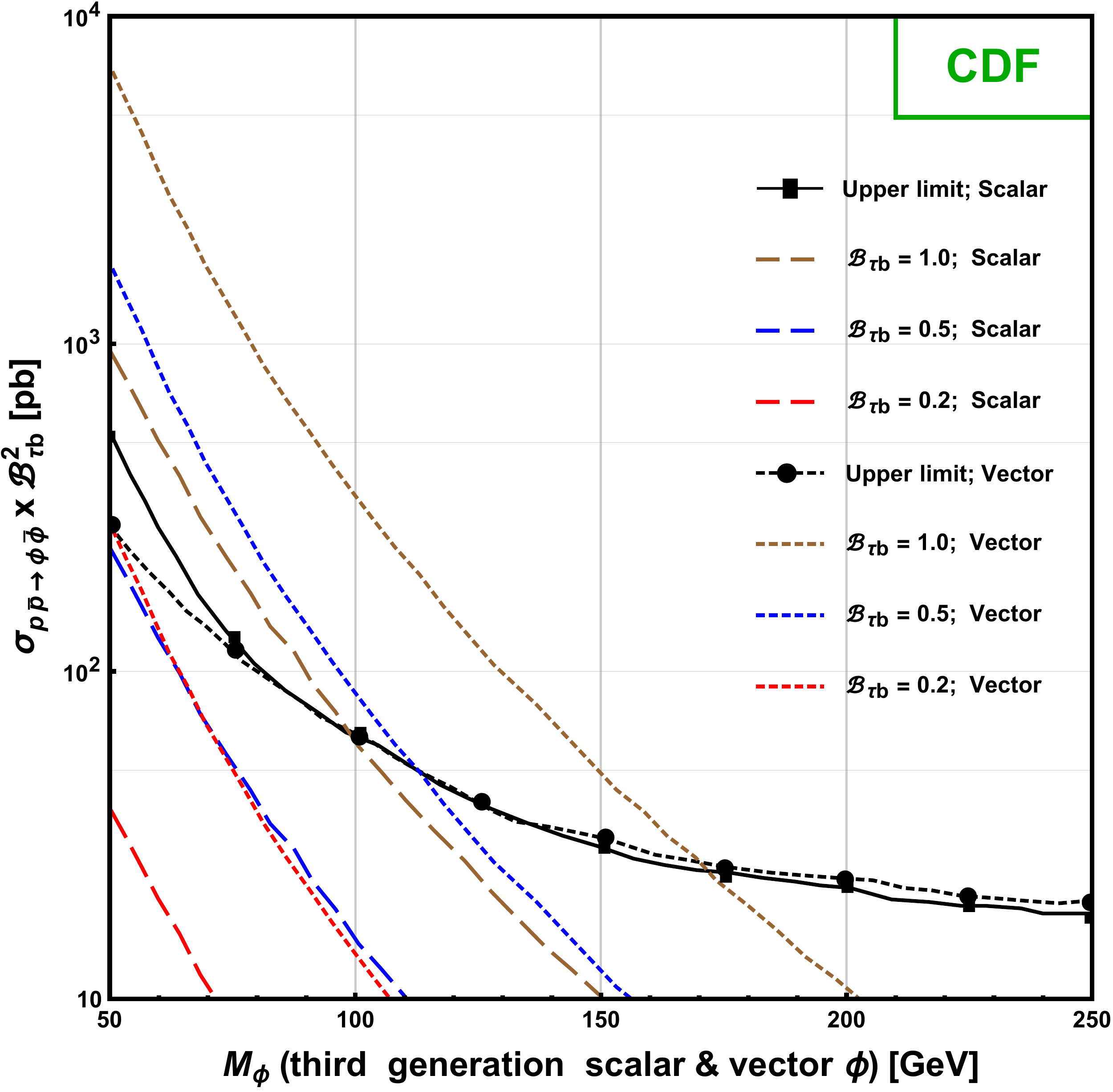}

\vspace*{0.5cm}

\includegraphics[scale=0.16]{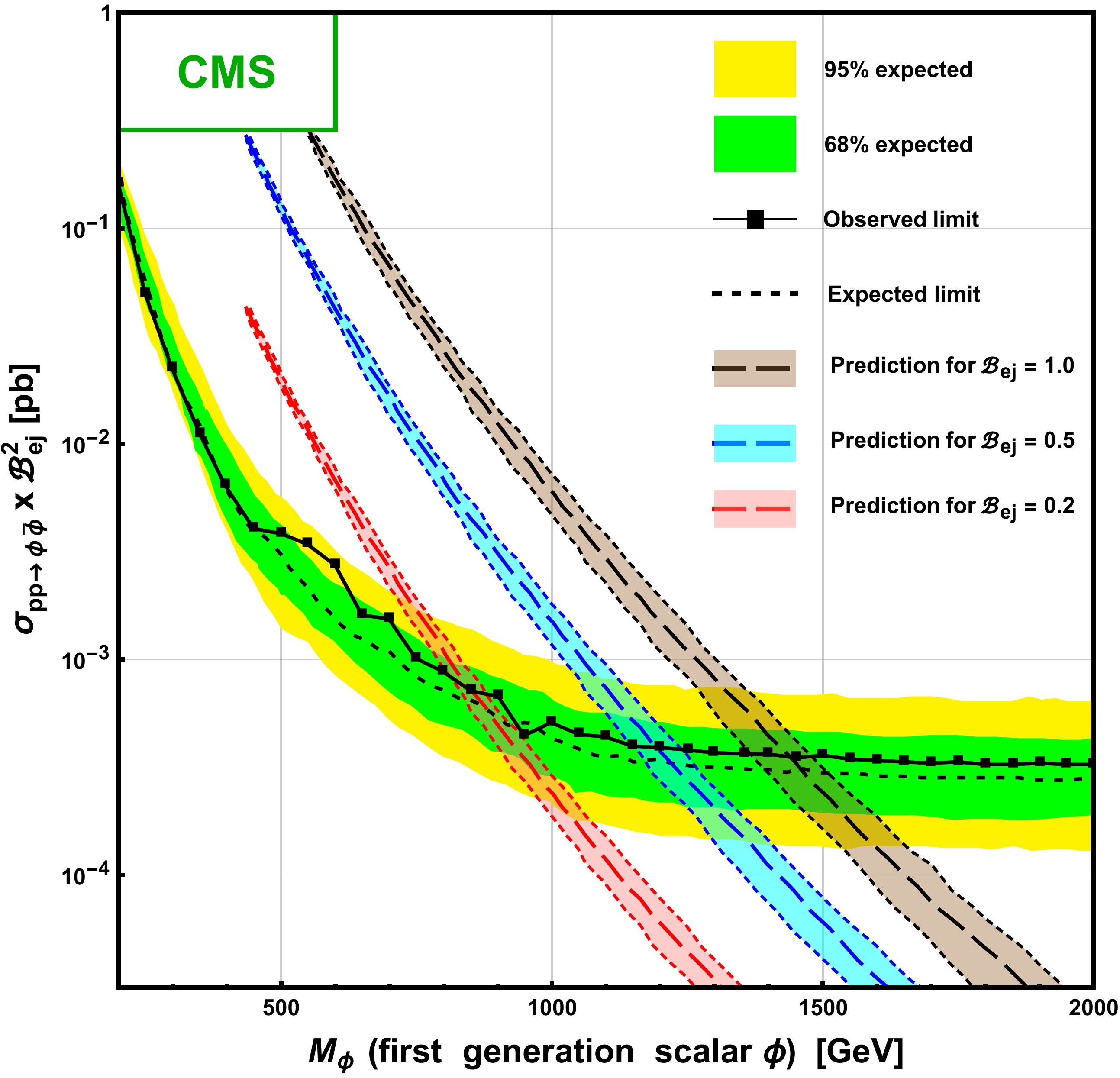}
\hfil
\includegraphics[scale=0.16]{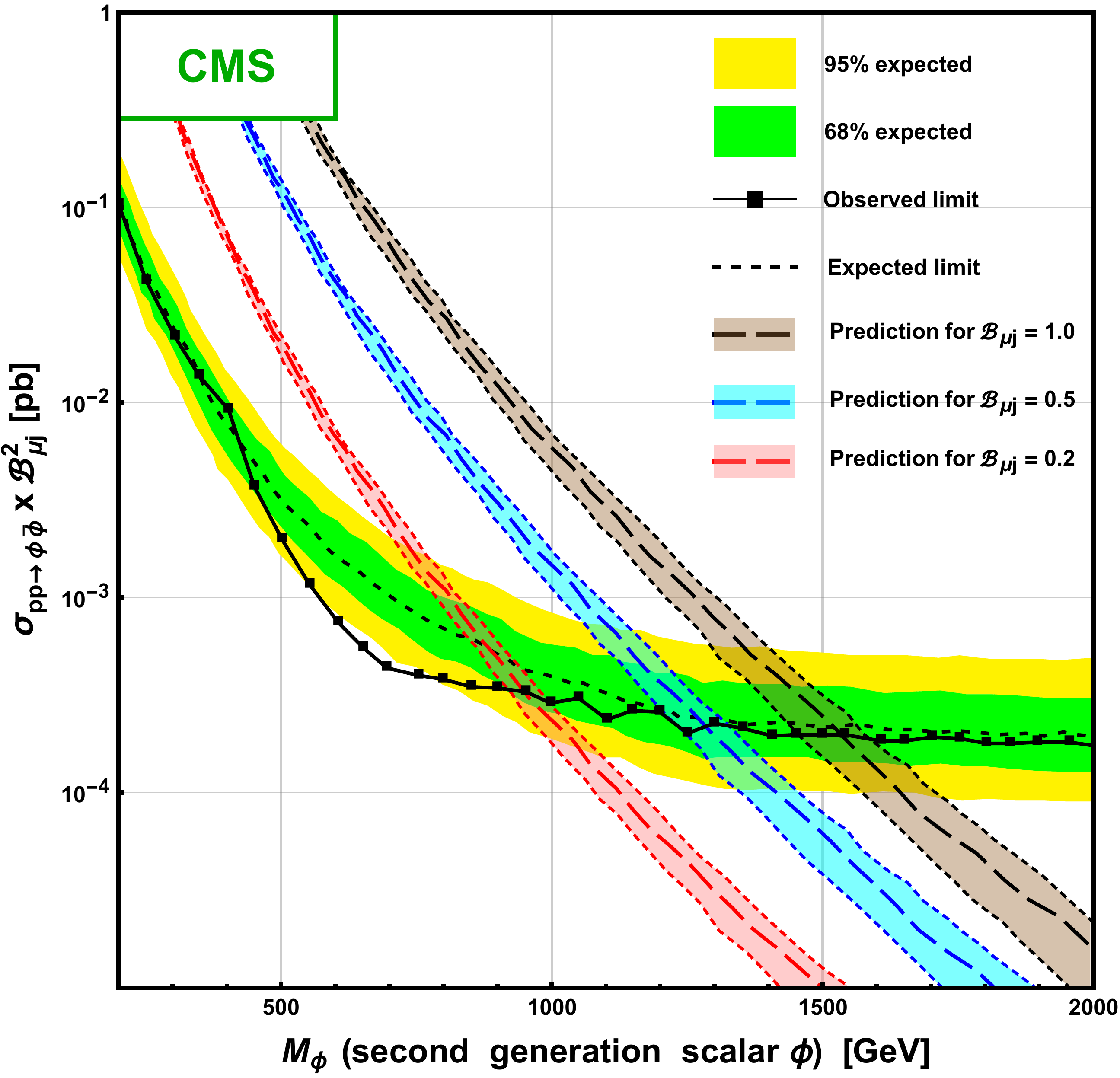}
\hfil
\includegraphics[scale=0.16]{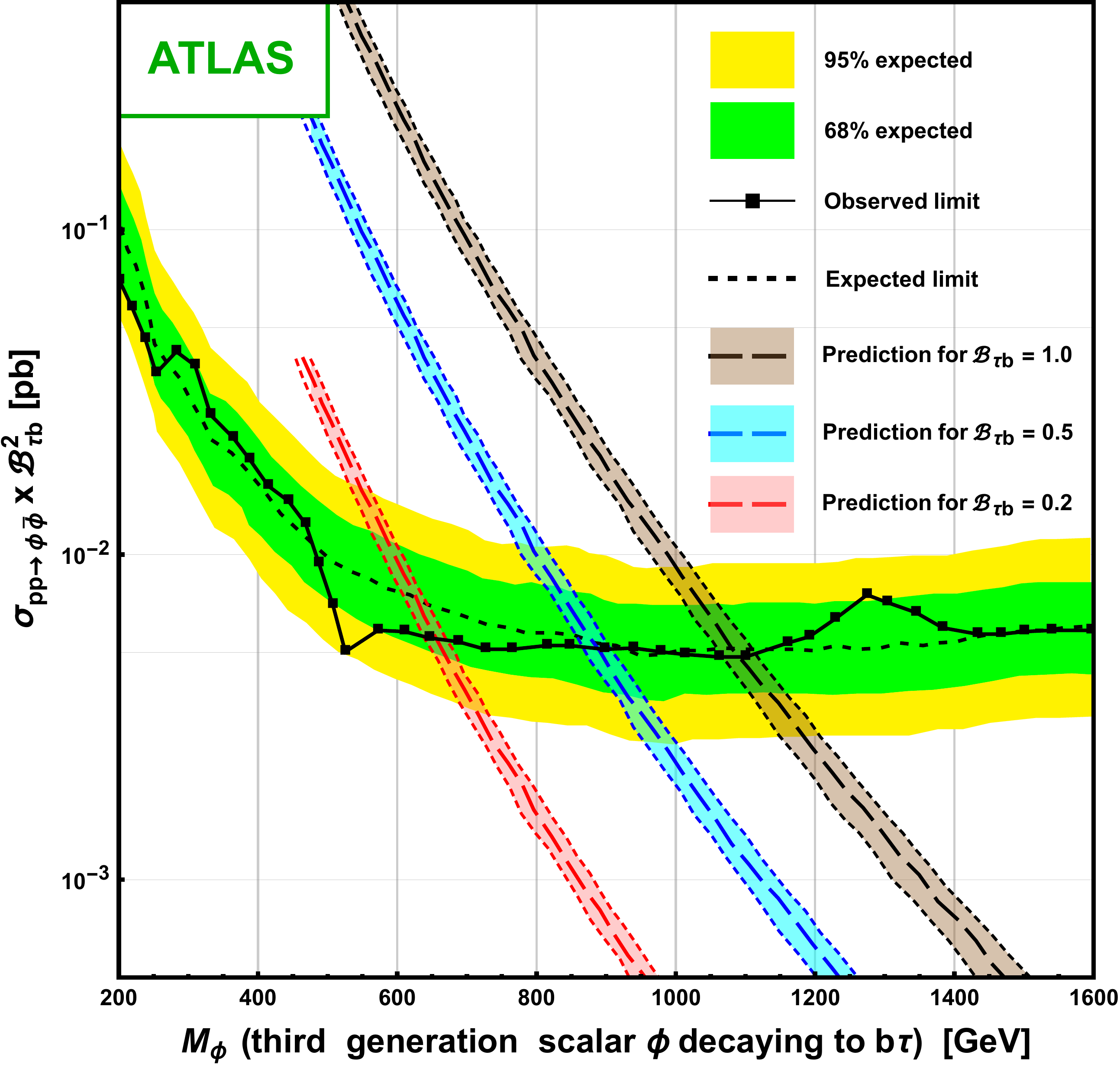}
\hfil
\includegraphics[scale=0.16]{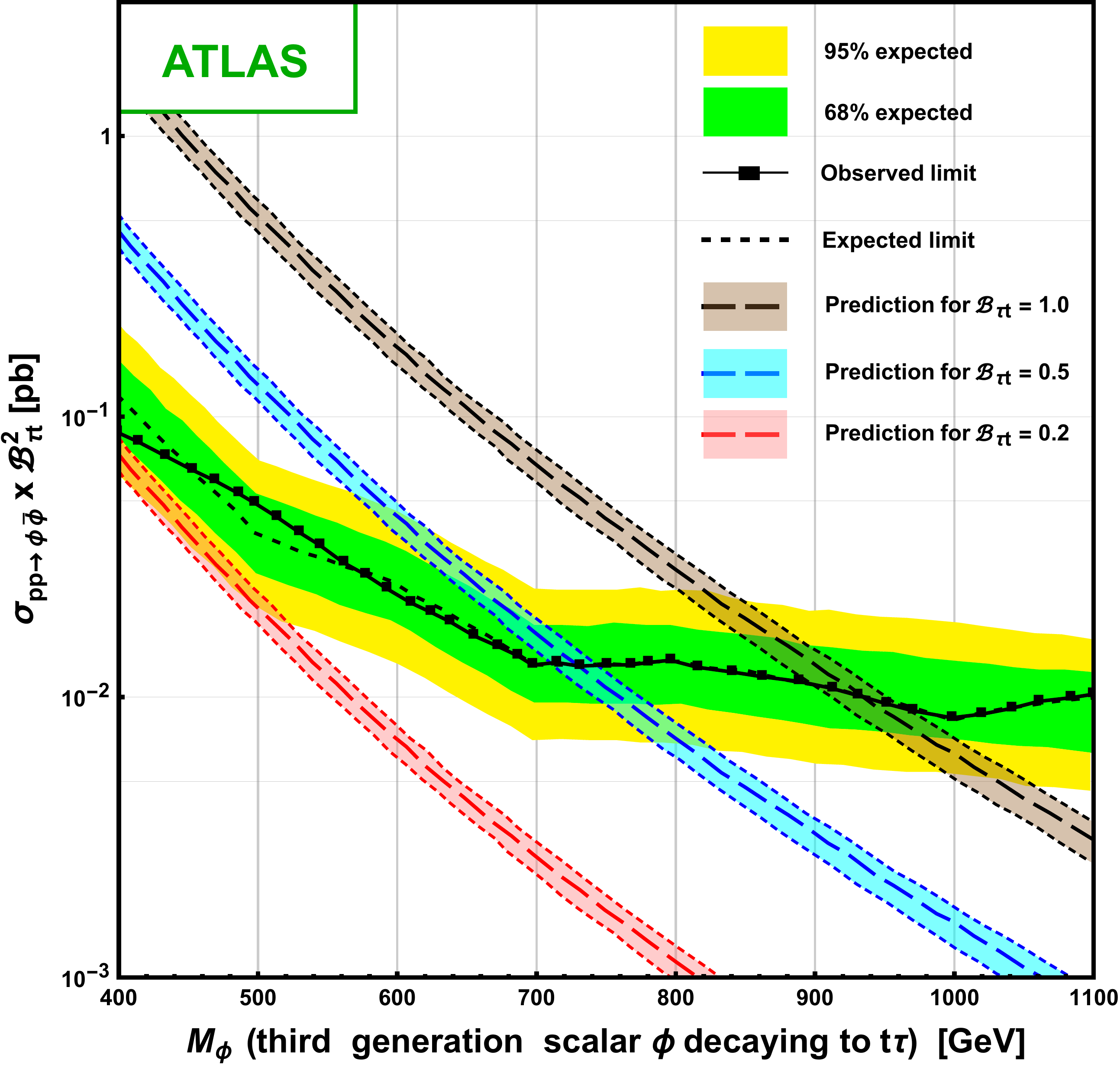}

\vspace*{0.5cm}

\includegraphics[scale=0.2]{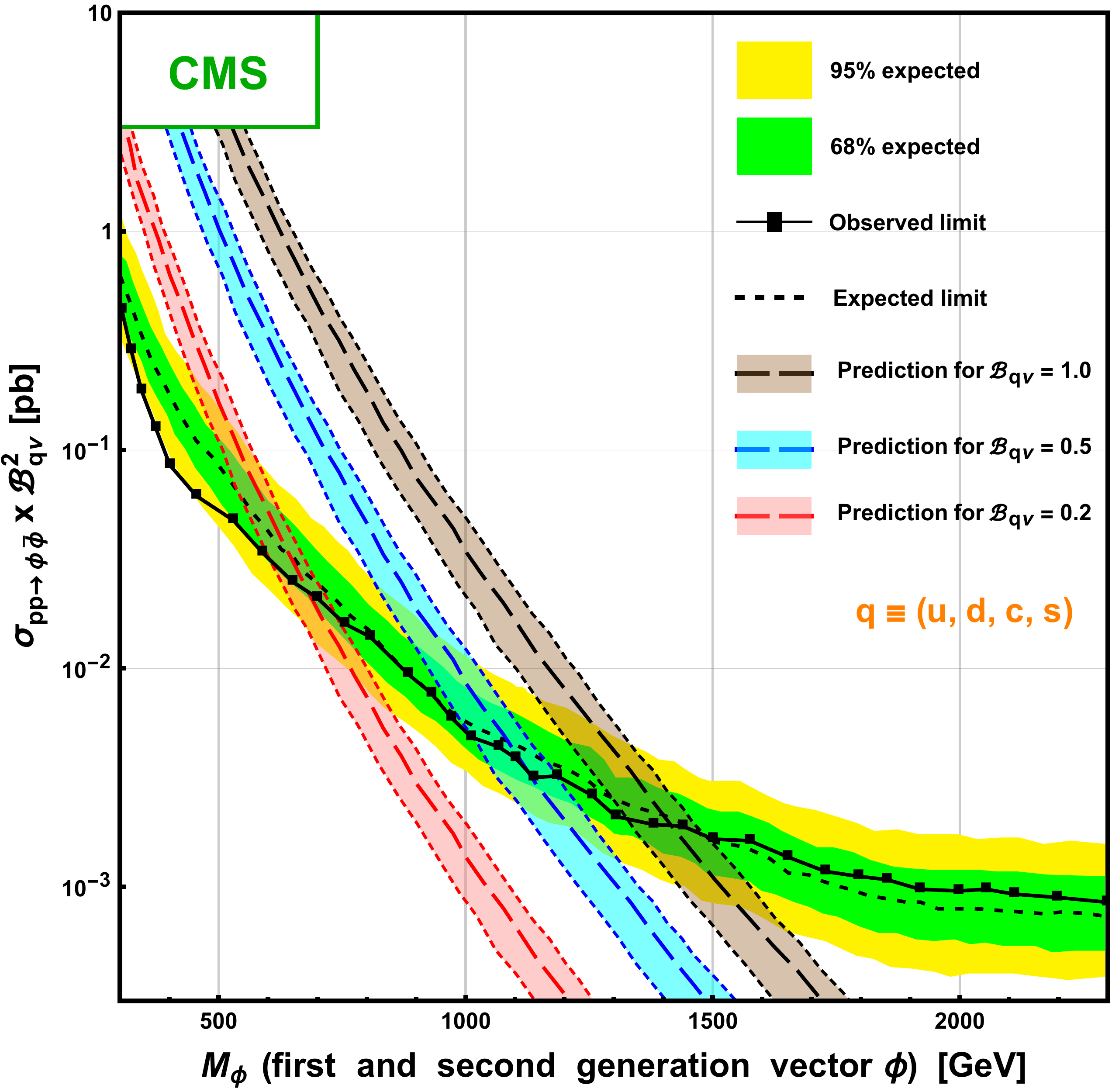}
\hfil
\includegraphics[scale=0.2]{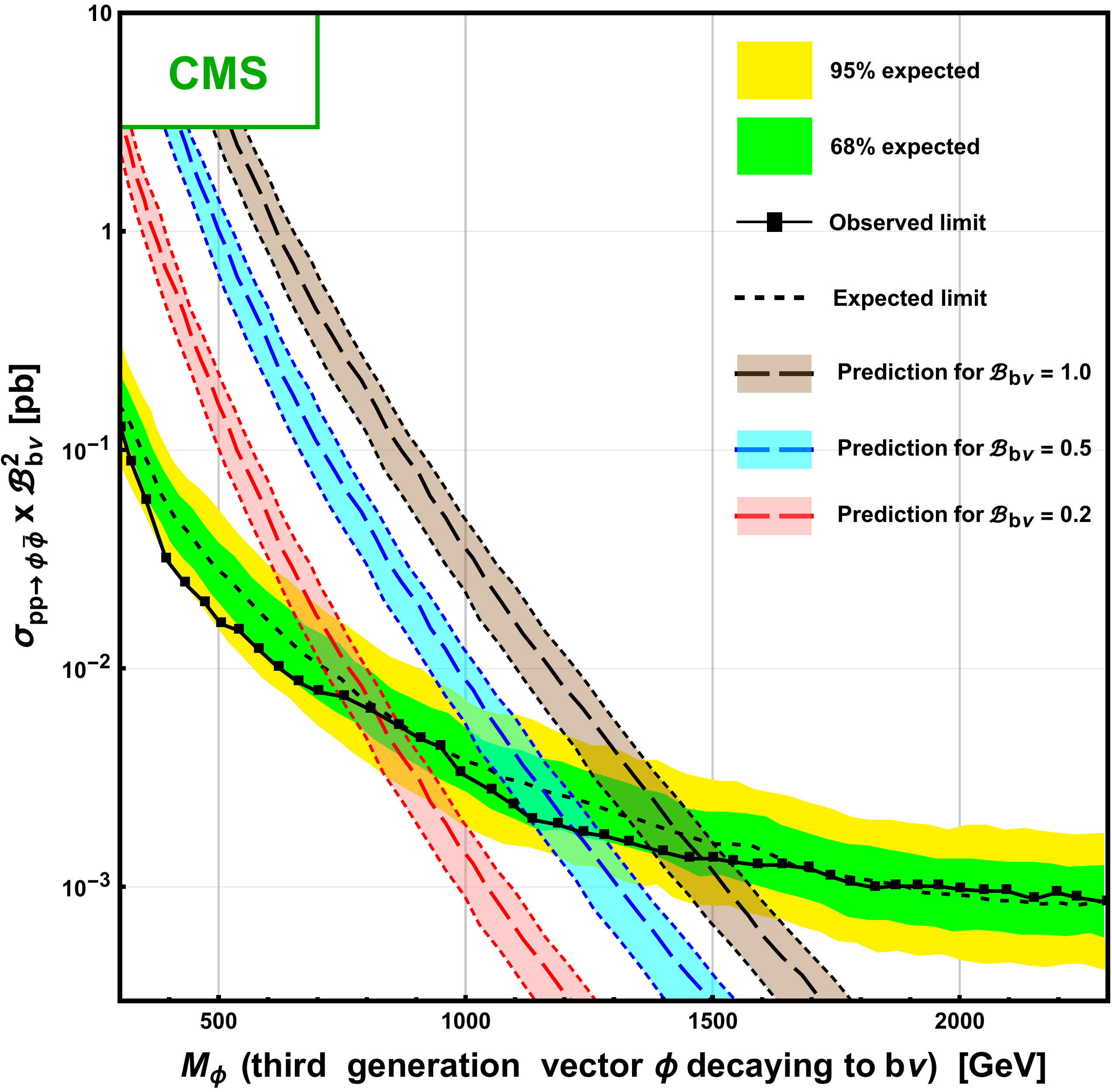}
\hfil
\includegraphics[scale=0.2]{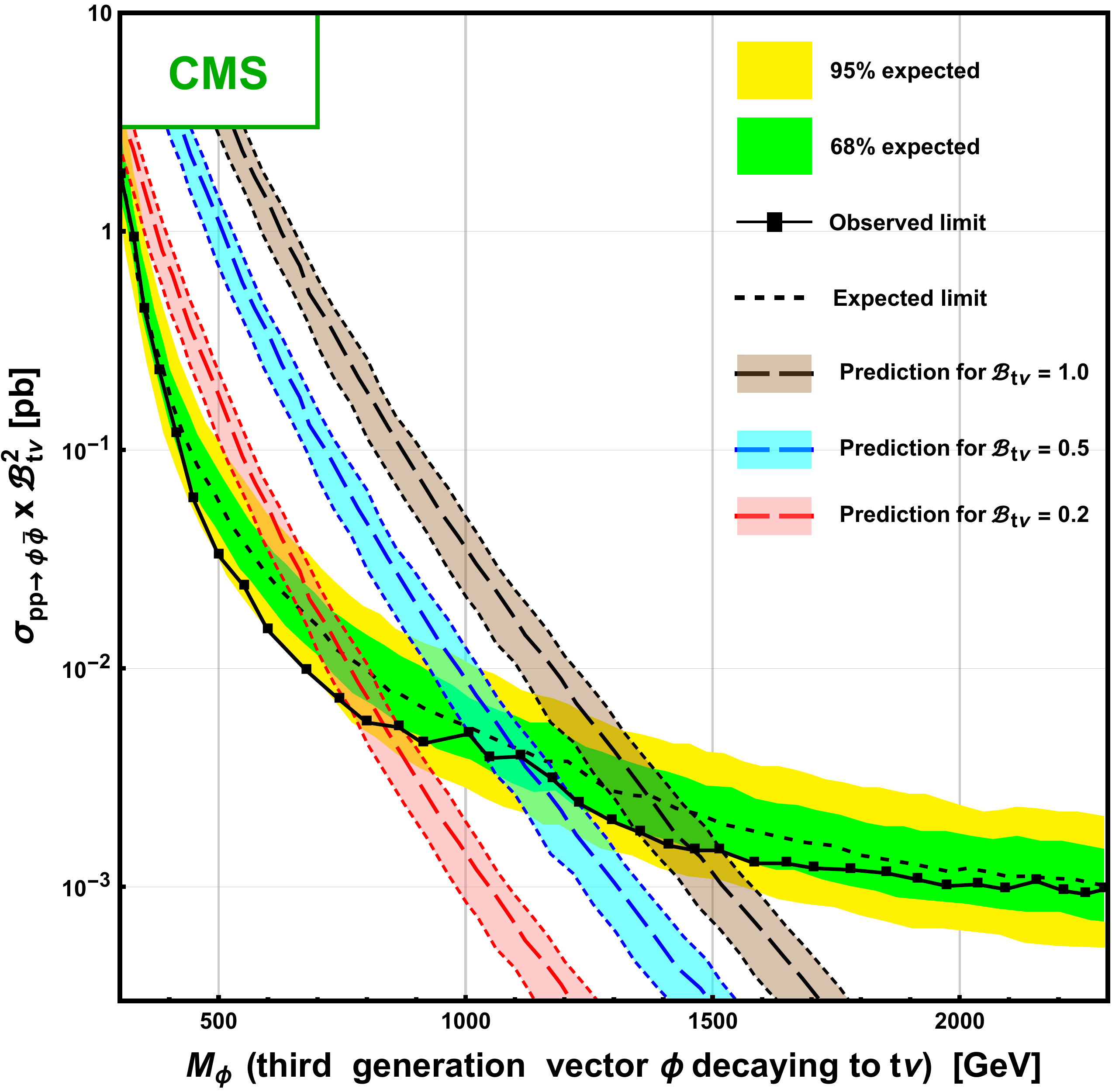}

\caption{Different experimental bounds on leptoquarks. The plots in the first row have been used for choosing the benchmark points for 70 GeV leptoquarks, the diagrams in the second row provide bounds on heavier scalar leptoquarks whereas the images in the third row serve the same purpose for heavier vector leptoquarks. In case of vectors, we have taken $\kappa=0$ only.}
\label{fig:limit}
\end{figure*}

This section deals with different experimental bounds on leptoquarks and choice of benchmark points for our simulation. There are several direct and indirect bounds on leptoquarks coming from different experiments. While the indirect constraints arise from restrictions on leptoquark-induced four-fermion interactions, testable at low-energy experiments, the direct bounds emerge from the possibility of their production at different high-energy colliders. The indirect constraints\footnote{Description of all the indirect bounds on leptoquarks is presented in the ``Indirect Limits for Leptoquarks'' section of Ref. \cite{10.1093/ptep/ptaa104}.} on leptoquarks have been studied extensively in the Refs. \cite{Davidson:1993qk,Leurer:1993em,Leurer:1993qx,Carpentier:2010ue,Mandal:2019gff}. For instance, it can be obtained from Ref. \cite{Carpentier:2010ue} that scalar leptoquarks coupling to the first generation of left-handed quark and lepton should satisfy the bound $(Y/\widetilde M_\phi)^2\leq 0.07$ and the similar vector leptoquarks should obey $(Y/\widetilde M_\phi)^2\leq 0.4$ where $\widetilde M_\phi=(M_\phi/1\,\text{TeV})$. However, for our present purpose we mainly focus on the collider bounds.

The search for leptoquarks at different colliders has a long history.  The first ever search was performed by the CELLO \cite{Behrend:1986jz} and JADE \cite{Bartel:1987de} Collaborations at the PETRA for pair production of leptoquarks through $e^+e^-$ collision with the centre of momentum energy $(\equiv \sqrt s)$ around 40 GeV, but no evidence was found in both the detectors, and the CELLO Collaboration excluded the leptoquark mass from 7 GeV to 20.5 GeV. A similar bound was also set up by the AMY Collaboration \cite{Kim:1989qz} when no signature of leptoquark was detected in $e^+e^-$ annihilation at the TRISTAN with 50 GeV $\leq \sqrt s\leq$ 60.8 GeV. The existence of leptoquark was looked for at another $e^+e^-$ collider, the LEP, by the ALEPH, L3, OPAL and DELPHI Collaborations. Results from the ALEPH \cite{Decamp:1991uy} and L3 \cite{Adriani:1993gk} Collaborations excluded the mass for each generation of leptoquark (coupling to one generation of quark and lepton only) below 44 GeV. The OPAL Collaboration \cite {Abbiendi:2003iv} analysed the LEP data with integrated luminosity $(\mathcal L_{int})$  of 596 $\text{pb}^{-1}$ and centre of momentum energy ranging from  189 GeV to 209 GeV to provide the lower limit\footnote{As the study suggests, this limit depends on the representation of leptoquark.} on $M_\phi$ being below 100 GeV.  On the other hand, the last update from the DELPHI Collaboration \cite{Abreu:1998fw} used the LEP-2 data with $\sqrt s=$ 183 GeV and integrated luminosity of 47.4 $\text{pb}^{-1}$ and constrained the  couplings of scalar and vector leptoquarks with varying $M_\phi$. According to their analysis, the lower limit on the masses of first generation scalar (vector) leptoquarks with coupling $Y=\e$ and 100\% branching to charged lepton mode $(\beta)$ were 161 GeV (171 GeV) for $Q_\phi=1/3,\,5/3$ and 134 GeV (150 GeV) for  $Q_\phi=2/3,\,4/3$ at 95\% confidence level (C.L.). 

A large number of investigations for the existence of both scalar and vector leptoquarks  have  been performed by the H1 and ZEUS detectors at the \ep collider HERA too. In the last update from the H1 Collaboration \cite{Collaboration:2011qaa}, full data with $\mathcal L_{int}=$ 446 $\text{pb}^{-1}$ has been used to rule out the first generation of leptoquark with mass up to 800 GeV at 95\% C.L. for leptoquark coupling $Y=\e$. Their previous study \cite{Aaron:2011zz} was performed for full data sample at $\sqrt s=319$ GeV with integrated luminosity of 245 \pbi for $e^+p$ and 166 \pbi for $e^-p$ collisions. No evidence for leptoquark compelled them to exclude second and third generations of leptoquarks with masses under 712 GeV and 479 GeV respectively for couplings to every generation of quark and lepton being $Y=0.3\,$. Both the surveys present the upper limit on the coupling $Y$ as a function of $M_\phi$; moreover, Ref. \cite{Collaboration:2011qaa} displays allowed branching of the leptoquark to electron mode with its varying mass. The ZEUS Collaboration \cite{Abramowicz:2012tg} also carried out similar analysis and excluded first generation of leptoquark for masses up to 699 GeV with the coupling $Y=0.3\,$. However, their recent result \cite{Abramowicz:2019uti}, using high-precision data from HERA with 1 $\text{fb}^{-1}$ of integrated luminosity, has pushed the limit in TeV range; additionally the upper limit on coupling to mass ratio for individual leptoquark has also been put up. Nevertheless, it is interesting to mention that an anomalous measurement at high values of squared four-momentum transfer was observed by both of the H1 \cite{Adloff:1997fg} and ZEUS \cite{Breitweg:1997ff} Collaborations.

Leptoquarks have been looked for in hadronic colliders as well. The first measurement in this context was done at the CERN $p\bar p$ collider $\text S\text p\bar {\text p}\text S$. The UA2 detector \cite{Alitti:1991dn} analysed 13 \pbi data from the CERN $\text S\text p\bar {\text p}\text S$ collider at $\sqrt s=630$ GeV and set up a lower bound on the mass of first generation leptoquark at 67 GeV (76 GeV) with 95\% C.L. for the branching into the electron mode being 50\% (100\%). The CDF and D$\slashed {\text{O}}$ Collaborations have tried extensively to probe the signature of leptoquarks at other $p \bar p$ collider, the Fermilab Tevatron. In their final updates, CDF Collaboration \cite{Acosta:2005ge,Abulencia:2005ua} has worked with data collected from the Fermilab Tevatron collider at $\sqrt s=1.96$ TeV with  integrated luminosities of 203 \pbi and 198 \pbi respectively for pair production of first and second generation of scalar leptoquarks. The lack of any significant signal event led to set the lower bounds on masses of first and second generation scalar leptoquarks being 236 GeV (205 GeV) and  226 GeV (208 GeV) for $\beta=1(0.5)$ respectively at 95\% C.L. The CDF Collaboration \cite{Aaltonen:2007rb} have also looked over the Fermilab Tevatron data with $\mathcal L_{int}=322$ \pbi in search for a third vector leptoquark decaying to a $b$-quark and $\tau$-lepton. Agreement of observation with SM prediction resulted in exclusion of third generation vector leptoquarks below 317 GeV (251 GeV) of mass, assuming Yang-Mills (minimal) couplings\footnote{Usually, a dimensionless parameter $\kappa(\equiv 1- \kappa_G^{})$ is introduced while describing the interaction of a vector leptoquark with gluons \cite{Blumlein:1996qp}. It is related to the \textit{anomalous chromo-magnetic moment} and \textit{anomalous chromo-electric dipole moment} of the vector leptoquark. The scenario of \textit{Yang-Mills} coupling is represented by $\kappa=1$ whereas $\kappa=0$ indicates \textit{minimal} coupling case.}. The D$\slashed {\text{O}}$ Collaboration  \cite{Abazov:2008np,Abazov:2010wq,Abazov:2011qj} too has looked for pair production of first, second and third generations of scalar leptoquarks separately, based on the data set form the Fermilab Tevatron at $\sqrt s=1.96$ TeV. Due to non-observance of any excess over the SM expectation, they constrain the three generations of leptoquarks to be heavier than 326 GeV ($\beta=0.5$), 316 GeV (for $\beta=1$ or 270 GeV for $\beta=0.5$) and 247 GeV ($Q_\phi=-1/3$) respectively at 95\% C.L.
 
However, the most stringent bounds on the leptoquarks so far have been provided by the ATLAS and CMS Collaborations, working with $pp$ collisions at the LHC. Using a data set corresponding to $\mathcal{L}_{int}=36.1$ \fbi and $\sqrt s=13$ TeV, ATLAS Collaboration \cite{Aaboud:2019jcc} has tried to probe the pair production of  scalar leptoquarks in first and second generation. Absence of any statistically significant evidence for excess over SM allowed for exclusion of first and second generation scalar leptoquarks lighter than 1400 GeV (1290 GeV) and 1560 GeV (1230 GeV) at 95\% C.L. presuming $\beta=1\,(0.5)\,$. The same data set has also been used to rule out the third generation of scalar leptoquarks having masses below 800 GeV independent of any branching ratio and masses under 1 TeV  for $\beta=0 \text{ or }1 $ \cite{Aaboud:2019bye}. On the other hand, CMS Collaboration \cite{Sirunyan:2018btu,Sirunyan:2018ryt} has also analysed the LHC data at same $\sqrt s$, but with an integrated luminosity of $35.9$ \fbi and restricted the first and second generation scalar leptoquark masses to higher than 1435 GeV (1270 GeV) and 1530 GeV (1285 GeV) respectively for $\beta=1\,(0.5)$ at 95\% C.L. With the help of data collected from LHC at $\sqrt s=$ 13 TeV and $\mathcal L_{int}=35.9$ \fbi CMS Collaboration \cite{Sirunyan:2018nkj,Sirunyan:2018vhk} has also explored the possibility of pair production for third generation scalar leptoquarks. Due to inadequacy of signal event over SM background, they declare at 95\% C.L. that the third generation scalar leptoquarks  decaying to a top-quark and a $\tau$-lepton should have masses greater than 900 GeV and the same decaying to a bottom-quark and a $\tau$-lepton should have masses above 1020 GeV for $\beta=1$. The CMS Collaboration \cite{Sirunyan:2018kzh} has looked into the neutrino modes of decay for both scalar and vector leptoquarks too. These invisible modes put strong limit on the vector leptoquarks. If the vector leptoquark has 50\% branching fraction to $t\nu$ mode and rest rest 50\% to $b\tau$ channel, then mass of it below 1530 GeV (1115 GeV) is excluded for $\kappa=1 \, (\kappa=0)$.

Most of the analysis mentioned above assumes the leptoquark to interact with one generation of quark and  lepton only. However, when the leptoquark is supposed to have interactions with all generations of quarks and leptons, the branching fraction to different modes get reduced and the bounds thus become more relaxed. In our simulation, we have searched for a 70 GeV leptoquark at HERA, a 900 GeV leptoquark at LHeC and a 1.5 TeV (2.0 TeV) leptoquark at FCC-he (run II) through the phenomenon of RAZ. For the light leptoquark, we choose the couplings to be significantly smaller than electromagnetic coupling in order to satisfy the cross-section limit by ZEUS Collaboration \cite{Chekanov:2003af}, as shown by the first diagram in Fig. \ref{fig:limit}. Furthermore, the couplings are adjusted to obey the CDF bounds \cite{Abe:1995fj,Abe:1996dn} in such a way that branching to the second generation of quark and lepton remains below 23\% and the same to $b\tau$ or $\bar b\tau$ mode\footnote{Actually, to obey the CDF plots in Fig \ref{fig:limit}, one has to satisfy $\sigma_{p\bar p\to \phi\bar \phi}\times \mathcal B_{b\tau/\bar b\tau} \leq$ 20 pb where the symbol $\mathcal B$ is used to represent branching fraction throughout the text.} stays under 22\%, as depicted by the second and third diagrams in the first row of Fig \ref{fig:limit}. For the scalar leptoquarks of mass 900 GeV, we pick the couplings respecting the bounds by ATLAS \cite{Aaboud:2019bye} and CMS \cite{Sirunyan:2018btu,Sirunyan:2018ryt} Collaborations coming from charged lepton decay modes  since the restrictions from neutrino modes \cite{Sirunyan:2018kzh} are already satisfied in these cases. These constraints are illustrated in the second row of Fig. \ref{fig:limit}. On the other hand, for the vector leptoquarks of same mass, we consider the bounds from invisible modes \cite{Sirunyan:2018kzh} with $\kappa=0$, as shown in the third row of Fig. \ref{fig:limit}. The theoretical prediction for the cross-section of vector leptoquark pair production in Yang-Mills scenario is so high that it is quite impossible to meet the observed limits of cross-section times branching fraction squared for all the three generations of neutrinos, and therefore, we deliberately omit the $\kappa=1$ case from our consideration. On the other hand, it is easily comprehensible from Fig \ref{fig:limit} that there is effectively no restriction on the branching fractions of scalar and vector leptoquarks beyond mass 1.5 TeV. So, we adopt equal coupling for all the allowed interactions in case of 1.5 TeV (2.0 TeV) scalar and vector leptoquarks. Table \ref{tab:BP} specifies all the benchmark points used in our simulation. It should also be noticed from Table \ref{tab:BP} that the interactions of  $\widetilde S_1$ and $\widetilde U_{1\mu}$ with quarks and leptons are entirely right-handed whereas the same for $\vec S_3$ and $\vec U_{3\mu}$ are fully left-handed. We emphasize on the fact that each coupling in our analysis is smaller than the electromagnetic coupling $\e$. It is worth mentioning that we have not introduced any cross-generation coupling, i.e. the $3\times3$ coupling matrices $Y_L$ and $Y_R$ are taken diagonal. The branching fractions for different leptoquarks produced at mentioned \ep colliders for the benchmark points specified in Table \ref{tab:BP} are exhibited in Table \ref{tab:BF}. We have not shown the branching fractions for BP4 scenario explicitly since they differ from BP3 cases by 0.1\% only. It should be noted that $\tau t$, $\tau \bar t$, $\nu_\tau t$ and $\nu_\tau \bar t$  decay modes are not accessible in case BP1 due to scarcity of enough phase space. The production cross-sections for all the leptoquarks associated with a photon are listed in Table \ref{tab:CS}. Actually, since the hard-scattering processes involve real photon emission the total cross-sections are divergent\footnote{Cross-section for any process incorporating real photon emission from massless charged particle suffers from infra-red (IR) and collinear divergences. Though the IR divergence can be controlled by considering a tiny but non-vanishing mass for the radiating charged particle, the collinear divergence still remains.}. To avoid the singularity, we choose a suitable cut on the transverse momentum of photon $(p_T^\gamma)$ being larger than 20 GeV for our entire simulation.


\begin{table}[h!]
	\begin{tabular*}{0.49\textwidth}{@{\hspace*{4mm}\extracolsep{\fill}}lllllll}
		\hline\\[-2mm]
		$\phi$&$Y_L^{11}$& $Y_L^{22}$& $Y_L^{33}$& $Y_R^{11}$& $Y_R^{22}$& $Y_R^{33}$\\[1mm]
		\hline\\[-1mm]
		
		\multicolumn{7}{l}{\textbf{BP1: HERA, $\bm{M_\phi=70}$ GeV}}\\[1.5mm]
		$R_2$&0.035&0.018&0.017&0.035&0.018&0.017\\[0.5mm]
		$\widetilde S_1$&---&---&---&0.035&0.022&0.022\\[0.5mm]
		$\vec S_3$&0.035&0.022&0.022&---&---&---\\[0.5mm]
		$V_{2\mu}$&0.035&0.022&0.022&0.035&0.022&0.022\\[0.5mm]
		$\widetilde U_{1\mu}$&---&---&---&0.035&0.018&0.017\\[0.5mm]
		$\vec U_{3\mu}$&0.035&0.018&0.017&---&---&---\\[3mm]
		
		\multicolumn{7}{l}{\textbf{BP2: LHeC, $\bm{M_\phi=900}$ GeV}}\\[1.5mm]
		$R_2$&0.200&0.200&0.200&0.200&0.200&0.200\\[0.5mm]
		$\widetilde S_1$&---&---&---&0.140&0.140&0.200\\[0.5mm]
		$\vec S_3$&0.140&0.140&0.200&---&---&---\\[0.5mm]
		$V_{2\mu}$&0.100&0.100&0.200&0.200&0.200&0.200\\[0.5mm]
		$\widetilde U_{1\mu}$&---&---&---&0.140&0.140&0.200\\[0.5mm]
		$\vec U_{3\mu}$&0.140&0.140&0.200&---&---&---\\[3mm]
		
		\multicolumn{7}{l}{\textbf{BP3 (BP4): FCC-I (II), $\bm{M_\phi=1.5\,(2.0)}$ TeV}}\\[1.5mm]
		$R_2$&0.200&0.200&0.200&0.200&0.200&0.200\\[0.5mm]
		$\widetilde S_1$&---&---&---&0.200&0.200&0.200\\[0.5mm]
		$\vec S_3$&0.200&0.200&0.200&---&---&---\\[0.5mm]
		$V_{2\mu}$&0.200&0.200&0.200&0.200&0.200&0.200\\[0.5mm]
		$\widetilde U_{1\mu}$&---&---&---&0.200&0.200&0.200\\[0.5mm]
		$\vec U_{3\mu}$&0.200&0.200&0.200&---&---&---\\[2mm]
		\hline
	\end{tabular*}
	\caption{Chosen couplings to different generations of quarks and leptons for various leptoquarks satisfying the experimental bounds discussed in Sect. \ref{Sec:BP}}
	\label{tab:BP}
\end{table}


\begin{table}
	\renewcommand{\arraystretch}{1}
		\setlength\tabcolsep{4pt}
			\setlength\extrarowheight{2pt}
	\begin{tabular*}{0.49\textwidth}{@{\extracolsep{\fill}}ccccccccc}
		\hline
		Decay &\multicolumn{3}{c}{Branching Fractions} &  &Decay
		&\multicolumn{3}{c}{Branching Fractions}\\
		\cline{2-4} \cline{7-9}
		modes&\textbf{BP1}&\textbf{BP2}&\textbf{BP3}&& 
		modes &\textbf{BP1}&\textbf{BP2}&\textbf{BP3}\\
		\cline{1-4} \cline{6-9}\\[-3mm]
		
		\multicolumn{4}{l}{\textbf{Scalar leptoquarks}}&&\multicolumn{4}{l}{\textbf{Vector leptoquarks}}\\[1mm]
		
		\multicolumn{4}{l}{Leptoquark $(\widetilde S_1^{\nicefrac{+4}{3}})^c$}&&\multicolumn{4}{l}{Leptoquark $(\widetilde U_1^{\nicefrac{+5}{3}})^c$}\\
		$e\,d$&0.560&0.247&0.333&&$e \, \bar u$&0.791&0.254&0.335\\
		$\mu \, s$&0.221&0.247&0.333&&$\mu \, \bar c$&0.209&0.254&0.335\\
		$\tau \,b$&0.219&0.505&0.333&&$\tau \, \bar t$&---&0.491&0.331\\[2mm]
		
		\multicolumn{4}{l}{Leptoquark $(R_2^{\nicefrac{+5}{3}})^c$}&&\multicolumn{4}{l}{Leptoquark $(V_2^{\nicefrac{+4}{3}})^c$}\\
		$e \, \bar u$&0.791&0.342&0.335&&$e\,d$&0.559&0.278&0.333\\
		$\mu \, \bar c$&0.209&0.342&0.335&&$\mu \, s$&0.221&0.278&0.333\\
		$\tau \, \bar t$&---&0.317&0.330&&$\tau \,b$&0.220&0.444&0.333\\[2mm]
		
		\multicolumn{4}{l}{Leptoquark $(R_2^{\nicefrac{+2}{3}})^c$}&&\multicolumn{4}{l}{Leptoquark $(V_2^{\nicefrac{+1}{3}})^c$}\\
		$e\,\bar d$&0.362&0.169&0.167&&$e\,u$&0.314&0.225&0.167\\
		$\mu\,\bar s$&0.096&0.169&0.167&&$\mu\,c$&0.124&0.225&0.167\\
		$\tau\,\bar b$&0.085&0.169&0.167&&$\tau\,t$&---&0.213&0.164\\
		$\nu_e\bar u$&0.362&0.169&0.167&&$\nu_e d$&0.314&0.056&0.167\\
		$\nu_\mu \bar c$&0.096&0.169&0.167&&$\nu_\mu s$&0.124&0.056&0.167\\
		$\nu_\tau \bar t$&---&0.156&0.165&&$\nu_\tau b$&0.123&0.225&0.167\\[2mm]
		
		\multicolumn{4}{l}{Leptoquark $(S_3^{\nicefrac{+4}{3}})^c$}&&\multicolumn{4}{l}{Leptoquark $(U_3^{\nicefrac{+5}{3}})^c$}\\
		$e\,d$&0.560&0.247&0.333&&$e \, \bar u$&0.791&0.254&0.335\\
		$\mu \, s$&0.221&0.247&0.333&&$\mu \, \bar c$&0.209&0.254&0.335\\
		$\tau \,b$&0.219&0.505&0.333&&$\tau \, \bar t$&---&0.491&0.331\\[2mm]
		
		\multicolumn{4}{l}{Leptoquark $(S_3^{\nicefrac{+1}{3}})^c$}&&\multicolumn{4}{l}{Leptoquark $(U_3^{\nicefrac{+2}{3}})^c$}\\
		$e\,u$&0.314&0.126&0.167&&$e\,\bar d$&0.362&0.125&0.167\\
		$\mu\,c$&0.124&0.126&0.167&&$\mu\,\bar s$&0.096&0.125&0.167\\
		$\tau\,t$&---&0.239&0.165&&$\tau\,\bar b$&0.085&0.256&0.167\\
		$\nu_e d$&0.314&0.126&0.167&&$\nu_e\bar u$&0.362&0.125&0.167\\
		$\nu_\mu s$&0.124&0.126&0.167&&$\nu_\mu \bar c$&0.096&0.125&0.167\\
		$\nu_\tau b$&0.123&0.257&0.167&&$\nu_\tau \bar t$&---&0.242&0.165\\[2mm]
		
		\multicolumn{4}{l}{Leptoquark $(S_3^{\nicefrac{-2}{3}})^c$}&&\multicolumn{4}{l}{Leptoquark $(U_3^{\nicefrac{-1}{3}})^c$}\\
		$\nu_e u$&0.717&0.257&0.335&&$\nu_e \bar d$&0.667&0.247&0.333\\
		$\nu_\mu c$&0.283&0.257&0.335&&$\nu_\mu\bar  s$&0.176&0.247&0.333\\
		$\nu_\tau t$&---&0.487&0.330&&$\nu_\tau\bar b$&0.157&0.505&0.333\\[1mm]
		\hline
	\end{tabular*}
	\caption{Branching fractions of different leptoquarks for masses and couplings specified in Table \ref{tab:BP}. Branching fractions for BP4 scenario have not been displayed explicitly because they differ from BP3 case at third decimal places only.}
	\label{tab:BF}
\end{table}

\renewcommand{\arraystretch}{1.2}
\begin{table*}[h!]
	\setlength\tabcolsep{0pt}
	\setlength\extrarowheight{2pt}
	\begin{tabular*}{\textwidth}{@{\hspace*{1mm}\extracolsep{\fill}}lcccccccccc}
		\hline
		&\multicolumn{5}{c}{\textbf{Scalar Leptoquarks}}&\multicolumn{5}{c}{\textbf{Vector Leptoquarks}}\\
		\cline{2-6}\cline{7-11}
		\multirow{2}{*}{Collider} & Singlet & \multicolumn{2}{c}{Doublet} & \multicolumn{2}{c}{Triplet} & Singlet & \multicolumn{2}{c}{Doublet} & \multicolumn{2}{c}{Triplet} \\
		\cline{2-2}\cline{3-4}\cline{5-6}\cline{7-7} \cline{8-9} \cline{10-11}
		& $(\widetilde{S}_1^{\nicefrac{+4}{3}})^c$ & $(R_2^{\nicefrac{+5}{3}})^c$ & $(R_2^{\nicefrac{+2}{3}})^c$ & $(S_3^{\nicefrac{+4}{3}})^c$ & $(S_3^{\nicefrac{+1}{3}})^c$  & $(\widetilde{U}_{1\mu}^{\nicefrac{+5}{3}})^c$ & $(V_{2\mu}^{\nicefrac{+4}{3}})^c$ & $(V_{2\mu}^{\nicefrac{+1}{3}})^c$ & $(U_{3\mu}^{\nicefrac{+5}{3}})^c$ & $(U_{3\mu}^{\nicefrac{+2}{3}})^c$ \\ 
		\cline{1-1}\cline{2-6}\cline{7-11} 
		
		BP1: HERA & 21.40 & 5.57 & 8.45 & 21.44 & 57.19   & 5.04 & 76.53 & 199.73 & 5.04 & 7.50   \\ [1mm]
		
		BP2: LHeC & 0.171 & 0.036 & 0.027 & 0.17 & 0.91   & 0.017 & 0.871 & 7.43 & 0.017 & 0.013   \\ [1mm]
		
		BP3: FCC $\mathrm{I}$ & 0.934 & 0.067 & 0.047 & 0.93 & 3.8 & 0.066 & 3.7 & 15.1 & 0.066 & 0.045   \\ [1mm]
		
		BP4: FCC $\mathrm{II}$ & 2.77 & 0.32 & 0.23 & 2.7 & 8.2   & 0.319 & 10.9 & 32.6 & 0.32 & 0.23   \\ [1mm]
		\hline
	\end{tabular*}
	\caption{Production cross-sections (in fb) of various leptoquarks associated with a photon at different \ep colliders for different benchmark points after applying the cut $p_T^\gamma>20$ GeV. As already mentioned earlier, $S_3^{\nicefrac{-2}{3}}$ and $U_{3\mu}^{\nicefrac{-1}{3}}$ can not be produced at \ep collider due to the gauge structure of the Lagrangian.}
	\label{tab:CS}
\end{table*}


\section{Set up for the collider simulation}
\label{Sec:collider}

As explained in Sect.~\ref{Sec:Th}, only six leptoquarks, i.e. $(\widetilde{S}_1^{\nicefrac{+4}{3}})^c$, $(R_2^{\nicefrac{+5}{3}})^c$, $(S_3^{\nicefrac{+4}{3}})^c$,   $(\widetilde{U}_{1\mu}^{\nicefrac{+5}{3}})^c$,  $(V_{2\mu}^{\nicefrac{+4}{3}})^c$ and $(U_{3\mu}^{\nicefrac{+5}{3}})^c$ have zeros within the observable limit in their differential angular distributions while getting produced at \ep collider associated with a photon. In this analysis we reconstruct these leptoquarks via their decay products in the visible channel, i.e. to a charged lepton and a quark. In a process of $e\, p \to \phi\,\gamma$, the final state requires to have  at least one hard photon, a hard charged lepton and a quark jet. In order to remove the SM background, we choose the charged lepton to be muon. It is also worth noticing that for $ \vec S_3^c, \,R_2^c, V_{2\mu}^c$  and $\vec{U}_{3\mu}^c$ there are two states which can give rise to the similar decay products, e.g. $\bar c\, \mu$ in case of $(R^{\nicefrac{+5}{3}}_2)^c$ and $\bar s\, \mu$ in case of $(R^{\nicefrac{+2}{3}}_2)^c$, while only one of them produces the zero in the angular distribution. To avoid such unwanted model backgrounds we need to separate the jets by reconstructing their charge. We emphasize here again that for models  $\widetilde{S}_1$ and  $\widetilde{U}_{1\mu}$ no such model background exists. 

In this section we will describe the set up for the collider analysis which will be carried out for three 
different colliders namely HERA \cite{19436,Klein:2008di},  LHeC and FCC-he \cite{Agostini:2020fmq,Bordry:2018gri}. 
For this purpose we first implement the models at SARAH-4.14.2 \cite{Staub:2013tta} and then prepare the model files for CalcHEP \cite{Belyaev:2012qa}. Determination of decay branching fractions for leptoquarks, estimation of production cross-sections associated with a photon and event generation are executed via CalcHEP. It is worth mentioning that cross-section estimation and event generation are done with the help of NNPDF parton distribution functions \cite{Nocera:2014gqa} at renormalization or factorization scale of  $\sqrt{\hat{s}}$ where $\hat s$ is the parton level centre of momentum energy. For the analysis, we fed the CalcHEP-generated ``.lhe'' files to PYTHIA8\cite{Sjostrand:2007gs,Sjostrand:2014zea} in order to simulate the events with FastJET-3.0.3 \cite{Cacciari:2011ma} with following criteria: 

\begin{enumerate}
	\item Although the detector extends up to $|\eta| < 4.5$, we are mostly interested in the central events for HERA and therefore consider all the stable particles with $|\eta| < 3.5$. However, as the lab frames for LHeC and FCC-he are highly boosted relative to the respective CM frames,  we select $|\eta| < 4.5$ for all the stable particles in those two experiments to ensure that the zeros of angular distributions do not lie inside the amputated region.
	\item The radius of the jet is chosen to be $R = 0.5$, with the following cuts:
	\begin{itemize}
		\item Minimum transverse momentum $p_{T,min}^{jet} = 20.0$ GeV.
		\item Jets are reconstructed solely from stable hadrons.
	\end{itemize}
	\item The detected stable leptons and photons satisfy the following cuts:
	\begin{itemize}
		\item Minimum transverse momentum $p_{T,min}^{} = 10.0$ GeV.
		\item Detected leptons are hadronically clean, \textit{i.e,} hadronic activity within a cone of $\Delta R < 0.3$ around each lepton is less than $15\%$ of the leptonic transverse momentum; they are distinctly registered from the simultaneously produced jets, $\Delta R_{lj} > 0.4$; and well distinguished from other stable leptons, if any, $\Delta R_{ll} > 0.2$, where $\Delta R_{i j}=\sqrt{\Delta \eta_{ij}^2 + \Delta \phi_{ij}^2} $. 
		\item Detected photons are hadronically clean, \textit{i.e,} hadronic activity within a cone of $\Delta R < 0.2$ around each photon is less than $15\%$ of the photonic transverse momentum; they are distinctly registered from the simultaneously produced jets, $\Delta R_{\gamma j} > 0.2$; and well distinguished from other stable leptons, $\Delta R_{\gamma l} > 0.2$. 
	\end{itemize}
\end{enumerate} 
In the next three sections we discuss the analysis including the angular distributions at HERA, LHeC and FCC-he along with additional cuts. 


\section{The Hadron-Electron Ring Accelerator (HERA)}
\label{Sec:hera}

Motivated by radiation amplitude zero which would give vanishing differential cross-section at some specific zones of phase space, we revisit the Hadron-Electron Ring Accelerator (HERA)\cite{19436,Klein:2008di} for low mass leptoquarks. We select only those SU(2) singlet, doublet and triplet representations of leptoquarks for which the zeros fall within the observed region of phase space. The analysis follows the prescription described before. HERA originally ran for the centre of momentum energy of 318.12 GeV \cite{Klein:2008di} where electron with energy 27.5 GeV collided with a proton of 920.0 GeV giving rise to an asymmetric collider as shown in Table~\ref{HERAspc}. Each of the H1 and ZEUS experiments collected data with integrated luminosities of around 120 pb$^{-1}$ for $e^+ p$ and 15 pb$^{-1}$ for $e^- p$ collisions till 2000. Since 2003, an upgraded version of HERA, HERA II, has recorded events with net integrated luminosity of 200 pb$^{-1}$ for both $e^+ p$ and $e^- p$ collisions. For the studies of the zeros of amplitude we project the required luminosity for the 5$\sigma$ signal significance over the total background which includes SM background and the model background both.

\begin{table}[!htb]
	\begin{center}
		\renewcommand{\arraystretch}{1.2}
		\begin{tabular*}{0.49\textwidth}{@{\hspace*{2mm}\extracolsep{\fill}}ccccc}\hline 
			$E_p$ & $E_{e^-}$ & $\sqrt{s}$ & $\mathcal{L}_{int}$ & $\mathcal{L}_{int}^\text{projected}$ \\ \hline 
			920 GeV & 27.5 GeV & 318.1 GeV & 400 pb$^{-1}$ & 100 fb$^{-1}$ \\ \hline
		\end{tabular*}
		\caption{Beam and Centre of Mass energies along with integrated luminosities at HERA.}\label{HERAspc}
	\end{center}
\end{table} 

In the following  subsections we describe different possible scenarios as mentioned in section~\ref{Sec:BP} with the corresponding benchmark points and their collider studies at HERA. As evoked before, each leptoquark $SU(2)_L$ multiplet exhibiting RAZ has been presented in the subsequent subsections with respective $SU(3)_C,~SU(2)_L$ representations and $U(1)_Y$ charge explicitly mentioned. Since scalar-doublet has two excitation among which one has zero within the observed range of $\cos{\theta}$ with possible model contamination from the other, we would address that scenario before hand. Here, we would also implement the jet-charge reconstruction to reduce such model contamination. 

\subsection{Scalar Doublet \texorpdfstring{$\bm{R_2}\,(\bm{3,2,}\frac{7}{3})$}{}}
\label{Sub:HeraScalDoub}

The scalar doublet leptoquark $R_2\, (\bm{3,2,}\frac{7}{3})$ couples with the SM fermions through Yukawa like couplings given by:
\begin{equation}
\label{eq:lagR2}
-\mathfrak{L}\supset Y_L\, \xbar{\uq}_R \left(i\sigma^2 R_2\right)^T \LL_L + Y_R\, \xbar{\Q}_L\, R_2\, \cl_R+ h.c.
\end{equation}
In our notation, $\Q_L$ and $\LL_L$ are weak isospin doublets for left-handed quarks and leptons, and $\uq_R$, $\dq_R$ and $\cl_R$ are weak isospin singlets for right-handed up-type quarks, down-type quarks and charged leptons respectively. It is evident from Eq. \eqref{eq:lagR2} that both the components of the doublet $R_2^c$, namely $(R_2^{\nicefrac{+5}{3}})^c$ and $(R_2^{\nicefrac{+2}{3}})^c$ will be produced at \ep collider simultaneously. However, during their associated production with a photon, the former one displays a zero in its angular distribution within the kinematically allowed region of phase space (to be specific, at $\cos\theta=-\frac{1}{5}$) while the later component does not, as shown in Table \ref{tab:LQ}.

Now, both the leptoquarks, produced after \ep collision, 
will eventually decay to a quark and a lepton. Based on the decay, three types of leptonic final states are possible:
\begin{itemize}
	\item The first one, leptoquark decays to first generation of quark and lepton. But this channel suffers from enormous irreducible Standard Model background. 
	
	\item The second one, leptoquark decays to third generation quark and lepton. But, in order to reconstruct the leptoquark in this case, the unstable $\tau$-lepton needs to be reconstructed from its hadronic decay products. This, along with the reconstruction of simultaneously produced top-quark from leptoquark decay will yield several stable jets and increase the combinatorial error. Moreover, reconstruction of top is a difficult task. Also, a 70 GeV leptoquark decaying to a top quark is kinematically forbidden.
	
	\item We hence consider the leptoquark decaying into second generation of quark and lepton as shown below:
	\begin{align}
	e\,\bar{u} \to ({R_2^{\nicefrac{+5}{3}}})^c\, \gamma \to \mu\, \bar{c} ~\gamma \,, \nonumber \\
	e\, \bar{d} \to ({R_2^{\nicefrac{+2}{3}}})^c\, \gamma \to \mu\,\bar{s} ~\gamma \,. 
	\label{eqn:HeraScDoublet}
	\end{align}
	It minimizes the SM Background, and the solo challenge remaining is to optimize the signal $(R_2^{\nicefrac{+5}{3}})^c$ over the model background $(R_2^{\nicefrac{+2}{3}})^c$. During the process of signal background separation, misidentification of $c$- and $s$-jets can lead to model contamination which can be resolve by determining the charge of the jets as discussed later in this section. 
\end{itemize}

\begin{figure*}[t]
\subfigure[]{\includegraphics[width=0.48\textwidth]{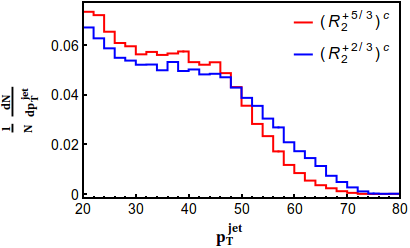}}
	\hfil
		\subfigure[]{\includegraphics[width=0.48\textwidth]{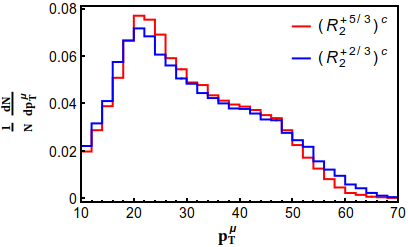}}

\subfigure[]{\includegraphics[width=0.48\textwidth]{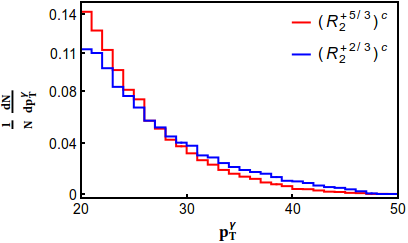}}
	\hfil
		\subfigure[]{\includegraphics[width=0.48\textwidth]{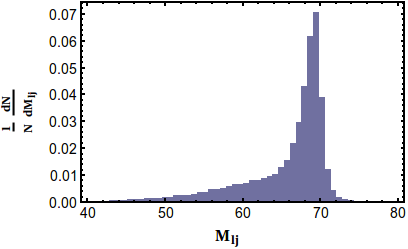}}
	\caption{$p^j_T$, $p^\ell_T$, $p^\gamma_T$ and $M_{\ell \, j}$ distributions at HERA with $\sqrt s=318.12$ GeV for the leptoquark $R_2^c$ with mass $70$ GeV. }\label{scalrdbDis}
\end{figure*} 

\begin{figure}[h]
	\centering
	\includegraphics[width=0.45\textwidth]{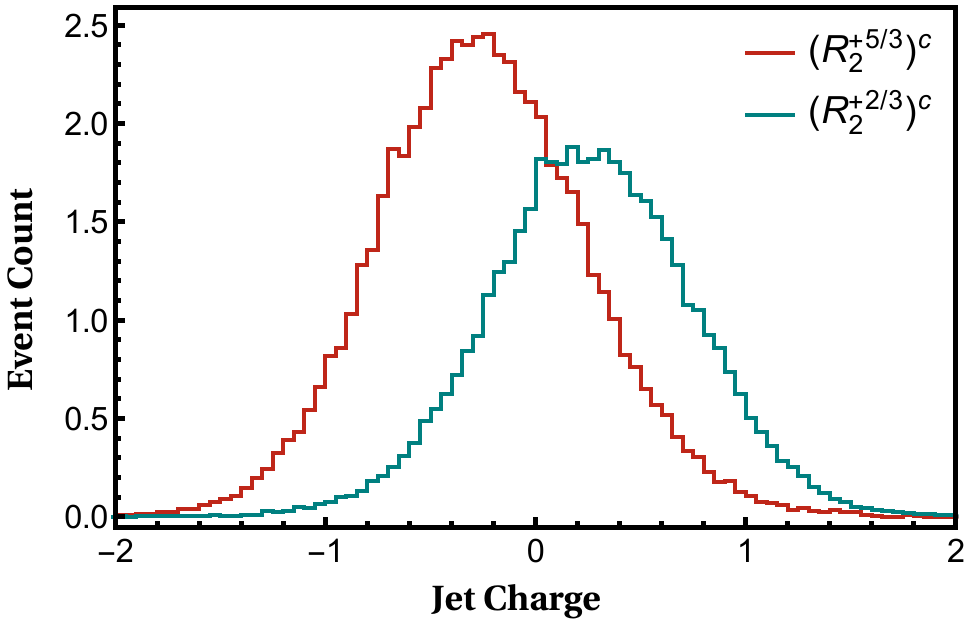}
	\caption{Determination of the charge (in the unit of $\e$) for the $c$- and $s$-jets coming from $(R^{\nicefrac{+5}{3}}_2)^c$ and $(R^{\nicefrac{+2}{3}}_2)^c$ respectively at HERA.}
	\label{HERAJCharge}
\end{figure}

\begin{figure}[!htb]
	\centering
		\subfigure[With a bin-width of 0.04]{\includegraphics[width=0.45\textwidth]{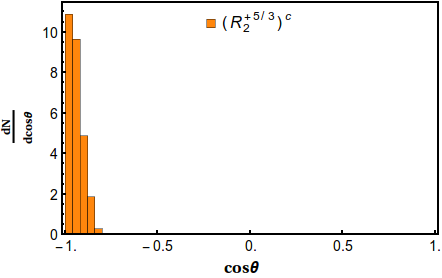}}  
		\subfigure[With a bin-width of 0.002]{\includegraphics[width=0.45\textwidth]{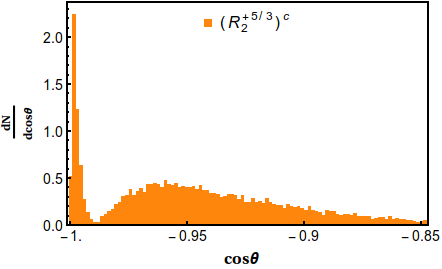}}  
	\caption{Distribution of the scattered photon, produced in association with $\left( R_2^{\nicefrac{+5}{3}} \right)^c$ with respect to the angle $(\theta)$ it makes with the incoming electron beam, at $\sqrt{s} = 318.12$ GeV and $\mathcal{L}_{int} = 100 \text{ fb}^{-1}$ in the Lab frame. We present here the same data with varying bin widths. The subfigure(a) presents the angular distribution with the bin-width of 0.04 for the variation of $\cos\theta$ used in plotting the distribution in CM frame for all cases in this paper. However this makes the signature of amplitude zero washed out due to high boost and we decrease the bin-width to 0.002 in subfigure(b) to obtain the manifestation of the amplitude zero, however, shifted due to the boost.}
	\label{HERARAZ0}
\end{figure}


\begin{figure}[h!]
	\centering
		\subfigure[$({R_2^{\nicefrac{+5}{3}}})^c$]{\includegraphics[width=0.45\textwidth]{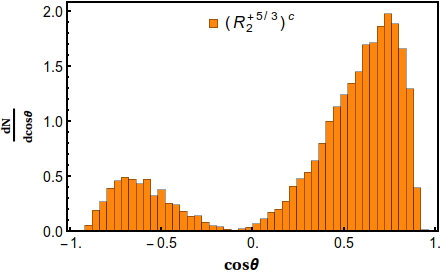}}  
		\subfigure[$({R_2^{\nicefrac{+2}{3}}})^c$]{\includegraphics[width=0.45\textwidth]{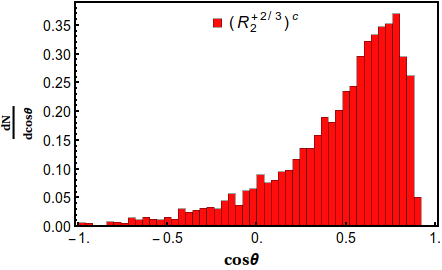}}  
		\subfigure[Combined (signal + background)]{\includegraphics[width=0.45\textwidth]{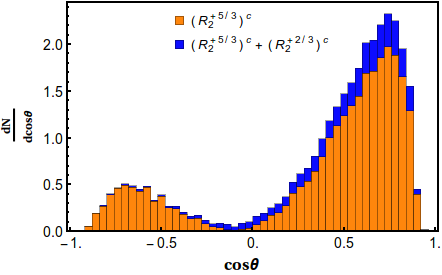}} 
	\caption{Angular distribution with respect to the angle $(\theta)$ made by the photon with the incoming electron beam, at $\sqrt{s} = 318.12$ GeV and $\mathcal{L}_{int} = 100 \text{ fb}^{-1}$ in the CM frame. The sub-figure (a) shows the angular distribution for the associate production of $({R_2^{\nicefrac{+5}{3}}})^c$, whereas the sub-figure (b) displays the same for $({R_2^{\nicefrac{+2}{3}}})^c$. In the sub-figure (c), we present the angular distributions of the signal and background together in the rest frame of interaction.}
	\label{HERAdoubCos}
\end{figure}

\begin{table}[h]
	\begin{center}
		\renewcommand{\arraystretch}{1.6}
		\begin{tabular*}{0.49\textwidth}{@{\hspace*{2mm}\extracolsep{\fill}}ccc}\hline 
			\multicolumn{3}{c}{$\mathcal{B}(R_2^c \to \mu ~\bar{c}/\bar{s})$: BP1} \\ \hline
			\multirow{2}{*}{Cuts} & Signal & SM + \\
			& $({R_2^{\nicefrac{+5}{3}}})^c$ & $({R_2^{\nicefrac{+2}{3}}})^c$  \\ \hline
			$\geqslant 1\mu + 1j + 1\gamma$ & 77.2 & 58.5 \\ 
			$|M_{\ell j}- M_{R_2}|\leq 10 \,\rm{GeV}$ & \multirow{2}{*}{59.2} & \multirow{2}{*}{44.7} \\ 
			$+ 1\gamma_{p_T > 20\text{GeV}}$ & &  \\
			$Q_{Jet} < -0.3$ & 27.7 & 5.3   \\ \hline
			$\sigma_{Sig} (\mathcal{L}_{int} = 100 \text{ fb}^{-1})$ & \multicolumn{2}{c}{4.8} \\ 
			$\sigma_{Sig} (\mathcal{L}_{int} = 400 \text{ pb}^{-1})$ & \multicolumn{2}{c}{0.3} \\ 
			$\mathcal{L}_{5\sigma} (\text{ in fb}^{-1})$ & \multicolumn{2}{c}{108.5} \\ \hline 
		\end{tabular*}
		\caption{Number of signal and background events after the cumulative cuts for the scalar doublet leptoquark $R_2^c$ at $\sqrt{s}=318.12$ GeV and $\mathcal{L}_{int} = 100 \text{ fb}^{-1}$. Significances with $\mathcal{L}_{int} = 100 \text{ fb}^{-1} \text{ and } 400 \text{ pb}^{-1} $ as well as integrated luminosity required for $5\sigma$ significance at HERA are also estimated.}\label{scaldubNum2}
	\end{center}
\end{table} 

For the collider study, we consider $(R_2^{\nicefrac{+5}{3}})^c \to \mu\, \bar c$  and $(R_2^{\nicefrac{+2}{3}})^c \to \mu\, \bar s $ with the corresponding benchmark point BP1 in Table~\ref{tab:BP}. The branching fractions and production cross-sections associated with a photon $(p_T^\gamma\geq 20\text { GeV})$ are mentioned in Table~\ref{tab:BF} and \ref{tab:CS} respectively. In Figure~\ref{scalrdbDis}(a) -- Figure~\ref{scalrdbDis}(c), we describe the kinematic distributions for the transverse momenta ($p_T$) of jet, lepton and photon respectively. It can be seen that jet, lepton and photon are hard enough to satisfy the basic cuts demanded in section~\ref{Sec:collider}. One important aspect to notice here is that the distributions for $(R^{\nicefrac{+5}{3}}_2)^c$ and $(R^{\nicefrac{+2}{3}}_2)^c$ with respect to the transverse momenta of jet, lepton and photon almost superimpose on each other and it is quite impossible to separate the signatures of these two components of the doublet by applying any cut on any of the transverse momenta.

To determine the four momenta of finalstate particles we consider only the visible decay modes of the leptoquarks i.e., to muon and quark, which forms jet. Information of the three-momenta of the finalstate particles enables us to determine the the boost axis which includes boost along the $z$-axis and in the traverse direction instrumental for reconstructing the CM frame. The RAZ information to distinguish different Leptoquark representations is valid only in the centre of mass frame. The boost effect of the Lab frame can smear or distort such distributions. Here we show for the case of  $\left( R_2^{\nicefrac{+5}{3}} \right)^c$ that how the distributions in the lab frame get smeared in comparison with the CM frame and for the rest of the examples we only show the boosted back distributions in the CM frame. 

In Fig.\ref{HERARAZ0} we show the angular distribution of the photon with the incoming electron in Lab frame produced in association with $\left( R_2^{\nicefrac{+5}{3}} \right)^c$. As we notice, due to substantial asymmetry of the colliding beams, photons thus produced are highly boosted and are all directed to the opposite direction of the electron beam. With normal bin-width (of 0.04 for the variation of $\cos\theta$) which is considered for showing the angular distribution of photon in CM frame, Figure~\ref{HERARAZ0}(a) implies that the zero in distribution gets washed away due to the high boost of the Lab frame. However, with a smaller bin-width of 0.002 in Figure~\ref{HERARAZ0}(b) we able to see the RAZ in a different position of $\cos{\theta}$. We recall here that the position of zero in the angular distribution is independent of the energy of interaction in the CM frame. Thus even though the parton distribution function changes the energy of the parton level interaction, the distribution in the partonic centre of mass frame should remain unaltered. Even so, the observed finalstates in the Lab frame are not in partonic rather a result of e-proton interaction governed by the parton distribution function and boosted along the the z-axis. The  small transverse boost can be estimated  by the uncertainty principle. This causes the smearing in the distributions as can be seen from Figure~\ref{HERARAZ0}(b), where the RAZ happens around $\cos{\theta} = -0.99$ in the Lab frame, whereas in the reconstructed the CM frame the corresponding distribution has RAZ  around $\cos{\theta} = -0.20$ as predicted by the theory. This effect is generic and manifests for all other leptoquarks considered in this study for different collision energies. We henceforth would present only the distributions of photon as observed in boosted-back centre of mass frame.
	

As already mentioned, the rest frame is reconstructed from tagging the 4-momenta of hard photon, produced in association with a leptoquark which is reconstructed from its decay product of muon and a quark-jet. The event topologies are principally comprised of exactly one hard photon, one hard lepton and jet. Hence the error in boosting back is minimal. We can propose a physical quantity to quantify the error in determining the deep in Lab/CM frame as follows  
\begin{equation}
\Delta_{\rm Lab/CM}= \frac{|\zeta^{\rm left }_{\rm Lab/CM} - \zeta^{\rm right }_{\rm Lab/CM} | }{ \zeta^{\rm left }_{\rm Lab/CM} + \zeta^{\rm right }_{\rm Lab/CM}}
\end{equation}
where $\zeta^{\rm left/right}$ refers to the number of events within the number (within the minimum and closest peak) of bins in the left/right of the minimum in the respective frames. For the associated production of $\left( R_2^{\nicefrac{+5}{3}} \right)^c$ at HERA with a data set of 100 fb$^{-1}$, we see that the smearing of the  boosted minima  is around $\sim 61\%$ in the Lab frame; which is reduced to $\sim 15 \%$ in the reconstructed CM frame with the expected position of the minima as predicted by the theory.

Next we consider the reconstruction of leptoquark mass via the invariant mass ($M_{lj}$) distribution of jet and lepton. For this purpose, we considered each event with at least one muon, one jet and one photon satisfying the above mentioned cuts. We took all possible combinations of jet and lepton to evaluate the invariant mass and plot them in Figure~\ref{scalrdbDis}(d). The peak around 70 GeV is clearly visible  and we select events within $\pm 10$ GeV  range of the peak. It is very crucial for the reconstruction of the CM frame \cite{Bandyopadhyay:2020klr,Bandyopadhyay:2020wfv}. 

However, as mentioned before $(R^{\nicefrac{+5}{3}}_2)^c$ and $(R^{\nicefrac{+2}{3}}_2)^c$ can be indistinguishable in case we cannot separate $c$- and $s$-jets properly. For this reason we need to determine the charge of the jets from their constituents  \cite{Krohn:2012fg,Tokar:2017syr,Sirunyan:2017tyr}. In Figure~\ref{HERAJCharge}
we show the charge distributions of the jets coming from $(R^{\nicefrac{+5}{3}}_2)^c$ and $(R^{\nicefrac{+2}{3}}_2)^c$. Though the jet charges are not peaking exactly at $-2\e/3$ or $\e/3$, it can be observed that the jet charge for the signal, $({R_2^{\nicefrac{+5}{3}}})^c \to \mu\,\bar{c}\,$, peaks at a negative charge while the  model background, $({R_2^{\nicefrac{+2}{3}}})^c \to \mu\,\bar{s}\,$ reaches the maximum around some positive charge. Both these maxima are well-separated and hence distinguishable. For our simulation, we have considered the events with jet charge $\leqslant-0.3\e$ to optimise the signal over the model background in case of $R_2^c$, as shown later.

Once the leptoquark is reconstructed, we determine the cosine of the angle between final state hard photon and the incoming electron beam in the lab frame and then boost it back to the CM frame of interaction. We impose three additional cuts as mentioned in Table~\ref{scaldubNum2}, i.e. $\big(\geqslant 1\mu + 1j + 1\gamma\big)\; + \; \big(|M_{\ell j}- M_{R_2}|\leq 10\, \rm{GeV}  \;+\; 1\gamma_{p_T > 20\text{GeV}}\big) \;+\; \big(Q_{Jet} < -0.3\big)$. The final state event numbers after each cumulative cuts for  $(R^{\nicefrac{+5}{3}}_2)^c$ (signal) and $\text{SM}+(R^{\nicefrac{+2}{3}}_2)^c$ (total background) at 
$\sqrt{s} = 318.12$ GeV  and $\mathcal L_{int}=$ 100 \fbi are presented in Table~\ref{scaldubNum2}. Significances with $\mathcal{L}_{int} = 100 \text{ fb}^{-1} \text{ and } 400 \text{ pb}^{-1} $  as well as the integrated luminosity required for $5\sigma$ significance at HERA are also estimated. It can be noticed that a signal significance of $4.8 \sigma$ can be achieved with an integrated luminosity of $100 \text{ fb}^{-1}$. 

Now, we look for the angular distribution with respect to the angle ($\theta$) between final state photon and initial state electron in CM frame. Figure~\ref{HERAdoubCos} illustrates the angular distributions for (a) $(R^{\nicefrac{+5}{3}}_2)^c$, (b) $(R^{\nicefrac{+2}{3}}_2)^c$ and (c) the combined one, i.e. $(R^{\nicefrac{+5}{3}}_2)^c + (R^{\nicefrac{+2}{3}}_2)^c $, in the CM frame of HERA at $\sqrt{s} = 318.12$ GeV with an integrated luminosity of 100 fb$^{-1}$. In Figure~\ref{HERAdoubCos}(a) a dip in the angular distribution for $(R^{\nicefrac{+5}{3}}_2)^c$ is clearly visible, whereas in Figure~\ref{HERAdoubCos}(b) no such minimum is observed for $(R^{\nicefrac{+2}{3}}_2)^c$. The combined plot in  Figure~\ref{HERAdoubCos} (c) still shows the minimum around the same point. However, it should be noticed that the dip is not exactly at $\cos\theta=-\frac{1}{5}$, as mentioned in Table \ref{tab:LQ}, rather it is a bit shifted. The uncertainty in constructing the CM frame due to parton distribution function is responsible for this minute alteration in the position of RAZ.

\subsection{Scalar Singlet \texorpdfstring{$\bm{\widetilde {S}_1\,(\bm{\bar {3},1,}\frac{8}{3})}$}{}}

With the scalar singlet leptoquark $\widetilde{S}_1 \,(\bm{\bar {3},1,}\frac{8}{3})$, the situation gets much simpler as we have only one excitation and there is no possibility of model contamination. This leptoquark couples to SM fermions through Yukawa like coupling given by,
\begin{equation}
\label{eq:lagS1t}
-\mathfrak{L} \supset Y_R\, \xbar{\dq}^c_R\, \widetilde{S}_1\,\cl_R  + h.c.
\end{equation} 
When this leptoquark is produced in association with a photon during the \ep collision, its differential distribution with respect to the angle ($\theta$) made by the photon with the incoming electron beam should exhibit a zero  at $\cos\theta = -\frac{1}{2}$. Similar to the previous case, here also we concentrate in the final state invoking muon in order to eliminate the SM backgrounds. The benchmark point (BP1) for the simulation is described in Table~\ref{tab:BP}. The branching fractions and the production cross-sections associated with a photon $(p_T^\gamma\geq 20\text { GeV})$ for the study at HERA are given in Table~\ref{tab:BF} and \ref{tab:CS} respectively.

Unlike the doublet case, here we have only one excitation leading to the final state topology in electron-hadron collision which constitutes of a muon, a $s$-quark and a photon:
\begin{equation}
e\, p \to (\widetilde{S}_1^{\nicefrac{+4}{3}})^c\, \gamma \to \mu\, s\, \gamma  .
\label{eqn:HeraScSinglet}
\end{equation}

\begin{figure}[h]
	\centering
	\includegraphics[width=0.45\textwidth]{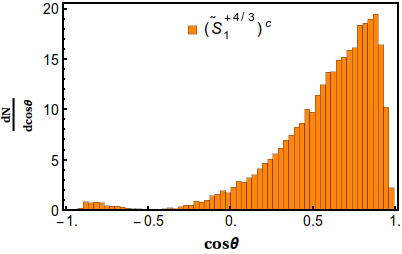}
	\caption{Angular distribution for the associated production of  reconstructed 70 GeV leptoquark $\widetilde{S}_1^c$ relative to photon angle $(\theta)$  made with electron beam, at $\sqrt{s} = 318.12$ GeV and $\mathcal{L}_{int} = 100 \text{ fb}^{-1}$ in the rest frame of interaction.}  \label{scLQdis}
\end{figure} 


Similar to the doublet leptoquark case, here also we reconstruct the scalar leptoquark mass via the invariant mass reconstruction of $\mu$-jet in order to go back to the CM frame. The angular distribution at the CM frame of HERA with $\mathcal L_{int}=$ 100 \fbi relative to the cosine of the angle between the hard photon and incoming electron  for the events with exactly one photon with $p^\gamma_T \geq 20$ GeV and exactly one reconstructed leptoquark is shown in Figure~\ref{scLQdis}. The minimum at $\cos{\theta}=-\frac{1}{2}$ is clearly visible. The numbers of events\footnote{Since this is a zero-background scenario, we do not mention the significance.} after each cuts for the above mentioned final states at an integrated luminosity 100 of fb$^{-1}$ are tabulated in Table~\ref{scalarLQHERA}.

\begin{table}[!htb]
	\begin{center}
		\renewcommand{\arraystretch}{1.6}
		
		\begin{tabular}{ccc}\hline
			\multicolumn{3}{c}{$\mathcal{B}(\widetilde{S}_1^c \to \mu~s)$: BP1} \\ \hline
			Cuts & Signal$(\widetilde{S}_1^{\nicefrac{+4}{3}})^c$ & Background \\ \hline
			$\geqslant 1\mu + 1j + 1\gamma$ & 326.7 & 0.0  \\
			$\lvert M_{lj} -M_{\widetilde{S}_1} \rvert \leq 10\text{ GeV}$ & 267.9 & 0.0   \\
			$ + 1\gamma_{(p_T > 20\text{GeV})}$ & 263.5 & 0.0   \\ 
			\hline
		\end{tabular}
		\caption{Number of  events after the cumulative cuts for the scalar singlet leptoquark $\widetilde{S}_1^c$ and SM background at $\sqrt{s}=318.12$ GeV with $\mathcal{L}_{int} = 100 \text{ fb}^{-1}$ at HERA.}\label{scalarLQHERA}
	\end{center}
\end{table} 

\subsection{Scalar Triplet \texorpdfstring{$\bm{\vec{S}_3}\,(\bm{\bar{3},3,}\frac{2}{3})$}{}}

We now consider the scalar triplet leptoquark, $\vec{S}_3 \,(\bm{\bar{3},3,}\frac{2}{3})$ for our analysis. This leptoquark couples to the SM fermions through Yukawa like couling given by,
\begin{equation}
\label{eq:lagS3}
-\mathfrak{L} \supset Y_L\,\xbar{\Q}_L^c \left(i\sigma^2 \,S_3^{adj}\right) \LL_L + h.c.\,,
\end{equation}
where, $S_3^{adj}=\begin{pmatrix}
\frac{S_3^{\nicefrac{+1}{3}}}{\sqrt 2}&S_3^{{\nicefrac{+4}{3}}}\\S_3^{{\nicefrac{-2}{3}}}&-\frac{S_3^{{\nicefrac{+1}{3}}}}{\sqrt 2}
\end{pmatrix}$ signifies the triplet in the adjoint representation.

Of the three members of the triplet, only two, namely $(S_3^{\nicefrac{+4}{3}})^c$ and $(S_3^{\nicefrac{+1}{3}})^c$, can be produced  from \ep collision, since $S_3^{\nicefrac{-2}{3}}$ does not couple to electron at all, as can be realised from Eq. \eqref{eq:lagS3}. Again, only $(S_3^{\nicefrac{+4}{3}})^c$ shows RAZ at $\cos\theta = -\frac{1}{2}$ in CM frame whereas $(S_3^{\nicefrac{+1}{3}})^c$ does not and acts as model contamination. Following the same argument as before, we look for the leptoquark decaying to second generation lepton and quark, i.e $\mu \text{ and } s/c\,$. The interaction at the \ep collider for the associated  production with a photon and the decay of these leptoquarks are shown below:

\begin{align}
e\,p \to ({S_3^{\nicefrac{+4}{3}}})^c \,\gamma \to \mu\,s\,\gamma\,, \nonumber \\
e\,p \to ({S_3^{\nicefrac{+1}{3}}})^c\,\gamma \to \mu\,c \,\gamma\,. 
\label{eqn:HeraScTriplet}
\end{align}

We follow the same approach for event generation and collider simulation with the allowed benchmark point (BP1) in Table~\ref{tab:BP}, the branching fractions in Table~\ref{tab:BF} and the cross-sections associated with a photon $(p_T^\gamma\geq 20\text { GeV})$ in Table~\ref{tab:CS}.

\begin{figure}[h]
	\centering
	\includegraphics[width=0.45\textwidth]{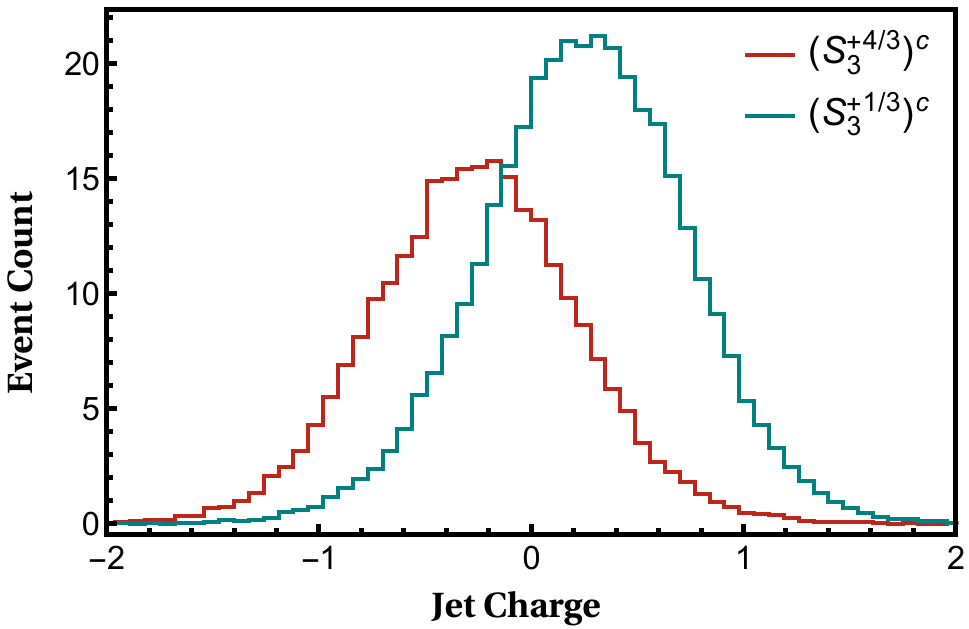}
	\caption{Reconstructed charge (in the unit of $\e$) of $c/s$-jets coming from the scalar triplet leptoquark decay.}\label{jetchsctrip}
\end{figure}
We see that while decaying, the signal leptoquark $(S_3^{\nicefrac{+4}{3}})^c$ produces a $c$-jet, whereas the background leptoquark $(S_3^{\nicefrac{+1}{3}})^c$ gives rise to a $s$-jet. So, we determine the jet-charge in order to separate the final states in a similar fashion as was done for scalar doublet $R_2^c$ in Sect \ref{Sub:HeraScalDoub}. From Figure~\ref{jetchsctrip} one can notice that the jets from the decay of $(S_3^{\nicefrac{+4}{3}})^c$ peaks at a negative charge whereas the same from $(S_3^{\nicefrac{+1}{3}})^c$ attains the maximum around some positive charge. Although the peaks of jet-charges are not exactly at $2\e/3$ or $-\e/3\,$, they are well-separated and hence can be recognized. For our analysis of $S_3^c$, we impose a cut and accept the events for which the jet-charge are negative, so that the signal can be optimized over the background.

After the leptoquark mass reconstruction from $\mu$-jet invariant mass, as explained before, we now determine the cosine of the angle of the hard photon with the incoming electron beam ($\cos\theta$) in the CM frame for the final state $\geq 1 \mu + 1j +1 \gamma\,$. In Table~\ref{scaltripHERA}, we tabulate the events coming from $(S_3^{\nicefrac{+4}{3}})^c$ and SM processes plus $(S_3^{\nicefrac{+1}{3}})^c$  at different stages of imposed cuts for $\mathcal L_{int}=$ 100 fb$^{-1}$. We determine the signal significance at integrated luminosities of 400 pb$^{-1}$ (HERA) and 100 fb$^{-1}$ and also predict the $\mathcal{L}_{int}$ required for a signal significance of 5$\sigma$. It is evident that the cuts substantially reduce the model background from the production of $(S_3^{\nicefrac{+1}{3}})^c$, whereas there is no SM background to be observed. 

\begin{figure}[!htb]
	\centering

		\subfigure[$(\vec{S}_3^{\nicefrac{+4}{3}})^c$]{\includegraphics[width=0.45\textwidth]{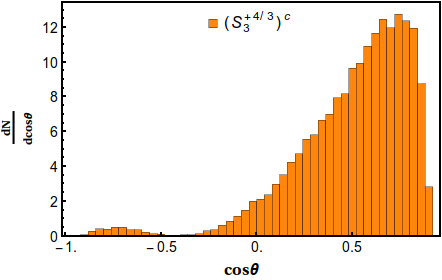}} 
		\subfigure[$(\vec{S}_3^{\nicefrac{+1}{3}})^c$]{\includegraphics[width=0.45\textwidth]{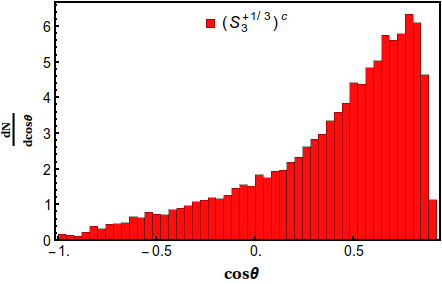}} 
\subfigure[Combined]{\includegraphics[width=0.45\textwidth]{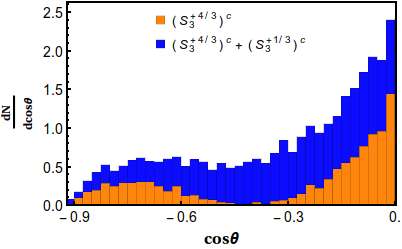}}
	\caption{Angular distribution for the associated production of reconstructed 70 GeV scalar triplet leptoquark $\vec S_3^c$ relative to the photon angle $(\theta)$ with electron beam, at $\sqrt{s} = 318.12$ GeV and $\mathcal{L}_{int} = 100 \text{ fb}^{-1}$ in CM frame. The sub-figure (a) shows the angular distribution for  $(S_3^{\nicefrac{+4}{3}})^c$, (b) shows the same for $(S_3^{\nicefrac{+1}{3}})^c$, and (c) depicts the combined distribution in the CM frame of interaction.}
	\label{HERAtripCos}
\end{figure}

After implementing the cuts of Table~\ref{scaltripHERA}, we plot the angular distributions for the leptoquark $\vec S_3^c$ in Figure~\ref{HERAtripCos}. Figure~\ref{HERAtripCos}(a) clearly shows a zero  around $\cos\theta = -\frac{1}{2}$ for $(S_3^{\nicefrac{+4}{3}})^c$, while Figure~\ref{HERAtripCos}(b) depicts no such zero for $(S_3^{\nicefrac{+1}{3}})^c$ in the angular distribution. Like the scalar doublet case, here also the zero is shifted from its theoretical position. We combine the signal and background together in Figure~\ref{HERAtripCos}(c) since both the components with same mass will be produced as parts of the triplet. As can be seen, the zero around $\cos\theta = -\frac{1}{2}$ is gone in case of combined plot, however, a dip is still noticeable. The analysis shows that a very early data of $\sim 21.8$ fb$^{-1}$ of integrated luminosity can probe the minimum in the angular distribution of photon for the scalar triplet leptoquark $\vec S_3^c$.

\begin{table}[!htb]
	\begin{center}
		\renewcommand{\arraystretch}{1.6}
		\begin{tabular*}{0.49\textwidth}{@{\hspace*{2mm}\extracolsep{\fill}}ccc}\hline
			\multicolumn{3}{c}{$\mathcal{B}(\vec{S}_3^c \to \mu\,s/c)$: BP1} \\ \hline
			\multirow{2}{*}{Cuts} & Signal & SM +  \\ 
			& $(S_3^{\nicefrac{+4}{3}})^c$ & $(S_3^{\nicefrac{+1}{3}})^c$ \\
			\hline
			$\geqslant 1\mu + 1j + 1\gamma$ & 328.5 & 520.2  \\ 
			$|M_{\ell j}- M_{\vec{S}_3}|\leq 10 \,\rm{GeV} $  & \multirow{2}{*}{263.3}  & \multirow{2}{*}{359.6}   \\ 
			$+ 1\gamma_{p_T > 20\text{GeV}}$ & & \\
			$Q_{Jet} < 0.0$ & 180.5 & 104.9  \\  \hline
			$\sigma_{Sig} (\mathcal{L}_{int} = 100 \text{ fb}^{-1})$ & \multicolumn{2}{c}{10.7} \\ 
			$\sigma_{Sig} (\mathcal{L}_{int} = 400 \text{ pb}^{-1})$ & \multicolumn{2}{c}{0.67} \\ 
			$\mathcal{L}_{5\sigma} (\text{ in fb}^{-1})$ & \multicolumn{2}{c}{21.8} \\ \hline
		\end{tabular*}
		\caption{Number of  events after the cumulative cuts for the scalar triplet leptoquark $\vec S_3^c$ with $\sqrt{s}=318.12$ GeV and $\mathcal{L}_{int} = 100 \text{ fb}^{-1}$. Significances with $\mathcal{L}_{int} = 100 \text{ fb}^{-1} \text{ and } 400 \text{ pb}^{-1} $ as well as integrated luminosity required for $5\sigma$ significance at HERA are also estimated. }\label{scaltripHERA}
	\end{center}
\end{table} 

\subsection{Vector Singlet \texorpdfstring{$\bm{\widetilde {U}_{1\mu}}\,(\bm{3,1,}\frac{10}{3})$}{}}

Having dealt with all the viable models of scalar leptoquarks, exhibiting zeros in their angular distributions, we now address the vector counterparts. We begin our discussion with the vector singlet $\widetilde{U}_{1\mu}\,(\bm{3,1,}\frac{10}{3})$. It couples with SM fermion through Yukawa like coupling given by,
\begin{equation}
-\mathfrak{L} \supset Y_R\,\xbar{\uq}_R\, \gamma^\mu \,\widetilde{U}_{1\mu}\,\cl_R + h.c.
\end{equation}
Here we have only one excitation, i.e. $\widetilde{U}_1$. This leptoquark, when gets produced in association with a photon from \ep collision, exhibits zero at $\cos\theta = -\frac{1}{5}\,$.

\begin{figure}[!htb]
	\centering
	\includegraphics[width=0.45\textwidth]{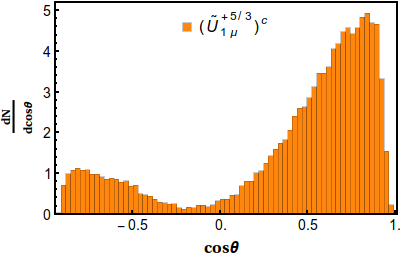}
	\caption{Angular distribution for the associated production of the reconstructed 70 GeV vector singlet leptoquark $\widetilde U_{1\mu}^c$ at $\sqrt{s} = 318.12$ GeV and $\mathcal{L}_{int} = 100 \text{ fb}^{-1}$ at the rest frame of interaction.}  \label{veclepdis}
\end{figure}

For collider study at HERA, we benchmark the scenario by BP1 in Tables \ref{tab:BP}, \ref{tab:BF} and \ref{tab:CS}. Following the reasoning in the preceding sections, we consider the leptoquark decay to only second generation lepton and quark. After \ep collision, this singlet will be produced in association with a photon and then decay to the desired fermions as follows:
\begin{align}
e\,p\to ({\widetilde{U}_{1\mu}^{\nicefrac{+5}{3}}})^c\, \gamma \to \mu \,\bar{c}\, \gamma. 
\label{Eq:HERAVcSinglet}
\end{align}
We simulated the events in PYTHIA8 and reconstitute the leptoquark via $c\,\mu$ invariant mass as before. The  angular distribution in the CM frame with respect to the cosine of the angle made by the hard photon with the incoming electron for  the events with $\big(\geqslant 1\mu + 1j + 1\gamma\big)\; + \; \big(|M_{\ell j}- M_{\widetilde{U}_{1\mu}}|\leq 10\, \rm{GeV}  \;+\; 1\gamma_{p_T > 20\text{GeV}}\big)$  is plotted  in Figure~\ref{veclepdis}. The existence of a minimum surrounding   $\cos\theta = -0.2$ is patently perceptible. The corresponding events\footnote{Like the scalar singlet case, here also we do not mention the significance, since it is a zero-background scenario.} at an integrated luminosity of 100 fb$^{-1}$ are then listed in Table~\ref{vecsingnum} with the cumulative cuts. The number of events seems very promising with a early data in the absence of any SM background.

\begin{table}[!htb]
	\begin{center}
		\renewcommand{\arraystretch}{1.6}
	\begin{tabular}{ccc}\hline 
			\multicolumn{3}{c}{$\mathcal{B}(\widetilde{U}_{1\mu}^c \to \mu^- ~\bar{c})$: BP1} \\ \hline
			Cuts & Signal$(\widetilde{U}_1^{\nicefrac{+5}{3}})^c$ & Background  \\ \hline
			$\geqslant 1\mu + 1j + 1\gamma$ & 74.8 & 0.0  \\ 
			$\lvert M_{lj} - M_{\widetilde{U}_{1\mu}} \rvert \leq 10\text{ GeV} $ & 54.9 & 0.0   \\
			$+ 1\gamma_{(p_T > 20\text{ GeV})}$ & 54.2 & 0.0  \\ \hline
		\end{tabular}
		\caption{Number of  events after the cumulative cuts for the vector singlet leptoquark $\widetilde U_{1\mu}^c$ with $\sqrt{s}=318$ GeV and $\mathcal{L}_{int} =  100 \text{ fb}^{-1}$ at HERA.}\label{vecsingnum}
	\end{center}
\end{table} 

\subsection{Vector Doublet \texorpdfstring{$\bm{V_{2\mu}}\,(\bm{\bar{3},2,}\frac{5}{3})$}{}}

We next take the vector doublet leptoquark $V_{2\mu}$ $(\bm{\bar{3},2,}\frac{5}{3})$. This leptoquark couples with the SM fermions through Yukawa like coupling given by,
\begin{equation}
-\mathfrak{L} \supset Y_L\xbar{\dq}_R^c\, \gamma^\mu \left(i\sigma^2 V_{2\mu}\right)^T \LL_L + Y_R\,\xbar{\Q}_L^c \,\gamma^\mu \left(i\sigma^2 V_{2\mu}\right)\cl_R + h.c.
\end{equation}
In this case, $(V_{2\mu}^{\nicefrac{+4}{3}})^c$, the member with isospin projection $+{\frac{1}{2}}$, shows zero in angular distribution about $\cos\theta = -\frac{1}{2}$ at CM frame, while $(V_{2\mu}^{\nicefrac{+1}{3}})^c$ exhibits a monotonically increasing graph of angular distribution with $\cos\theta$. The couplings, decay branching fractions and cross-sections $(p_T^\gamma\geq 20\text { GeV})$ are given in Tables \ref{tab:BP}, \ref{tab:BF} and \ref{tab:CS} respectively. 

The production channel associated with a photon and the decay final sates for $(V_{2\mu}^{\nicefrac{+4}{3}})^c$ and 
$(V_{2\mu}^{\nicefrac{+1}{3}})^c$ at \ep collider are given below:
\begin{align}
e\, p\to ({V_{2\mu}^{\nicefrac{+4}{3}}})^c\, \gamma \to \mu \,s \,\gamma \,, \nonumber \\
e\, p \to ({V_{2\mu}^{\nicefrac{+1}{3}}})^c \gamma \to \mu\, c\, \gamma\,. 	 		  
\label{eqn:HeraVcDoublet}
\end{align}
Similarly to previous cases, here also we determine the charge of the jets from the distributions given below in Figure~\ref{chdisvecdub}. We see in Table~\ref{v2HERAnum} that a demand of $Q_{Jet} < -0.3$  reduces the model contamination of $(V_{2\mu}^{\nicefrac{+1}{3}})^c$  substantially. 
\begin{figure}[!htb]
	\centering
	\includegraphics[width=0.45\textwidth]{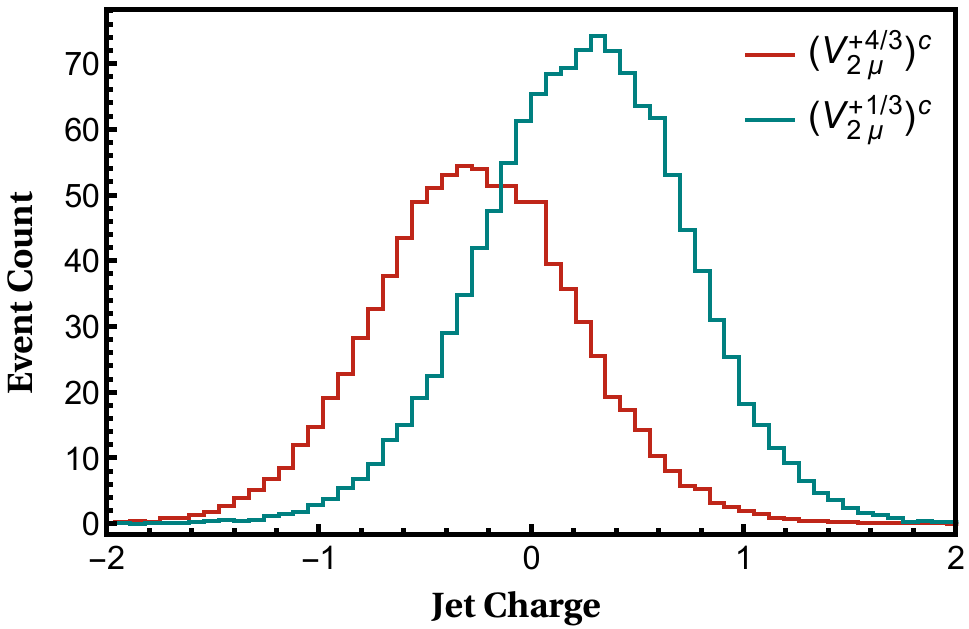}
	\caption{Charge (in the unit of $\e$) reconstruction of $c$- and $s$-jets from the vector doublet leptoquark decay.}\label{chdisvecdub}
\end{figure}

\begin{figure}[!htb]
	\centering
\subfigure[ $(V_{2\mu}^{\nicefrac{+4}{3}})^c$]{\includegraphics[width=0.45\textwidth]{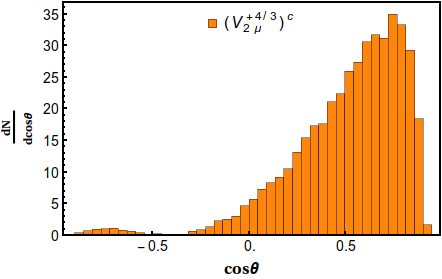}} \subfigure[$(V_{2\mu}^{\nicefrac{+1}{3}})^c$]{\includegraphics[width=0.45\textwidth]{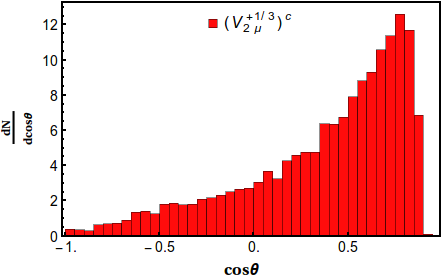}}
\subfigure[Combined]{\includegraphics[width=0.45\textwidth]{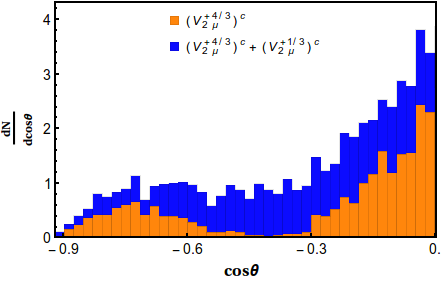}} 
	\caption{Angular distribution of photon with respect to the electron beam, at $\sqrt{s} = 318.12$ GeV and $\mathcal{L}_{int} = 100 \text{ fb}^{-1}$ for the associated production of  70 GeV vector doublet leptoquark $\vec V_{2\mu}^c$ in CM frame. The first plot  shows the angular distribution of $(V_{2\mu}^{\nicefrac{+4}{3}})^c$, the second one shows the same for $(V_{2\mu}^{\nicefrac{+1}{3}})^c$, and the last one exhibits the angular distribution for signal and background together in the rest  frame of interaction.}
	\label{HERAV2Cos}
\end{figure}
Equipped with all the cuts we then reconstruct the leptoquark mass from the $j\,\mu$ invariant mass distribution as explained before and plot the angular distribution of the hard photon  with the incoming electron axis in the CM frame for the final sate with $\geqslant 1\mu + 1j + 1\gamma$. The angular distributions for  $(V_{2\mu}^{\nicefrac{+4}{3}})^c$, $(V_{2\mu}^{\nicefrac{+1}{3}})^c$ and the combined scenario are depicted in Figure~\ref{HERAV2Cos} by the sub-figures (a), (b) and (c) respectively. It is apparent that for $(V_{2\mu}^{\nicefrac{+4}{3}})^c$ we see a zero in the neighbourhood of $\cos\theta = -\frac{1}{2}$ which is still reflected in the combined plot as a minimum of the distribution. 

The number of events for such final states with cumulative cuts are given in Table~\ref{v2HERAnum} at integrated luminosity of 400 pb$^{-1}$ and 100 fb$^{-1}$. It can be seen a very early data of $\sim 8$ fb$^{-1}$ can give us a $5\sigma$ signal significance. 
\begin{table}
	\begin{center}
		\renewcommand{\arraystretch}{1.6}
		\begin{tabular*}{0.49\textwidth}{@{\hspace*{2mm}\extracolsep{\fill}}ccc}\hline
			\multicolumn{3}{c}{$\mathcal{B}(V_{2\mu}^c \to \mu\,s/c)$: BP1} \\ \hline
			\multirow{2}{*}{Cuts} & Signal & SM + \\
			& $({V_{2\mu}^{\nicefrac{+4}{3}}})^c$ & $({V_{2\mu}^{\nicefrac{+1}{3}}})^c$  \\\hline
			$\geqslant 1\mu + 1j + 1\gamma$ & 1167.6 & 1818.5  \\ 
			$+ \lvert M_{lj} - M_{V_{2\mu}} \rvert \leq 10\text{ GeV} $ & \multirow{2}{*}{929.8} & \multirow{2}{*}{1257.8} \\ 
			$+ 1\gamma_{p_T > 20\text{GeV}}$ & &  \\
			$Q_{Jet} < -0.3$ & 435.5 &  157.0  \\ \hline 
			$\sigma_{Sig} (\mathcal{L}_{int} = 100 \text{ fb}^{-1})$ & \multicolumn{2}{c}{17.9} \\
			$\sigma_{Sig} (\mathcal{L}_{int} = 400 \text{ pb}^{-1})$ & \multicolumn{2}{c}{1.1} \\
			$\mathcal{L}_{5\sigma} (\text{ in fb}^{-1})$ & \multicolumn{2}{c}{7.8} \\ \hline 
		\end{tabular*}
		\caption{Number of  events after the cumulative cuts for the vector doublet leptoquark $V_{2\mu}^c$ at $\sqrt{s}=318.12$ GeV and $\mathcal{L}_{int} = 100 \text{ fb}^{-1}$. Significances with $\mathcal{L}_{int} = 100 \text{ fb}^{-1} \text{ and } 400 \text{ pb}^{-1} $ as well as integrated luminosity required for $5\sigma$ significance at HERA are also estimated. }\label{v2HERAnum}
	\end{center}
\end{table} 

\subsection{Vector Triplet \texorpdfstring{$\bm{\vec {U}_{3\mu}}\,(\bm{3,3,}\frac{4}{3})$}{}}

We finally consider the last model for our analysis, the vector triplet leptoquark $\vec{U}_{3\mu}\,(\bm{3,3,}\frac{4}{3})$. Two out of three members of the triplet, with isospin projections +1 and 0, can be produced at \ep collider. This leptoquark couples with the SM fermions through Yukawa like coupling given by,
\begin{equation}
-\mathfrak{L} \supset Y_L\,\xbar{\Q}_L \,\gamma^\mu\, U_{3\mu}^{adj}\, \LL_L +h.c.\,,
\end{equation}
where, $U_{3\mu}^{adj}=\begin{pmatrix}
\frac{U_{3\mu}^{{\nicefrac{+2}{3}}}}{\sqrt 2}&U_{3\mu}^{{\nicefrac{+5}{3}}}\\U_{3\mu}^{{\nicefrac{-1}{3}}}&-\frac{U_{3\mu}^{{\nicefrac{+2}{3}}}}{\sqrt 2}
\end{pmatrix}$ symbolizes the triplet in adjoint representation.

The interaction at \ep collider for the production and decay of these leptoquarks are,
\begin{align}
e\, p \to (U_{3\mu}^{\nicefrac{+5}{3}})^c \,\gamma \to \mu\, \bar{c}\, \gamma\,,  \nonumber \\
e\, p\to (U_{3\mu}^{\nicefrac{+2}{3}})^c \,\gamma \to \mu\, \bar{s}\, \gamma\,. 
\label{eqn:HeraVcTriplet}
\end{align}
\begin{figure}[h!]
	\centering
	\includegraphics[width=0.45\textwidth]{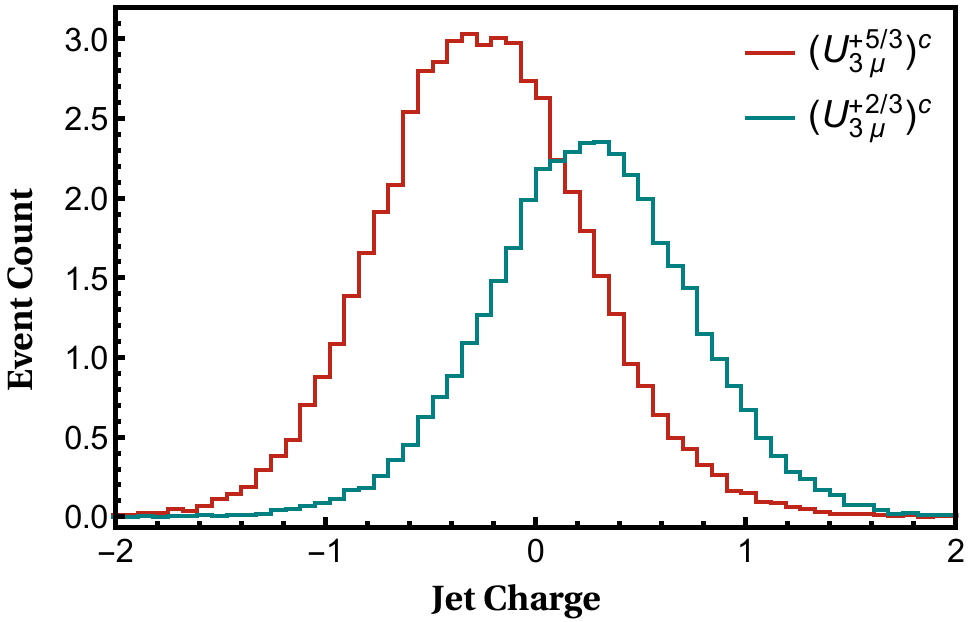}
	\caption{Charge distribution (in the unit of $\e$) of $c,\, s$-jets coming from the vector triplet leptoquark decays.}\label{chjetvectrip}
\end{figure}
The benchmark point (BP1) for the collider study along with the decay branching fractions and cross-sections with $ p_T^\gamma\geq 20\text { GeV}$ are given by Tables~\ref{tab:BP}, \ref{tab:BF} and \ref{tab:CS} respectively. Like the scalar triplet case, here also we have two excitations which can be produced i.e. $(U_{3\mu}^{\nicefrac{+5}{3}})^c$ and  $(U_{3\mu}^{\nicefrac{+2}{3}})^c$. Similar to the previous cases here also we need to identify the charges of jets in order to distinguish $(U_{3\mu}^{\nicefrac{+5}{3}})^c$ from $(U_{3\mu}^{\nicefrac{+2}{3}})^c$. For this purpose the jet charge distributions are plotted in Figure~\ref{chjetvectrip}.

\begin{figure}[h!]
	\centering
		\subfigure[$(U_{3\mu}^{\nicefrac{+5}{3}})^c$]{\includegraphics[width=0.45\textwidth]{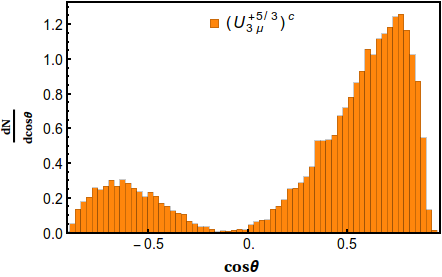}}
		\subfigure[$(U_{3\mu}^{\nicefrac{+2}{3}})^c$]{\includegraphics[width=0.45\textwidth]{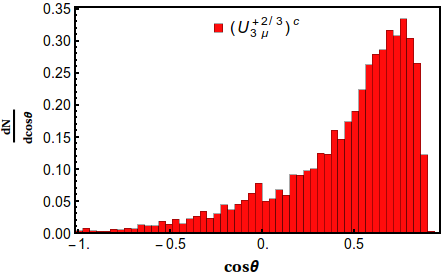}}
		\subfigure[Combined]{\includegraphics[width=0.45\textwidth]{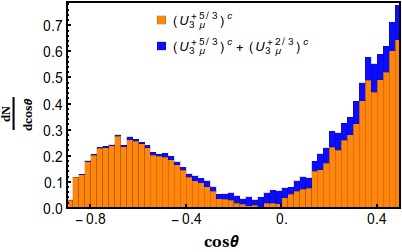} }
	\caption{Angular distribution of photon with respect to the electron beam, at $\sqrt{s} = 318.12$ GeV and $\mathcal{L}_{int} = 100 \text{ fb}^{-1}$ for the associated production of  70 GeV vector triplet leptoquark $\vec U_{3\mu}^c$ in CM frame. The first plot  shows the angular distribution of $(U_{3\mu}^{\nicefrac{+5}{3}})^c$, the second one shows the same for $(U_{3\mu}^{\nicefrac{+2}{3}})^c$, and the last one exhibits the angular distribution for signal and background together in the rest  frame of interaction.}
	\label{HERAU3Cos}
\end{figure}
For the analysis of signal and background we again chose $\geqslant 1\mu + 1j + 1\gamma$ final sate and reconstruct the leptoquark via the peak in invariant mass distribution of $\mu j$.  We have implemented the first three cuts of Table~\ref{NumvectripHERA} where $Q_{Jet} < -0.3$ is instrumental in reducing the model contamination coming from $(U_{3\mu}^{\nicefrac{+2}{3}})^c$. 

Ready with the set up we plot the differential distribution with respect to the angle between the hard photon and the electron in CM frame, shown in Figure~\ref{HERAU3Cos}. Figure~\ref{HERAU3Cos}(a) depicts the minimum of distribution for $(U_{3\mu}^{\nicefrac{+5}{3}})^c$ about $\cos{\theta}=-\frac{1}{5}$, whereas Figure~\ref{HERAU3Cos}(b) for $(U_{3\mu}^{\nicefrac{+2}{3}})^c$ does not show any, as expected. We combine the distributions in Figure~\ref{HERAU3Cos}(c) and the dip can still be observed.  

\begin{table}[h!]
	\begin{center}
		\renewcommand{\arraystretch}{1.5}
		\begin{tabular*}{0.49\textwidth}{@{\hspace*{2mm}\extracolsep{\fill}}ccc}\hline 
			\multicolumn{3}{c}{$\mathcal{B}(\vec{U}_{3\mu}^c \to \mu\,\bar{c}/\bar{s})$: BP1} \\ \hline
			\multirow{2}{*}{Cuts} & Signal & SM + \\
			&  $(U_{3\mu}^{\nicefrac{+5}{3}})^c$&$+(U_{3\mu}^{\nicefrac{+2}{3}})^c$ \\ \hline
			$\geqslant 1\mu + 1j + 1\gamma$ & 69.4 & 52.1 \\ 
			$ \lvert M_{lj} - M_{V_2} \rvert \leq 10\text{ GeV} $& \multirow{2}{*}{53.6}  & \multirow{2}{*}{39.6}  \\ 
			$ 1\gamma_{p_T > 20\text{GeV}}$ & &  \\
			$Q_{Jet} < -0.3$ & 24.9  & 4.8  \\
			\hline
			$\sigma_{Sig} (\mathcal{L}_{int} = 100 \text{ fb}^{-1})$ & \multicolumn{2}{c}{4.6} \\ 
			$\sigma_{Sig} (\mathcal{L}_{int} = 400 \text{ pb}^{-1})$ & \multicolumn{2}{c}{0.28} \\ 
			$\mathcal{L}_{5\sigma} (\text{ in fb}^{-1})$ & \multicolumn{2}{c}{118.1} \\ \hline 
		\end{tabular*}
		\caption{Number of  events after the cumulative cuts for the vector triplet leptoquark $\vec{U}_{3\mu}^c$ at $\sqrt{s}=318.12$ GeV and $\mathcal{L}_{int} = 100 \text{ fb}^{-1}$. Significances with $\mathcal{L}_{int} = 100 \text{ fb}^{-1} \text{ and } 400 \text{ pb}^{-1} $ as well as integrated luminosity required for $5\sigma$ significance at HERA are also estimated. }\label{NumvectripHERA}
	\end{center}
\end{table} 

In Table~\ref{NumvectripHERA}, we present the number of events after the cumulative cuts for $(U_{3\mu}^{\nicefrac{+5}{3}})^c$ and SM plus  $(U_{3\mu}^{\nicefrac{+2}{3}})^c$. The significance for $(U_{3\mu}^{\nicefrac{+5}{3}})^c$ are calculated at  $\sqrt{s}=318.12$ GeV: at $\mathcal{L}_{int} = 100 \text{ fb}^{-1},\,  400 \text{ pb}^{-1}$ at HERA. It is noticed that the signature of $(U_{3\mu}^{\nicefrac{+5}{3}})^c$  can be determined over the model contamination from $(U_{3\mu}^{\nicefrac{+2}{3}})^c$ at an integrated luminosity of $\sim 118$ fb$^{-1}$.

Clearly, as it follows from the Signal-Background table of events for all the leptoquark models exhibiting zeros in amplitude, with the size of the combined dataset of both the collaborations, H1 and ZEUS amounting to 400 pb$^{-1}$ for $e^- p$ collision, that it is not sufficient to observe a significant deviation for events exhibiting amplitude zeros. The highest significance obtained with this dataset is around 1.1$\sigma$ for the vector doublet leptoquark ($V_{2\mu}^c$) probe. However, for a futuristic collider operating with the same collision energy as HERA, if provide a dataset of size 100 fb$^{-1}$, the leptoquark signatures for $(\vec S_3)^c$ and $(V_{2\mu})^c$ will definitely be measured with more than $5\sigma$ significance through RAZ; the other two leptoquarks $(R_2)^c$ and $(\vec U_{3\mu})^c$ can also be observed with nearly $5\sigma$ significance for the same integrated luminosity.

\section{Large Hadron Electron Collider (LHeC)}
\label{Sec:LHeC}

\begin{figure*}[t!]
	\centering
\subfigure[]{\includegraphics[width=0.45\textwidth,height=5cm]{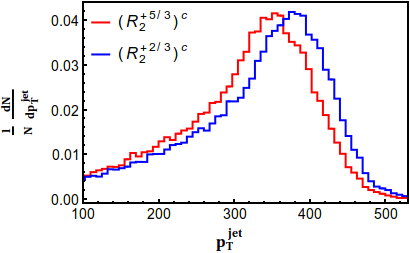}}
\hfil
\subfigure[]{\includegraphics[width=0.45\textwidth,height=5cm]{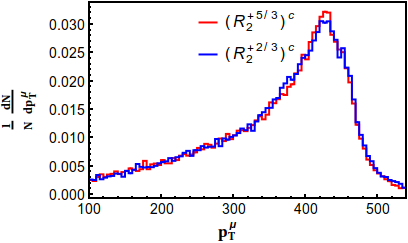}}

\subfigure[]{\includegraphics[width=0.45\textwidth,height=5cm]{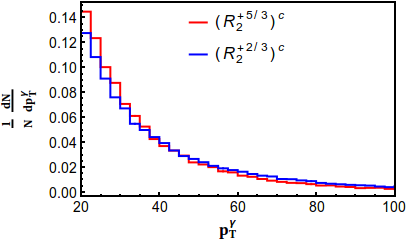}}
\hfil
\subfigure[]{\includegraphics[width=0.45\textwidth,height=5cm]{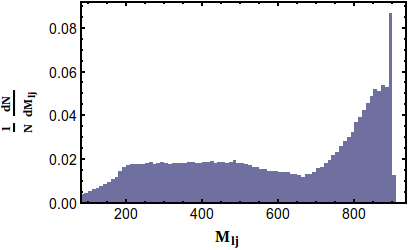}}

	\caption{Distributions of $p^j_T$, $p^\mu_T$, $p^\gamma_T$ and $M_{\ell \, j}$ at LHeC with $\sqrt s=1183.2$ GeV for the scalar-doublet leptoquark $R_2^c$ with mass 900 GeV. }
	\label{LHECSLQKin}
\end{figure*}

The Large Hadron Electron Collider (LHeC) is proposed at CERN, Geneva \cite{Agostini:2020fmq,Bordry:2018gri}. We present in Table~\ref{LHEcsp} the technicalities for Run 6 of LHeC, where a beam of $7$ TeV proton will be collided with a $50$ GeV electron beam giving rise to collisions at centre of momentum energy $\sim1.2$ TeV with a projected luminosity of $2000$ \fbi.
%
\begin{table}[h!]
	\begin{center}
		\renewcommand{\arraystretch}{1.2}
		\begin{tabular*}{0.49\textwidth}{@{\hspace*{2mm}\extracolsep{\fill}}cccc}\hline 
			$E_p$ & $E_{e}$ & $\sqrt{s}$ & $\mathcal{L}_{int}^\text{projected}$ \\ \hline 
			7 TeV & 50 GeV & 1183.2 GeV & 2000 fb$^{-1}$ \\ \hline
		\end{tabular*}
		\caption{Beam and Centre of Mass energies along with integrated luminosities at LHeC.}\label{LHEcsp}
	\end{center}
\end{table} 

For the collider simulation at LHeC, we choose the mass of leptoquarks to be 900 GeV with the couplings specified by BP2 scenario in Table \ref{tab:BP}. The corresponding branching fractions and production cross-sections associated with a photon $(p_T^\gamma\geq 20 \text{ GeV})$ for different leptoquarks are gathered in Table \ref{tab:BF} and \ref{tab:CS} respectively. It is important to mention that unlike HERA we had to select $|\eta|<4.5$ for all stable particles to confirm the occurrence of RAZ inside the detector since the lab frame of LHeC is highly boosted relative to the CM frame. As the position of RAZ in \ep collider depends only on the charge of produced leptoquark and the production channels as well as the decay modes, taken for the simulation at LHeC, are same  with the preceding section we present here a comparative analysis of all the viable scalar and vector leptoquark models instead of presenting them in an elaborated repetitive discussion.

Similar to the earlier collider, we start with the kinematical distributions of scalar doublet leptoquark $R_2^c$ shown in Figure \ref{LHECSLQKin}. The sub-figures (a), (b) and (c) portray
the distributions of $p_T$ for jet, muon and photon respectively. As anticipated, the distributions of $p_T^j$ and $p_T^\mu$ show peaks roughly around half the mass of leptoquark. On the other hand, the sub-figure (d) displays the distribution for invariant mass of jet-muon pair which peaks at the leptoquark mass. However, in this case also the transverse momentum distributions for $(R_2^{\nicefrac{+5}{3}})^c$ and $(R_2^{\nicefrac{+2}{3}})^c$  approximately lie on top of each other and hence one should go for the jet-charge determination for the separation of the signatures of these two components, as described in the last section. 

\begin{table*}[h!]
	\begin{center}
		\renewcommand{\arraystretch}{1.5}
		\scalebox{0.9}{
			\begin{tabular}{cccccccccccc}\hline 
				&\multicolumn{5}{c}{Scalar Leptoquark}&&\multicolumn{5}{c}{Vector Leptoquark}\\
				\cline{2-6}\cline{8-12}
				\multirow{2}{*}{Cuts} & Singlet & \multicolumn{2}{c}{Doublet} & \multicolumn{2}{c}{Triplet} && Singlet & \multicolumn{2}{c}{Doublet} & \multicolumn{2}{c}{Triplet} \\
				\cline{2-6}\cline{8-12} 
				& Signal &Signal& SM+ &Signal& SM+ &&Signal&Signal& SM+ &Signal& SM+\\
				& $(\widetilde{S}_1^{\nicefrac{+4}{3}})^c$ & $(R_2^{\nicefrac{+5}{3}})^c$ & $(R_2^{\nicefrac{+2}{3}})^c$ & $(S_3^{\nicefrac{+4}{3}})^c$ & $(S_3^{\nicefrac{+1}{3}})^c$ && $(\widetilde{U}_{1\mu}^{\nicefrac{+5}{3}})^c$ & $(V_{2\mu}^{\nicefrac{+4}{3}})^c$ & $(V_{2\mu}^{\nicefrac{+1}{3}})^c$ & $(U_{3\mu}^{\nicefrac{+5}{3}})^c$ & $(U_{3\mu}^{\nicefrac{+2}{3}})^c$ \\ 
				\cline{1-6}\cline{8-12} 
				$\geqslant 1\mu + 1j + 1\gamma$ & 76.3 & 22.7 & 8.5 & 79.3 & 215.8 && 7.9 & 454.2 & 3110.3 & 8.2 & 3.1 \\ 
				$\lvert M_{\ell j}- M_{\phi} \rvert \leq 10$ GeV & \multirow{2}{*}{40.4} & \multirow{2}{*}{12.1} & \multirow{2}{*}{4.7} & \multirow{2}{*}{41.8} & \multirow{2}{*}{108.9} && \multirow{2}{*}{3.6} & \multirow{2}{*}{239.4} & \multirow{2}{*}{1568.1} & \multirow{2}{*}{4.1} & \multirow{2}{*}{1.6}  \\ 
				$+ 1\gamma_{p_T > 20\text{GeV}}$ & & & &\\
				$Q_{Jet} < Q_{Jet}^\text{threshold}$ & --- & 5.1 & 0.53 & 18.8 & 14.8 && --- & 108.1 & 217.1 & 2.8 & 0.46  \\ \cline{1-1}\cline{2-6}\cline{8-12} 
				$\sigma_{Sig} (\mathcal{L}_{int} = 2000 \text{ fb}^{-1})$ & --- & \multicolumn{2}{c}{2.2} & \multicolumn{2}{c}{3.2} && --- & \multicolumn{2}{c}{6.0} & \multicolumn{2}{c}{1.5} \\ 
				$\mathcal{L}_{5\sigma} (\text{ in fb}^{-1})$ & --- & \multicolumn{2}{c}{10300} & \multicolumn{2}{c}{4900}&& --- & \multicolumn{2}{c}{1400} & \multicolumn{2}{c}{22200} \\ \hline 
			\end{tabular}}
			\caption{Signal-background analysis for associated production of photon and different leptoquarks at LHeC with $\sqrt{s}=1183.2$ GeV and $\mathcal{L}_{int} = 2000 \text{ fb}^{-1}$. The value of $Q_{Jet}^\text{threshold}$ has been chosen $-0.3$ for all the doublet and triplet leptoquarks.}
			\label{tab:LHeC}
		\end{center}
	\end{table*}

	\begin{figure*}[h!]
		\centering
		\mbox {
			\subfigure[$R_2^c$]{\includegraphics[width=0.3\textwidth]{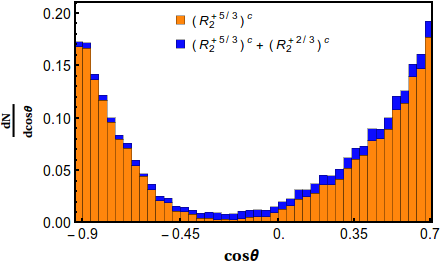}}
			\subfigure[$\widetilde{S}_1^c$]{\includegraphics[width=0.3\textwidth]{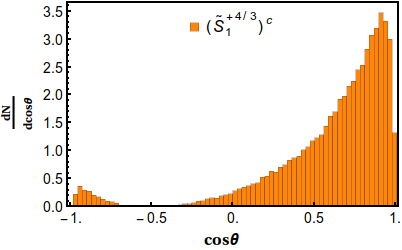}}  
			\hfill
			\subfigure[$\vec{S}_3^c$]{\includegraphics[width=0.3\textwidth]{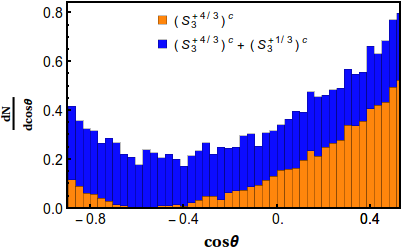}} }
		\mbox  {
			\subfigure[$\widetilde{U}_{1\mu}^c$]{\includegraphics[width=0.3\textwidth]{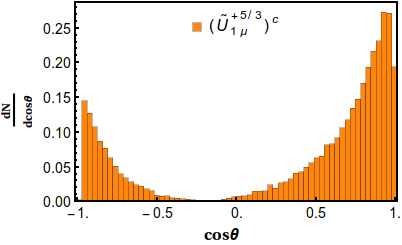}}
			\subfigure[$V_{2\mu}^c$]{\includegraphics[width=0.3\textwidth]{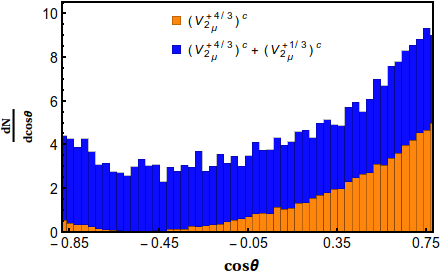}}
			\subfigure[$\vec{U}_{3\mu}^c$]{\includegraphics[width=0.3\textwidth]{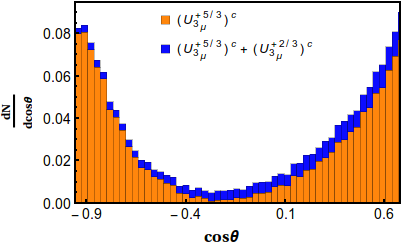}} }
		\caption{Distribution of cosine of the angle made by the photon with the electron beam, at $\sqrt{s} = 1183.2$ GeV and $\mathcal{L}_{int} = 2000$ fb$^{-1}$. The sub-figures (a)$R_2^c$, (b) $\widetilde{S}_1^c$, (c)  $\vec{S}_3^c$, (d) $\widetilde{U}_{1\mu}^c$, (e) $V_{2\mu}^c$ and (f)  $\vec{U}_{3\mu}^c$ show the angular distribution for associated production of the corresponding leptoquarks along with a photon at LHeC.}
		\label{LHeCTripCos}
	\end{figure*}
	
	In this collider also, we probe the final state $\geqslant 1\mu + 1j + 1\gamma_{p_T > 20\text{GeV}}$ with a reconstructed leptoquark satisfying $\lvert M_{\ell j}- M_{\phi} \rvert \leq 10 \text{ GeV}$. We show the angular distributions of photon with respect to its angle with electron beam in the CM frame for different leptoquark scenarios in Figure \ref{LHeCTripCos}. The first row portrays the angular distributions for all the scalar leptoquark scenarios whereas the second indicates the same for the vector ones. As has been seen for HERA, in case of  doublet and triplet leptoquarks, the highest isospin component exhibits RAZ while the second excitation acts as model background showing no zero in the angular distribution. However, there still exits a dip in the combined plot of signal and background  as can easily be observed from Figure \ref{LHeCTripCos}.
	
	Lastly, we collect the final state event numbers after the implication of different cuts for all the leptoquarks in Table \ref{tab:LHeC} considering $\mathcal L_{int}=2000$ \fbi and $\sqrt s=1183.2$ GeV. A significance of $5\sigma$ is reachable for the vector doublet $V_{2\mu}^c$ only, additionally, the scalar triplet $S_3^c$ can also be observed with more than $3\sigma$ significance.


	\section{FCC-he}
	\label{Sec:FCC}
	Finally we move to Future Circular Collider hadron-electron (FCC-he) \cite{Agostini:2020fmq,Bordry:2018gri} probe the TeV scale leptoquark at higher energy as compared to LHeC in section~\ref{Sec:LHeC} where the cross-sections are relatively small (See Table~\ref{tab:CS}). FCC-he is proposed in two phases, i.e. Phase I and Phase II. The FCC-I will have the centre of momentum energy around $\sim 2.2$ TeV, whereas FCC-II will have reach till $\sim 3.5$ TeV as can be seen from Table~\ref{beamFCC}. The projected luminosity is around $2000$ fb$^{-1}$.
	
	\begin{table}[h!]
		\begin{center}
			\renewcommand{\arraystretch}{1.2}
			\begin{tabular*}{0.49\textwidth}{@{\hspace*{2mm}\extracolsep{\fill}}ccccc}\hline 
				\multirow{2}{*}{Stage} & $E_p$ & $E_{e}$ & $\sqrt{s}$ & $\mathcal{L}_{int}^\text{projected}$ \\
				& (in TeV) & (in GeV) & (in GeV) &  (in fb$^{-1}$) \\ \hline
				I & 20 & 60 & 2190.2 & 2000 \\ \hline  
				II & 50 & 60 & 3464.1 & 2000 \\ \hline
			\end{tabular*}
			\caption{Beam and Centre of Mass energies along with projected integrated luminosities at FCC-he.}\label{beamFCC}
		\end{center}
	\end{table} 
	
	\subsection{FCC I}\label{FCCI}
\begin{figure*}[t!]
	\subfigure[]{\includegraphics[width=0.45\textwidth]{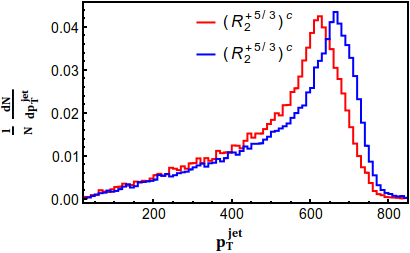}}\hfil
			\subfigure[]{\includegraphics[width=0.45\textwidth]{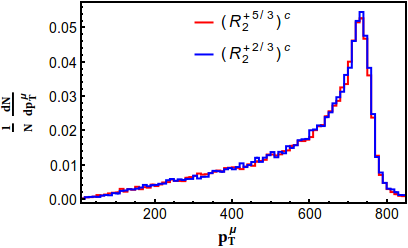}}
			
	\subfigure[]{\includegraphics[width=0.45\textwidth]{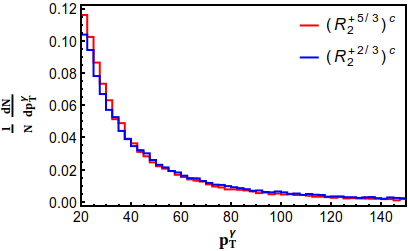}}\hfil
			\subfigure[]{\includegraphics[width=0.45\textwidth]{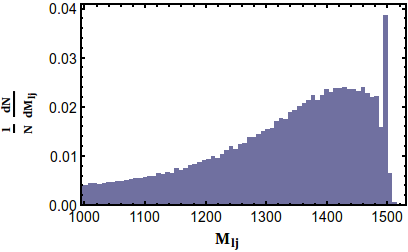}}
		\caption{$p^j_T$, $p^\ell_T$, $p^\gamma_T$ and $M_{\ell \, j}$ distributions at FCC-I with $\sqrt s=2190.2$ GeV for the scalar-doublet leptoquark $R_2^c$ with mass $1.5$ TeV. }\label{fig:DisFCCI}
	\end{figure*}

	For FCC-I the leptoquark mass is chosen as 1.5 TeV and like previous case, we present the kinematical distributions of scalar doublet leptoquark $R_2^c$ in Figure~\ref{fig:DisFCCI}. The sub-figures (a), (b), (c) describe the distribution of transverse momenta for jet, muon and photon respectively. It can be seen that jet and muon $p_T$ peak around half of the leptoquark mass i.e. $\sim 750$ GeV and certainly are very highly energetic. On the other hand, sub-figure (d) represent the invariant mass of the jet-muon pair and it peaks around the leptoquark mass $\sim 1.5$ TeV. 
	
The benchmark points for the collider study are described in Table~\ref{tab:BP} as BP3. The respective decay branching fractions of different leptoquark models are shown in Table~\ref{tab:BF}  with the cross-sections for the leptoquarks production in association with a photon $(p_T^\gamma\geq 20 \text{ GeV})$ given by Table~\ref{tab:CS}.  
	
	\begin{figure*}[h!]
		\centering
		\mbox {\subfigure[$R_2^c$]{\includegraphics[width=0.3\textwidth]{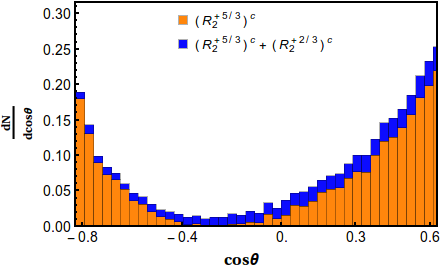}}
			\subfigure[$\widetilde{S}_1^c$]{\includegraphics[width=0.3\textwidth]{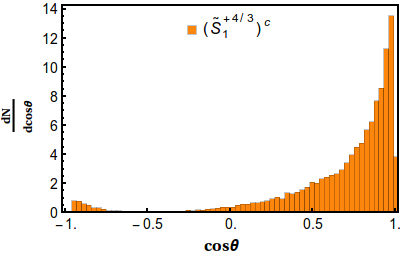}}  
			\hfill
			\subfigure[$\vec{S}_3^c$]{\includegraphics[width=0.3\textwidth]{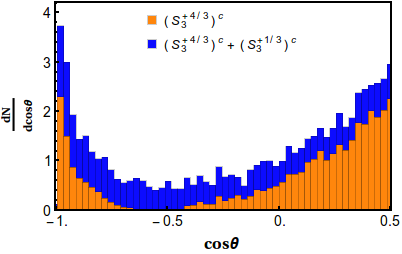}} }
		\mbox  { 	\subfigure[$\widetilde{U}_{1\mu}^c$]{\includegraphics[width=0.3\textwidth]{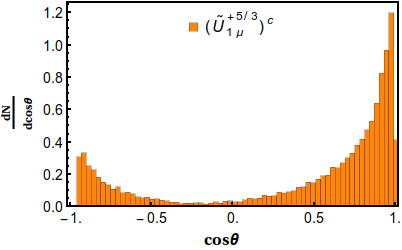}}
			\subfigure[$V_{2\mu}^c$]{\includegraphics[width=0.3\textwidth]{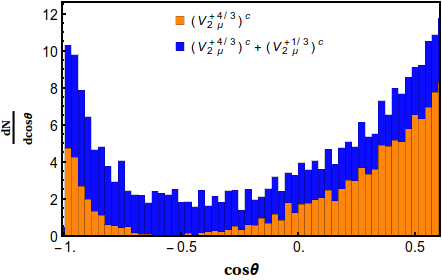}}
			\subfigure[$\vec{U}_{3\mu}^c$]{\includegraphics[width=0.3\textwidth]{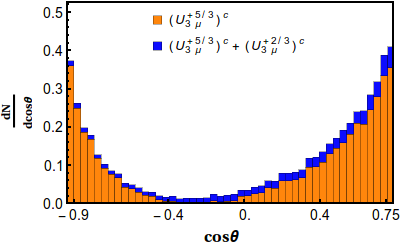}} }
		\caption{Distribution of cosine of the angle made by the photon with the electron beam, at $\sqrt{s} = 2190.2$ GeV and $\mathcal{L}_{int} = 2000$ fb$^{-1}$. The sub-figure (a)$R_2^c$, (b) $\widetilde{S}_1^c$, (c)  $\vec{S}_3^c$, (d) $\widetilde{U}_{1\mu}^c$, (e) $V_{2\mu}^c$ and (f)  $\vec{U}_{3\mu}^c$ show the angular distribution for associated leptoquark production with a photon at FCC-I.}
		\label{FCCITripCos}
	\end{figure*}
	
	Like earlier scenarios, we look for the final state $\big(\geqslant 1\mu + 1j + 1\gamma_{p_T > 20\text{GeV}}\big)$ with a reconstructed leptoquark ,i.e. $\big(\lvert M_{\ell j}- M_{\phi} \rvert \leq 10 \text{ GeV}\big)$, and plot the angular distribution with respect to the angle between the hard photon and electron in the CM frame, as illustrated in Figure~\ref{FCCITripCos}. Here the upper panel shows the distributions from the scalar leptoquarks whereas lower panel describes the same for vector ones.  Similar to HERA and LHeC, here also the zeros in there cross-sections are visible as minima in their angular distributions combined with SM plus the model backgrounds. Doublet and triplet leptoquarks which have more than one excitation
	get model contaminations from the other component of the multiplet that does not show any trace of zero, as explained before. Generally such different excitations have unalike decay modes leading to separate charged jets, i.e. up- or down-type, and thus can be distinguished by reconstructing the jet-charge. 
	
	We then gather the final state event numbers guided by the different cuts and present them in Table~\ref{tab:FCCI} at 2000 fb$^{-1}$ of integrated luminosity and $2190$ GeV of the centre of momentum energy. It is evident that a signal significance of $\sim 3\sigma$ is achievable for most of the scenarios while more than $5\sigma$ is for a few.

	\begin{table*}[h!]
		\begin{center}
			\renewcommand{\arraystretch}{1.5}
			\scalebox{0.9}{
				\begin{tabular}{cccccccccccc}\hline 
					&\multicolumn{5}{c}{Scalar Leptoquark}&&\multicolumn{5}{c}{Vector Leptoquark}\\
					\cline{2-6}\cline{8-12}
					\multirow{2}{*}{Cuts} & Singlet & \multicolumn{2}{c}{Doublet} & \multicolumn{2}{c}{Triplet} && Singlet & \multicolumn{2}{c}{Doublet} & \multicolumn{2}{c}{Triplet} \\
					\cline{2-6}\cline{8-12} 
					& Signal &Signal& SM+ &Signal& SM+ &&Signal&Signal& SM+ &Signal& SM+\\
					& $(\widetilde{S}_1^{\nicefrac{+4}{3}})^c$ & $(R_2^{\nicefrac{+5}{3}})^c$ & $(R_2^{\nicefrac{+2}{3}})^c$ & $(S_3^{\nicefrac{+4}{3}})^c$ & $(S_3^{\nicefrac{+1}{3}})^c$ && $(\widetilde{U}_{1\mu}^{\nicefrac{+5}{3}})^c$ & $(V_{2\mu}^{\nicefrac{+4}{3}})^c$ & $(V_{2\mu}^{\nicefrac{+1}{3}})^c$ & $(U_{3\mu}^{\nicefrac{+5}{3}})^c$ & $(U_{3\mu}^{\nicefrac{+2}{3}})^c$ \\ 
					\cline{1-6}\cline{8-12} 
					$\geqslant 1\mu + 1j + 1\gamma$ & 569.8 & 42.4 & 15.2 & 600.9 & 1223.9 && 39.9 & 2397.5 & 4886.9 & 42.4 & 15.2\\ 
					$\lvert M_{\ell j}- M_{\phi} \rvert \leq 10$ GeV & \multirow{2}{*}{166.8} & \multirow{2}{*}{14.0} & \multirow{2}{*}{5.1} & \multirow{2}{*}{201.6} & \multirow{2}{*}{401.9} && \multirow{2}{*}{12.2} & \multirow{2}{*}{800.2} & \multirow{2}{*}{1621.4} & \multirow{2}{*}{14.0} & \multirow{2}{*}{5.1}  \\ 
					$+ 1\gamma_{p_T > 20\text{GeV}}$ & & & &\\
					$Q_{Jet} < Q_{Jet}^\text{threshold}$ & --- & 9.7 & 1.4 & 91.2 & 52.5 && --- & 364.4 & 209.3 & 9.7 & 1.5 \\ \cline{1-6}\cline{8-12} 
					$\sigma_{Sig} (\mathcal{L}_{int} = 2000 \text{ fb}^{-1})$ &  --- & \multicolumn{2}{c}{2.9} & \multicolumn{2}{c}{7.6} && --- & \multicolumn{2}{c}{15.2} & \multicolumn{2}{c}{2.9} \\ 
					$\mathcal{L}_{5\sigma} (\text{ in fb}^{-1})$ & --- & \multicolumn{2}{c}{5900} & \multicolumn{2}{c}{900}&& --- & \multicolumn{2}{c}{200} & \multicolumn{2}{c}{5900} \\ \hline 
				\end{tabular}}
				\caption{Signal-background analysis for associated production of photon and different leptoquarks at FCC $\mathrm{I}$ with $\sqrt{s}=2190.2$ GeV and $\mathcal{L}_{int} = 2000 \text{ fb}^{-1}$. The value of $Q_{Jet}^\text{threshold}$ has been chosen 0 for $R_2^c$ and $\vec U_{3\mu}^c$ while the same for $\vec S_3^c$ and $V_{2\mu}^c$ is chosen to be $-0.3\,$.}
				\label{tab:FCCI}
			\end{center}
		\end{table*} 
		

		\subsection{FCC-II}
		\label{FCCII}
		
	As FCC-II has elevated centre of momentum energy $\sim 3464$ GeV, we choose a relatively higher mass for the leptoquark i.e. $M_{\phi}=2$ TeV keeping all the couplings same as FCC-I. The benchmark points (BP4) for this scenario are already mentioned in Table \ref{tab:BP} and the cross-sections for production of different leptoquarks in association with a photon $(p_T^\gamma\geq 20 \text{ GeV})$ are enlisted in Table \ref{tab:CS}. The branching fractions for different leptoquarks under this scenario are almost same with those in case of FCC-I differing only at third decimal place and hence we do not show it explicitly in Table \ref{tab:BF}.  
	
\begin{figure*}[t!]
		\subfigure[]{\includegraphics[width=0.45\textwidth]{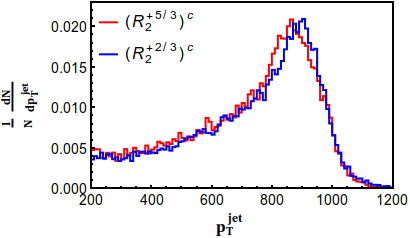}}\hfil
				\subfigure[]{\includegraphics[width=0.45\textwidth]{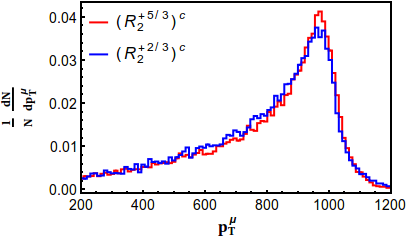}}
				
		\subfigure[]{\includegraphics[width=0.45\textwidth]{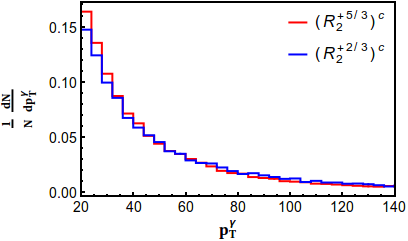}}\hfil
				\subfigure[]{\includegraphics[width=0.45\textwidth]{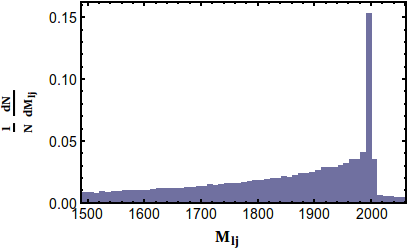}}
			\caption{$p^j_T$, $p^\ell_T$, $p^\gamma_T$ and $M_{\ell \, j}$ distributions at FCC II with $\sqrt s=3464.1$ GeV for the scalar-doublet leptoquark $R_2^c$ with mass $2.0$ TeV. }\label{scalrdbDisFCCII}
		\end{figure*}

	Due to heavier mass, the kinematical distributions for FCC-II will  have tails at the higher momentum spectrum as depicted in Figure~\ref{scalrdbDisFCCII}. Figure~\ref{scalrdbDisFCCII}(a) and (b) show the jet and muon $p_T$ distributions which have their peaks around half of the leptoquark mass i.e. $\sim 1$ TeV, as during the on-shell production of leptoquarks the momenta are equally shared between the jets and the muons. Figure~\ref{scalrdbDisFCCII}(c) and (d) describe the photon $p_T$ and invariant mass distributions of the jet-muon pair and the later peaks around the leptoquark mass of $\sim 2 $ TeV.

\begin{figure*}[t!]
			\centering
			\mbox {\subfigure[$R_2^c$]{\includegraphics[width=0.3\textwidth]{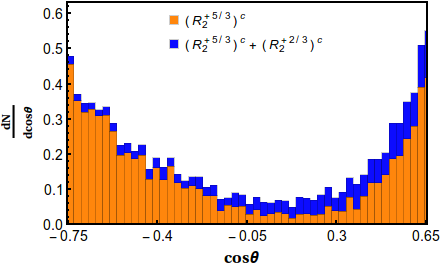}}
				\subfigure[$\widetilde{S}_1^c$]{\includegraphics[width=0.3\textwidth]{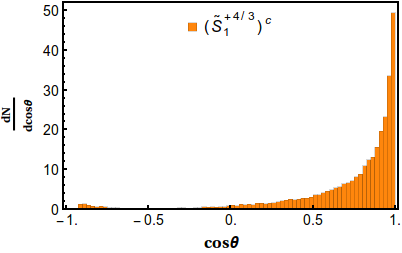}}  
				\hfill
				\subfigure[$\vec{S}_3^c$]{\includegraphics[width=0.3\textwidth]{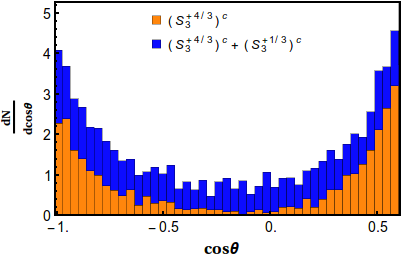}} }
			\mbox  { 	\subfigure[$\widetilde{U}_{1\mu}^c$]{\includegraphics[width=0.3\textwidth]{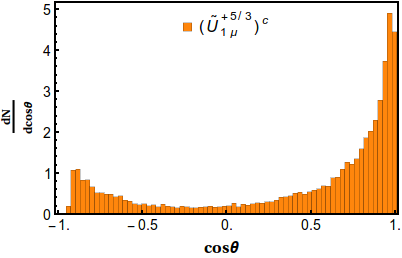}}
				\subfigure[$V_{2\mu}^c$]{\includegraphics[width=0.3\textwidth]{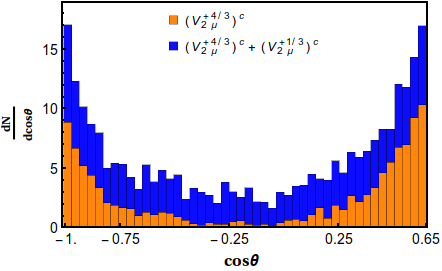}}
				\subfigure[$\vec{U}_{3\mu}^c$]{\includegraphics[width=0.3\textwidth]{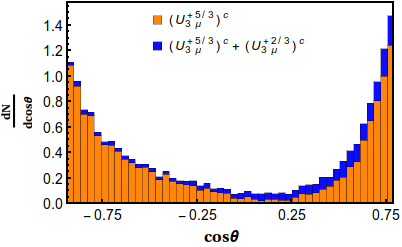}} }
			\caption{Distribution of cosine of the angle made by the photon with the electron beam, at $\sqrt{s} = 3464.1$ GeV and $\mathcal{L}_{int} = 2000$ fb$^{-1}$. The sub-figure (a)$R_2^c$, (b) $\widetilde{S}_1^c$, (c)  $\vec{S}_3^c$, (d) $\widetilde{U}_{1\mu}^c$, (e) $V_{2\mu}^c$ and (f)  $\vec{U}_{3\mu}^c$ show the angular distribution for associated leptoquark production with a photon at FCC-II.}
			\label{FCCIITripCos}
		\end{figure*}

		Similar to FCC-I, we present the angular distributions  in CM frame of FCC-II for the different leptoquark scenarios with respect to the angle between the hard photon and the incoming electron in Figure~\ref{FCCIITripCos}.  The minima are visible, as expected, in different leptoquark scenarios, combined with SM plus model backgrounds, for the final state satisfying $\geqslant 1\mu + 1j + 1\gamma_{p_T > 20\text{GeV}}$ with the reconstructed leptoquark obeying $\lvert M_{\ell j}- M_{\phi} \rvert \leq 10$ GeV. Later we present the final state numbers at FCC-II with centre of momentum energy of $3464.1$ GeV at an integrated luminosity of 2000 fb$^{-1}$ in Table~\ref{tab:FCC-II}. All the scenarios have signal significance over $5\sigma$ in that given integrated luminosity.  
		
		\begin{table*}[h!]
			\begin{center}
				\renewcommand{\arraystretch}{1.5}
				\scalebox{0.9}{
					\begin{tabular}{cccccccccccc}\hline 
						&\multicolumn{5}{c}{Scalar Leptoquark}&&\multicolumn{5}{c}{Vector Leptoquark}\\
						\cline{2-6}\cline{8-12}
						\multirow{2}{*}{Cuts} & Singlet & \multicolumn{2}{c}{Doublet} & \multicolumn{2}{c}{Triplet} && Singlet & \multicolumn{2}{c}{Doublet} & \multicolumn{2}{c}{Triplet} \\
						\cline{2-6}\cline{8-12} 
						& Signal &Signal& SM+ &Signal& SM+ &&Signal&Signal& SM+ &Signal& SM+\\
						& $(\widetilde{S}_1^{\nicefrac{+4}{3}})^c$ & $(R_2^{\nicefrac{+5}{3}})^c$ & $(R_2^{\nicefrac{+2}{3}})^c$ & $(S_3^{\nicefrac{+4}{3}})^c$ & $(S_3^{\nicefrac{+1}{3}})^c$ && $(\widetilde{U}_{1\mu}^{\nicefrac{+5}{3}})^c$ & $(V_{2\mu}^{\nicefrac{+4}{3}})^c$ & $(V_{2\mu}^{\nicefrac{+1}{3}})^c$ & $(U_{3\mu}^{\nicefrac{+5}{3}})^c$ & $(U_{3\mu}^{\nicefrac{+2}{3}})^c$ \\ 
						\cline{1-6}\cline{8-12} 
						$\geqslant 1\mu + 1j + 1\gamma$ & 1636.7 & 205.8 & 75.5 & 1781.6 & 2663.4 && 186.4 & 7051.4 & 10604.2 & 205.8 & 76.3\\ 
						$\lvert M_{\ell j}- M_{\phi} \rvert \leq 10$ GeV &  \multirow{2}{*}{569.2} & \multirow{2}{*}{74.1} & \multirow{2}{*}{24.5} & \multirow{2}{*}{575.4} & \multirow{2}{*}{874.9} && \multirow{2}{*}{69.9} & \multirow{2}{*}{2322.6} & \multirow{2}{*}{3496.6} & \multirow{2}{*}{75.1} & \multirow{2}{*}{29.3}  \\ 
						$+ 1\gamma_{p_T > 20\text{GeV}}$ & & & &\\
						$Q_{Jet} < Q_{Jet}^\text{threshold}$ & --- & 43.0 & 6.6 & 339.6 & 81.5 && ---  & 779.4 & 329.6 & 43.7 & 6.1 \\ 
						\cline{1-6}\cline{8-12} 
						$\sigma_{Sig} (\mathcal{L}_{int} = 2000 \text{ fb}^{-1})$ & ---  & \multicolumn{2}{c}{6.1} & \multicolumn{2}{c}{16.5} && --- & \multicolumn{2}{c}{23.4} & \multicolumn{2}{c}{6.2} \\ 
						$\mathcal{L}_{5\sigma} (\text{ in fb}^{-1})$ & --- &  \multicolumn{2}{c}{1300} & \multicolumn{2}{c}{200} && --- & \multicolumn{2}{c}{90} & \multicolumn{2}{c}{1300}\\ \hline 
					\end{tabular}}
					\caption{Signal-background analysis for associated production of photon and different leptoquarks at FCC-II with $\sqrt{s}=3464.1$ GeV and $\mathcal{L}_{int} = 2000 \text{ fb}^{-1}$. The value of $Q_{Jet}^\text{threshold}$ has been chosen 0 for $R_2^c$ and $\vec U_{3\mu}^c$ while the same for $\vec S_3^c$ and $V_{2\mu}^c$ is chosen to be $-0.3\,$.}
					\label{tab:FCC-II}
				\end{center}
			\end{table*}
			

			\section{Discussion and Conclusion}
			\label{Sec:concl}
			
			Radiation amplitude zero is a well-established phenomenon in flavour Physics. It has mostly been used in determining the electromagnetic properties of $W$-boson. In this article, we use it to probe the signatures of the proposed particles leptoquarks in electron-proton collider. For our purpose, we explored the zeros in the angular distribution of the photon, produced in association with the leptoquark, in different electron proton colliders namely HERA, LHeC, FCC-I and FCC-II. In our simulation, we have worked with leptoquarks of different mass scales, viz. 70 GeV to 2000 TeV, which are still allowed by various present and past collider bounds if proper couplings and and branching fractions are chosen.

			Scalar and vector singlet leptoquarks come with only one excitation thus there is no model contamination, which makes such typical angular distribution with zero at some kinematical point very easily detectable compared to the other scenarios. In case of doublets and triplets, the situation  is little complicated. It is interesting to note that some of the components for these multiplets cannot be produced at any electron-handron collider due to the particular gauge structure of the interactions, whereas some of the other components, though get produced in \ep collision, do not show RAZ within the physically allowed phase space. These second type of excitations perplex the circumstances for doublets and triplets since they tend to obscure the zeros in the distribution for the desired components. Such cases are handled by applying appropriate cuts which effectively reduce the model background. It is worth mentioning that since we have looked for the muons in the final state as the charged lepton, the channel becomes SM background free absolutely. 
			
			The reconstruction of leptoquark via jet-muon invariant  mass is crucial in order to establish the centre of momentum frame where we look for the zeros in angular distributions. The results at  the centre of momentum energy of HERA are very promising and early data would give $5\sigma$ signal significance for all the scenarios except scalar doublet and vector triplet ones which would take integrated luminosity of 100 fb$^{-1}$.
			
			On the other hand, the situation is a bit tight for LHeC. Due to various bounds from ATLAS and CMS, it is very difficult to have a leptoquark within the mass range 200 GeV to 900 GeV. But centre of momentum energy  available at LHeC is around 1.2 TeV and therefore, due to phase space suppression, it is quite challenging to achieve a good significance for the associated production of leptoquark with a photon. 
			
			 In FCC-I,  we have investigated the leptoquarks with mass 1.5 TeV where the cross-sections are not very encouraging  and except scalar triplet and vector doublet scenarios a $5\sigma$ reach is quite not possible with the integrated luminosity of 2000 fb$^{-1}$. The situation however changes when we upgrade to FCC-II and search for leptoquark with mass 2 TeV. In this collider $5\sigma$ significance is achievable within the projected luminosity of  2000 fb$^{-1}$ for all the scenarios. 
			 
			 Finally, we find that the leptoquarks showing RAZ in \ep collider do not exhibit zero in $e$-$\gamma$ collider while getting produced in association with a quark or anti-quark. Hence, in order to probe leptoquarks through the zeros of single photon tree level amplitude, it is necessary to study their production in both \ep and $e$-$\gamma$ colliders.  
			
			\section*{Acknowledgements} PB and AK acknowledge SERB CORE Grant CRG/2018/004971 and MATRICS Grant MTR/2020/000668 for the financial support towards the work. 
			
			\bibliography{References}

\providecommand{\href}[2]{#2}\begingroup\begin{thebibliography}{100}

\bibitem{Mikaelian:1979nr}
K.~Mikaelian, M.~Samuel and D.~Sahdev, \textit{{The Magnetic Moment of Weak
  Bosons Produced in $p p$ and $p\bar{p}$ Collisions}},
  \href{https://doi.org/10.1103/PhysRevLett.43.746}{\textit{Phys. Rev. Lett.}
  {\bfseries 43} (1979) 746}.

\bibitem{Brown:1979ux}
R.~Brown, D.~Sahdev and K.~Mikaelian, \textit{{$W^\pm Z^0$ and $W^\pm \gamma$
  Pair Production in $\nu e$, $pp$, and $\bar p p$ Collisions}},
  \href{https://doi.org/10.1103/PhysRevD.20.1164}{\textit{Phys. Rev. D}
  {\bfseries 20} (1979) 1164}.

\bibitem{Goebel:1980es}
C.~Goebel, F.~Halzen and J.~Leveille, \textit{{Angular zeros of Brown,
  Mikaelian, Sahdev, and Samuel and the factorization of tree amplitudes in
  gauge theories}},
  \href{https://doi.org/10.1103/PhysRevD.23.2682}{\textit{Phys. Rev. D}
  {\bfseries 23} (1981) 2682--2685}.

\bibitem{Zhu:1980sz}
D.~P. Zhu, \textit{{Zeros in Scattering Amplitudes and the Structure of
  Nonabelian Gauge Theories}},
  \href{https://doi.org/10.1103/PhysRevD.22.2266}{\textit{Phys. Rev. D}
  {\bfseries 22} (1980) 2266}.

\bibitem{Deshpande:1994vf}
N.~Deshpande, X.-G. He and S.~Oh, \textit{{Amplitude zeros in radiative decays
  of scalar particles}},
  \href{https://doi.org/10.1103/PhysRevD.51.2295}{\textit{Phys. Rev. D}
  {\bfseries 51} (1995) 2295--2301},
  [\href{https://arxiv.org/abs/hep-ph/9410373}{{\ttfamily hep-ph/9410373}}].

\bibitem{Brodsky:1982sh}
S.~J. Brodsky and R.~W. Brown, \textit{{Zeros in Amplitudes: Gauge Theory and
  Radiation Interference}},
  \href{https://doi.org/10.1103/PhysRevLett.49.966}{\textit{Phys. Rev. Lett.}
  {\bfseries 49} (1982) 966}.

\bibitem{Brown:1982xx}
R.~W. Brown, K.~Kowalski and S.~J. Brodsky, \textit{{Classical Radiation Zeros
  in Gauge Theory Amplitudes}},
  \href{https://doi.org/10.1103/PhysRevD.29.2100}{\textit{Phys. Rev. D}
  {\bfseries 28} (1983) 624}. [Addendum: Phys. Rev. D 29, 2100--2104 (1984)].

\bibitem{Passarino:1982hh}
G.~Passarino, \textit{{Physical Null Zones and Radiation Representation}},
  \href{https://doi.org/10.1016/0550-3213(83)90005-6}{\textit{Nucl. Phys. B}
  {\bfseries 224} (1983) 265--288}.

\bibitem{Samuel:1983eg}
M.~Samuel, \textit{{Amplitude Zeros}},
  \href{https://doi.org/10.1103/PhysRevD.27.2724}{\textit{Phys. Rev. D}
  {\bfseries 27} (1983) 2724--2731}.

\bibitem{Samuel:1984ru}
M.~Samuel, A.~Sen, G.~Sylvester and M.~Laursen, \textit{{General Criteria for
  Radiation Amplitude Zeros}},
  \href{https://doi.org/10.1103/PhysRevD.29.994}{\textit{Phys. Rev. D}
  {\bfseries 29} (1984) 994--999}.

\bibitem{Stroughair:1984pj}
J.~Stroughair and C.~Bilchak, \textit{{The Determination of the $W$ Anomalous
  Magnetic Moment in $p \bar{p} \to W \gamma X$}},
  \href{https://doi.org/10.1007/BF01452568}{\textit{Z. Phys. C} {\bfseries 26}
  (1984) 415--419}.

\bibitem{Baur:1988qt}
U.~Baur and D.~Zeppenfeld, \textit{{Probing the $W W \gamma$ Vertex at Future
  Hadron Colliders}},
  \href{https://doi.org/10.1016/0550-3213(88)90045-4}{\textit{Nucl. Phys. B}
  {\bfseries 308} (1988) 127--148}.

\bibitem{Cortes:1985yi}
J.~Cortes, K.~Hagiwara and F.~Herzog, \textit{{Testing the $W W \gamma$
  Coupling of the Standard Model at $p \bar{p}$ Colliders}},
  \href{https://doi.org/10.1016/0550-3213(86)90105-7}{\textit{Nucl. Phys. B}
  {\bfseries 278} (1986) 26}.

\bibitem{Valenzuela:1985dp}
G.~Valenzuela and J.~Smith, \textit{{Experimental Implications of Finite Width
  and $p_T$ Effects in $W$ Boson Radiative Production}},
  \href{https://doi.org/10.1103/PhysRevD.31.2787}{\textit{Phys. Rev. D}
  {\bfseries 31} (1985) 2787}.

\bibitem{Laursen:1982iv}
M.~L. Laursen, M.~A. Samuel, A.~Sen and G.~Tupper, \textit{{Do Amplitude Zeros
  Persist in Higher Order?}},
  \href{https://doi.org/10.1016/0550-3213(83)90201-8}{\textit{Nucl. Phys. B}
  {\bfseries 226} (1983) 429--436}.

\bibitem{Laursen:1983kw}
M.~Laursen, M.~Samuel and A.~Sen, \textit{{On the Spoiling of Amplitude
  (Radiation) Zeros at the One Loop Level and Infrared Finiteness}},
  \href{https://doi.org/10.1103/PhysRevD.28.650}{\textit{Phys. Rev. D}
  {\bfseries 28} (1983) 650}.

\bibitem{Smith:1989xz}
J.~Smith, D.~Thomas and W.~van Neerven, \textit{{QCD Corrections to the
  Reaction $p \bar{p} \to W \gamma X$}},
  \href{https://doi.org/10.1007/BF01557332}{\textit{Z. Phys. C} {\bfseries 44}
  (1989) 267}.

\bibitem{Ohnemus:1992jn}
J.~Ohnemus, \textit{{Order $\alpha_s$ calculations of hadronic $W^\pm \gamma$
  and $Z \gamma$ production}},
  \href{https://doi.org/10.1103/PhysRevD.47.940}{\textit{Phys. Rev. D}
  {\bfseries 47} (1993) 940--955}.

\bibitem{Baur:1989gk}
U.~Baur and E.~L. Berger, \textit{{Probing the $W W \gamma$ Vertex at the
  Tevatron Collider}},
  \href{https://doi.org/10.1103/PhysRevD.41.1476}{\textit{Phys. Rev. D}
  {\bfseries 41} (1990) 1476}.

\bibitem{Baur:1993ir}
U.~Baur, T.~Han and J.~Ohnemus, \textit{{QCD corrections to hadronic $W \gamma$
  production with nonstandard $W W \gamma$ couplings}},
  \href{https://doi.org/10.1103/PhysRevD.48.5140}{\textit{Phys. Rev. D}
  {\bfseries 48} (1993) 5140--5161},
  [\href{https://arxiv.org/abs/hep-ph/9305314}{{\ttfamily hep-ph/9305314}}].

\bibitem{Baur:1994sa}
U.~Baur, S.~Errede and G.~L. Landsberg, \textit{{Rapidity correlations in $W
  \gamma$ production at hadron colliders}},
  \href{https://doi.org/10.1103/PhysRevD.50.1917}{\textit{Phys. Rev. D}
  {\bfseries 50} (1994) 1917--1930},
  [\href{https://arxiv.org/abs/hep-ph/9402282}{{\ttfamily hep-ph/9402282}}].

\bibitem{Capdevilla:2019zbx}
R.~M. Capdevilla, R.~Harnik and A.~Martin, \textit{{The radiation valley and
  exotic resonances in $W\gamma$ production at the LHC}},
  \href{https://doi.org/10.1007/JHEP03(2020)117}{\textit{JHEP} {\bfseries 03}
  (2020) 117}, [\href{https://arxiv.org/abs/1912.08234}{{\ttfamily
  1912.08234}}].

\bibitem{Pati:1973uk}
J.~C. Pati and A.~Salam, \textit{{Unified Lepton-Hadron Symmetry and a Gauge
  Theory of the Basic Interactions}},
  \href{https://doi.org/10.1103/PhysRevD.8.1240}{\textit{Phys. Rev. D}
  {\bfseries 8} (1973) 1240--1251}.

\bibitem{Pati:1974yy}
J.~C. Pati and A.~Salam, \textit{{Lepton Number as the Fourth Color}},
  \href{https://doi.org/10.1103/PhysRevD.10.275}{\textit{Phys. Rev. D}
  {\bfseries 10} (1974) 275--289}. [Erratum: Phys. Rev. D 11, 703--703 (1975)].

\bibitem{Georgi:1974my}
H.~Georgi, \textit{{The State of the Art\textemdash{}Gauge Theories}},
  \href{https://doi.org/10.1063/1.2947450}{\textit{AIP Conf. Proc.} {\bfseries
  23} (1975) 575--582}.

\bibitem{Georgi:1974sy}
H.~Georgi and S.~Glashow, \textit{{Unity of All Elementary Particle Forces}},
  \href{https://doi.org/10.1103/PhysRevLett.32.438}{\textit{Phys. Rev. Lett.}
  {\bfseries 32} (1974) 438--441}.

\bibitem{Dimopoulos:1979es}
S.~Dimopoulos and L.~Susskind, \textit{{Mass Without Scalars}}, .

\bibitem{Farhi:1980xs}
E.~Farhi and L.~Susskind, \textit{{Technicolor}},
  \href{https://doi.org/10.1016/0370-1573(81)90173-3}{\textit{Phys. Rept.}
  {\bfseries 74} (1981) 277}.

\bibitem{Schrempp:1984nj}
B.~Schrempp and F.~Schrempp, \textit{{Light Leptoquarks}},
  \href{https://doi.org/10.1016/0370-2693(85)91450-9}{\textit{Phys. Lett. B}
  {\bfseries 153} (1985) 101--107}.

\bibitem{Wudka:1985ef}
J.~Wudka, \textit{{Composite Leptoquarks}},
  \href{https://doi.org/10.1016/0370-2693(86)90356-4}{\textit{Phys. Lett. B}
  {\bfseries 167} (1986) 337--342}.

\bibitem{Nilles:1983ge}
H.~P. Nilles, \textit{{Supersymmetry, Supergravity and Particle Physics}},
  \href{https://doi.org/10.1016/0370-1573(84)90008-5}{\textit{Phys. Rept.}
  {\bfseries 110} (1984) 1--162}.

\bibitem{Haber:1984rc}
H.~E. Haber and G.~L. Kane, \textit{{The Search for Supersymmetry: Probing
  Physics Beyond the Standard Model}},
  \href{https://doi.org/10.1016/0370-1573(85)90051-1}{\textit{Phys. Rept.}
  {\bfseries 117} (1985) 75--263}.

\bibitem{Hewett:1997ce}
J.~L. Hewett and T.~G. Rizzo, \textit{{Much ado about leptoquarks: A
  Comprehensive analysis}},
  \href{https://doi.org/10.1103/PhysRevD.56.5709}{\textit{Phys. Rev. D}
  {\bfseries 56} (1997) 5709--5724},
  [\href{https://arxiv.org/abs/hep-ph/9703337}{{\ttfamily hep-ph/9703337}}].

\bibitem{Dorsner:2016wpm}
I.~Dor\v{s}ner, S.~Fajfer, A.~Greljo, J.~Kamenik and N.~Ko\v{s}nik,
  \textit{{Physics of leptoquarks in precision experiments and at particle
  colliders}},
  \href{https://doi.org/10.1016/j.physrep.2016.06.001}{\textit{Phys. Rept.}
  {\bfseries 641} (2016) 1--68},
  [\href{https://arxiv.org/abs/1603.04993}{{\ttfamily 1603.04993}}].

\bibitem{Behrend:1986jz}
{\scshape CELLO} collaboration, H.~Behrend et~al., \textit{{Search for Light
  Leptoquark Bosons}},
  \href{https://doi.org/10.1016/0370-2693(86)91410-3}{\textit{Phys. Lett. B}
  {\bfseries 178} (1986) 452--456}. [Addendum: Phys.Lett.B 184, 417 (1987)].

\bibitem{Bartel:1987de}
{\scshape JADE} collaboration, W.~Bartel et~al., \textit{{Search for
  Leptoquarks and Other New Particles With Lepton - Hadron Signature in $e^+
  e^-$ Interactions}}, \href{https://doi.org/10.1007/BF01556160}{\textit{Z.
  Phys. C} {\bfseries 36} (1987) 15}.

\bibitem{Kim:1989qz}
{\scshape AMY} collaboration, G.~Kim et~al., \textit{{A search for leptoquark
  and colored lepton pair production in $e^+e^-$ annihilations at TRISTAN}},
  \href{https://doi.org/10.1016/0370-2693(90)90442-9}{\textit{Phys. Lett. B}
  {\bfseries 240} (1990) 243--249}.

\bibitem{Decamp:1991uy}
{\scshape ALEPH} collaboration, D.~Decamp et~al., \textit{{Searches for new
  particles in $Z$ decays using the ALEPH detector}},
  \href{https://doi.org/10.1016/0370-1573(92)90177-2}{\textit{Phys. Rept.}
  {\bfseries 216} (1992) 253--340}.

\bibitem{Adriani:1993gk}
{\scshape L3} collaboration, O.~Adriani et~al., \textit{{Results from the L3
  experiment at LEP}},
  \href{https://doi.org/10.1016/0370-1573(93)90027-B}{\textit{Phys. Rept.}
  {\bfseries 236} (1993) 1--146}.

\bibitem{Abbiendi:2003iv}
{\scshape OPAL} collaboration, G.~Abbiendi et~al., \textit{{Search for pair
  produced leptoquarks in $e^{+} e^{-}$ interactions at $\sqrt{s} \simeq$ 189
  GeV -- 209 GeV}},
  \href{https://doi.org/10.1140/epjc/s2003-01325-y}{\textit{Eur. Phys. J. C}
  {\bfseries 31} (2003) 281--305},
  [\href{https://arxiv.org/abs/hep-ex/0305053}{{\ttfamily hep-ex/0305053}}].

\bibitem{Abreu:1998fw}
{\scshape DELPHI} collaboration, P.~Abreu et~al., \textit{{Search for
  Leptoquarks and FCNC in $e^{+} e^{-}$ annihilations at $\sqrt {s}$ = 183
  GeV}}, \href{https://doi.org/10.1016/S0370-2693(98)01525-1}{\textit{Phys.
  Lett. B} {\bfseries 446} (1999) 62--74},
  [\href{https://arxiv.org/abs/hep-ex/9903072}{{\ttfamily hep-ex/9903072}}].

\bibitem{Aaron:2011zz}
{\scshape H1} collaboration, F.~Aaron et~al., \textit{{Search for Lepton
  Flavour Violation at HERA}},
  \href{https://doi.org/10.1016/j.physletb.2011.05.023}{\textit{Phys. Lett. B}
  {\bfseries 701} (2011) 20--30},
  [\href{https://arxiv.org/abs/1103.4938}{{\ttfamily 1103.4938}}].

\bibitem{Collaboration:2011qaa}
{\scshape H1} collaboration, F.~Aaron et~al., \textit{{Search for first
  generation leptoquarks in $ep$ collisions at HERA}},
  \href{https://doi.org/10.1016/j.physletb.2011.09.017}{\textit{Phys. Lett. B}
  {\bfseries 704} (2011) 388--396},
  [\href{https://arxiv.org/abs/1107.3716}{{\ttfamily 1107.3716}}].

\bibitem{Abramowicz:2012tg}
{\scshape ZEUS} collaboration, H.~Abramowicz et~al., \textit{{Search for
  first-generation leptoquarks at HERA}},
  \href{https://doi.org/10.1103/PhysRevD.86.012005}{\textit{Phys. Rev. D}
  {\bfseries 86} (2012) 012005},
  [\href{https://arxiv.org/abs/1205.5179}{{\ttfamily 1205.5179}}].

\bibitem{Abramowicz:2019uti}
{\scshape ZEUS} collaboration, H.~Abramowicz et~al., \textit{{Limits on contact
  interactions and leptoquarks at HERA}},
  \href{https://doi.org/10.1103/PhysRevD.99.092006}{\textit{Phys. Rev. D}
  {\bfseries 99} (2019) 092006},
  [\href{https://arxiv.org/abs/1902.03048}{{\ttfamily 1902.03048}}].

\bibitem{Alitti:1991dn}
{\scshape UA2} collaboration, J.~Alitti et~al., \textit{{A Search for scalar
  leptoquarks at the CERN $\bar{p}p$ collider}},
  \href{https://doi.org/10.1016/0370-2693(92)92024-B}{\textit{Phys. Lett. B}
  {\bfseries 274} (1992) 507--512}.

\bibitem{Acosta:2005ge}
{\scshape CDF} collaboration, D.~Acosta et~al., \textit{{Search for
  first-generation scalar leptoquarks in $p\bar{p}$ collisions at $\sqrt{s} =
  1.96$ TeV}}, \href{https://doi.org/10.1103/PhysRevD.72.051107}{\textit{Phys.
  Rev. D} {\bfseries 72} (2005) 051107},
  [\href{https://arxiv.org/abs/hep-ex/0506074}{{\ttfamily hep-ex/0506074}}].

\bibitem{Abulencia:2005ua}
{\scshape CDF} collaboration, A.~Abulencia et~al., \textit{{Search for
  second-generation scalar leptoquarks in $p\bar{p}$ collisions at $\sqrt{s} =$
  1.96 TeV.}}, \href{https://doi.org/10.1103/PhysRevD.73.051102}{\textit{Phys.
  Rev. D} {\bfseries 73} (2006) 051102},
  [\href{https://arxiv.org/abs/hep-ex/0512055}{{\ttfamily hep-ex/0512055}}].

\bibitem{Aaltonen:2007rb}
{\scshape CDF} collaboration, T.~Aaltonen et~al., \textit{{Search for Third
  Generation Vector Leptoquarks in $p \bar{p}$ Collisions at $\sqrt{s}$ = 1.96
  TeV}}, \href{https://doi.org/10.1103/PhysRevD.77.091105}{\textit{Phys. Rev.
  D} {\bfseries 77} (2008) 091105},
  [\href{https://arxiv.org/abs/0706.2832}{{\ttfamily 0706.2832}}].

\bibitem{Abazov:2008np}
{\scshape D0} collaboration, V.~Abazov et~al., \textit{{Search for pair
  production of second generation scalar leptoquarks}},
  \href{https://doi.org/10.1016/j.physletb.2008.12.017}{\textit{Phys. Lett. B}
  {\bfseries 671} (2009) 224--232},
  [\href{https://arxiv.org/abs/0808.4023}{{\ttfamily 0808.4023}}].

\bibitem{Abazov:2010wq}
{\scshape D0} collaboration, V.~M. Abazov et~al., \textit{{Search for Scalar
  Bottom Quarks and Third-Generation Leptoquarks in $p\bar{p}$ Collisions at
  $\sqrt{s}=1.96$ TeV}},
  \href{https://doi.org/10.1016/j.physletb.2010.08.028}{\textit{Phys. Lett. B}
  {\bfseries 693} (2010) 95--101},
  [\href{https://arxiv.org/abs/1005.2222}{{\ttfamily 1005.2222}}].

\bibitem{Abazov:2011qj}
{\scshape D0} collaboration, V.~M. Abazov et~al., \textit{{Search for first
  generation leptoquark pair production in the electron + missing energy + jets
  final state}},
  \href{https://doi.org/10.1103/PhysRevD.84.071104}{\textit{Phys. Rev. D}
  {\bfseries 84} (2011) 071104},
  [\href{https://arxiv.org/abs/1107.1849}{{\ttfamily 1107.1849}}].

\bibitem{Aaboud:2019bye}
{\scshape ATLAS} collaboration, M.~Aaboud et~al., \textit{{Searches for
  third-generation scalar leptoquarks in $\sqrt{s}$ = 13 TeV pp collisions with
  the ATLAS detector}},
  \href{https://doi.org/10.1007/JHEP06(2019)144}{\textit{JHEP} {\bfseries 06}
  (2019) 144}, [\href{https://arxiv.org/abs/1902.08103}{{\ttfamily
  1902.08103}}].

\bibitem{Aaboud:2019jcc}
{\scshape ATLAS} collaboration, M.~Aaboud et~al., \textit{{Searches for scalar
  leptoquarks and differential cross-section measurements in dilepton-dijet
  events in proton-proton collisions at a centre-of-mass energy of $\sqrt{s}$ =
  13 TeV with the ATLAS experiment}},
  \href{https://doi.org/10.1140/epjc/s10052-019-7181-x}{\textit{Eur. Phys. J.
  C} {\bfseries 79} (2019) 733},
  [\href{https://arxiv.org/abs/1902.00377}{{\ttfamily 1902.00377}}].

\bibitem{Sirunyan:2018btu}
{\scshape CMS} collaboration, A.~M. Sirunyan et~al., \textit{{Search for pair
  production of first-generation scalar leptoquarks at $\sqrt{s} =$ 13 TeV}},
  \href{https://doi.org/10.1103/PhysRevD.99.052002}{\textit{Phys. Rev. D}
  {\bfseries 99} (2019) 052002},
  [\href{https://arxiv.org/abs/1811.01197}{{\ttfamily 1811.01197}}].

\bibitem{Sirunyan:2018kzh}
{\scshape CMS} collaboration, A.~M. Sirunyan et~al., \textit{{Constraints on
  models of scalar and vector leptoquarks decaying to a quark and a neutrino at
  $\sqrt{s}=$ 13 TeV}},
  \href{https://doi.org/10.1103/PhysRevD.98.032005}{\textit{Phys. Rev. D}
  {\bfseries 98} (2018) 032005},
  [\href{https://arxiv.org/abs/1805.10228}{{\ttfamily 1805.10228}}].

\bibitem{Sirunyan:2018nkj}
{\scshape CMS} collaboration, A.~M. Sirunyan et~al., \textit{{Search for
  third-generation scalar leptoquarks decaying to a top quark and a $\tau$
  lepton at $\sqrt{s}=$ 13 TeV}},
  \href{https://doi.org/10.1140/epjc/s10052-018-6143-z}{\textit{Eur. Phys. J.
  C} {\bfseries 78} (2018) 707},
  [\href{https://arxiv.org/abs/1803.02864}{{\ttfamily 1803.02864}}].

\bibitem{Sirunyan:2018ryt}
{\scshape CMS} collaboration, A.~M. Sirunyan et~al., \textit{{Search for pair
  production of second-generation leptoquarks at $\sqrt{s}=$ 13 TeV}},
  \href{https://doi.org/10.1103/PhysRevD.99.032014}{\textit{Phys. Rev. D}
  {\bfseries 99} (2019) 032014},
  [\href{https://arxiv.org/abs/1808.05082}{{\ttfamily 1808.05082}}].

\bibitem{Sirunyan:2018vhk}
{\scshape CMS} collaboration, A.~M. Sirunyan et~al., \textit{{Search for heavy
  neutrinos and third-generation leptoquarks in hadronic states of two $\tau$
  leptons and two jets in proton-proton collisions at $\sqrt{s} =$ 13 TeV}},
  \href{https://doi.org/10.1007/JHEP03(2019)170}{\textit{JHEP} {\bfseries 03}
  (2019) 170}, [\href{https://arxiv.org/abs/1811.00806}{{\ttfamily
  1811.00806}}].

\bibitem{Buchmuller:1986iq}
W.~Buchmuller and D.~Wyler, \textit{{Constraints on SU(5) Type Leptoquarks}},
  \href{https://doi.org/10.1016/0370-2693(86)90771-9}{\textit{Phys. Lett. B}
  {\bfseries 177} (1986) 377--382}.

\bibitem{Buchmuller:1986zs}
W.~Buchmuller, R.~Ruckl and D.~Wyler, \textit{{Leptoquarks in Lepton - Quark
  Collisions}},
  \href{https://doi.org/10.1016/0370-2693(87)90637-X}{\textit{Phys. Lett. B}
  {\bfseries 191} (1987) 442--448}. [Erratum: Phys.Lett.B 448, 320--320
  (1999)].

\bibitem{Hewett:1987yg}
J.~Hewett and S.~Pakvasa, \textit{{Leptoquark Production in Hadron Colliders}},
  \href{https://doi.org/10.1103/PhysRevD.37.3165}{\textit{Phys. Rev. D}
  {\bfseries 37} (1988) 3165}.

\bibitem{Belyaev:2005ew}
A.~Belyaev, C.~Leroy, R.~Mehdiyev and A.~Pukhov, \textit{{Leptoquark single and
  pair production at LHC with CalcHEP/CompHEP in the complete model}},
  \href{https://doi.org/10.1088/1126-6708/2005/09/005}{\textit{JHEP} {\bfseries
  09} (2005) 005}, [\href{https://arxiv.org/abs/hep-ph/0502067}{{\ttfamily
  hep-ph/0502067}}].

\bibitem{Bandyopadhyay:2018syt}
P.~Bandyopadhyay and R.~Mandal, \textit{{Revisiting scalar leptoquark at the
  LHC}}, \href{https://doi.org/10.1140/epjc/s10052-018-5959-x}{\textit{Eur.
  Phys. J. C} {\bfseries 78} (2018) 491},
  [\href{https://arxiv.org/abs/1801.04253}{{\ttfamily 1801.04253}}].

\bibitem{Bhaskar:2020kdr}
A.~Bhaskar, D.~Das, B.~De and S.~Mitra, \textit{{Enhancing scalar productions
  with leptoquarks at the LHC}},
  \href{https://doi.org/10.1103/PhysRevD.102.035002}{\textit{Phys. Rev. D}
  {\bfseries 102} (2020) 035002},
  [\href{https://arxiv.org/abs/2002.12571}{{\ttfamily 2002.12571}}].

\bibitem{Chandak:2019iwj}
K.~Chandak, T.~Mandal and S.~Mitra, \textit{{Hunting for scalar leptoquarks
  with boosted tops and light leptons}},
  \href{https://doi.org/10.1103/PhysRevD.100.075019}{\textit{Phys. Rev. D}
  {\bfseries 100} (2019) 075019},
  [\href{https://arxiv.org/abs/1907.11194}{{\ttfamily 1907.11194}}].

\bibitem{Mandal:2015vfa}
T.~Mandal, S.~Mitra and S.~Seth, \textit{{Single Productions of Colored
  Particles at the LHC: An Example with Scalar Leptoquarks}},
  \href{https://doi.org/10.1007/JHEP07(2015)028}{\textit{JHEP} {\bfseries 07}
  (2015) 028}, [\href{https://arxiv.org/abs/1503.04689}{{\ttfamily
  1503.04689}}].

\bibitem{Alves:2018krf}
A.~Alves, O.~J.~t. Eboli, G.~Grilli Di~Cortona and R.~R. Moreira,
  \textit{{Indirect and monojet constraints on scalar leptoquarks}},
  \href{https://doi.org/10.1103/PhysRevD.99.095005}{\textit{Phys. Rev. D}
  {\bfseries 99} (2019) 095005},
  [\href{https://arxiv.org/abs/1812.08632}{{\ttfamily 1812.08632}}].

\bibitem{Dorsner:2019vgp}
I.~Dor\v{s}ner, S.~Fajfer and M.~Patra, \textit{{A comparative study of the
  $S_1$ and $U_1$ leptoquark effects in the light quark regime}},
  \href{https://doi.org/10.1140/epjc/s10052-020-7754-8}{\textit{Eur. Phys. J.
  C} {\bfseries 80} (2020) 204},
  [\href{https://arxiv.org/abs/1906.05660}{{\ttfamily 1906.05660}}].

\bibitem{Mandal:2018qpg}
S.~Mandal, M.~Mitra and N.~Sinha, \textit{{Probing leptoquarks and heavy
  neutrinos at the LHeC}},
  \href{https://doi.org/10.1103/PhysRevD.98.095004}{\textit{Phys. Rev. D}
  {\bfseries 98} (2018) 095004},
  [\href{https://arxiv.org/abs/1807.06455}{{\ttfamily 1807.06455}}].

\bibitem{Padhan:2019dcp}
R.~Padhan, S.~Mandal, M.~Mitra and N.~Sinha, \textit{{Signatures of
  $\tilde{R}_2$ class of Leptoquarks at the upcoming $ep$ colliders}},
  \href{https://doi.org/10.1103/PhysRevD.101.075037}{\textit{Phys. Rev. D}
  {\bfseries 101} (2020) 075037},
  [\href{https://arxiv.org/abs/1912.07236}{{\ttfamily 1912.07236}}].

\bibitem{Fuentes-Martin:2019ign}
J.~Fuentes-Mart\'\i{}n, G.~Isidori, M.~K\"onig and N.~Selimovi\'c,
  \textit{{Vector Leptoquarks Beyond Tree Level}},
  \href{https://doi.org/10.1103/PhysRevD.101.035024}{\textit{Phys. Rev. D}
  {\bfseries 101} (2020) 035024},
  [\href{https://arxiv.org/abs/1910.13474}{{\ttfamily 1910.13474}}].

\bibitem{Baker:2019sli}
M.~J. Baker, J.~Fuentes-Mart\'\i{}n, G.~Isidori and M.~K\"onig, \textit{{High-
  $p_T$ signatures in vector\textendash{}leptoquark models}},
  \href{https://doi.org/10.1140/epjc/s10052-019-6853-x}{\textit{Eur. Phys. J.
  C} {\bfseries 79} (2019) 334},
  [\href{https://arxiv.org/abs/1901.10480}{{\ttfamily 1901.10480}}].

\bibitem{Bhattacharyya:1994ig}
G.~Bhattacharyya, J.~R. Ellis and K.~Sridhar, \textit{{Bounds on the masses and
  couplings of leptoquarks from leptonic partial widths of the $Z$}},
  \href{https://doi.org/10.1016/0370-2693(94)00927-9}{\textit{Phys. Lett. B}
  {\bfseries 336} (1994) 100--106},
  [\href{https://arxiv.org/abs/hep-ph/9406354}{{\ttfamily hep-ph/9406354}}].
  [Erratum: Phys.Lett.B 338, 522--523 (1994)].

\bibitem{Hewett:1987bh}
J.~Hewett and T.~Rizzo, \textit{{Leptoquark Signals at $e^+ e^-$ Colliders}},
  \href{https://doi.org/10.1103/PhysRevD.36.3367}{\textit{Phys. Rev. D}
  {\bfseries 36} (1987) 3367}.

\bibitem{Plehn:1997az}
T.~Plehn, H.~Spiesberger, M.~Spira and P.~Zerwas, \textit{{Formation and decay
  of scalar leptoquarks / squarks in ep collisions}},
  \href{https://doi.org/10.1007/s002880050426}{\textit{Z. Phys. C} {\bfseries
  74} (1997) 611--614}, [\href{https://arxiv.org/abs/hep-ph/9703433}{{\ttfamily
  hep-ph/9703433}}].

\bibitem{Kramer:1997hh}
M.~Kramer, T.~Plehn, M.~Spira and P.~Zerwas, \textit{{Pair production of scalar
  leptoquarks at the Tevatron}},
  \href{https://doi.org/10.1103/PhysRevLett.79.341}{\textit{Phys. Rev. Lett.}
  {\bfseries 79} (1997) 341--344},
  [\href{https://arxiv.org/abs/hep-ph/9704322}{{\ttfamily hep-ph/9704322}}].

\bibitem{Cuypers:1995ax}
F.~Cuypers, \textit{{Leptoquark production in $e^-\gamma$ scattering}},
  \href{https://doi.org/10.1016/0550-3213(96)00270-2}{\textit{Nucl. Phys. B}
  {\bfseries 474} (1996) 57--71},
  [\href{https://arxiv.org/abs/hep-ph/9508397}{{\ttfamily hep-ph/9508397}}].

\bibitem{Eboli:1993qx}
O.~J. Eboli, E.~Gregores, M.~Magro, P.~Mercadante and S.~Novaes,
  \textit{{Searching for leptoquarks in electron - photon collisions}},
  \href{https://doi.org/10.1016/0370-2693(93)90547-U}{\textit{Phys. Lett. B}
  {\bfseries 311} (1993) 147--152},
  [\href{https://arxiv.org/abs/hep-ph/9306229}{{\ttfamily hep-ph/9306229}}].

\bibitem{Nadeau:1993zv}
H.~Nadeau and D.~London, \textit{{Leptoquarks at e gamma colliders}},
  \href{https://doi.org/10.1103/PhysRevD.47.3742}{\textit{Phys. Rev. D}
  {\bfseries 47} (1993) 3742--3749},
  [\href{https://arxiv.org/abs/hep-ph/9303238}{{\ttfamily hep-ph/9303238}}].

\bibitem{Atag:1994hk}
S.~Atag and O.~Cakir, \textit{{Pair production of scalar leptoquarks at TeV
  energy gamma p colliders}},
  \href{https://doi.org/10.1103/PhysRevD.49.5769}{\textit{Phys. Rev. D}
  {\bfseries 49} (1994) 5769--5772}.

\bibitem{Atag:1994np}
S.~Atag, A.~Celikel and S.~Sultansoy, \textit{{Scalar leptoquark production at
  TeV energy gamma p colliders}},
  \href{https://doi.org/10.1016/0370-2693(94)91212-2}{\textit{Phys. Lett. B}
  {\bfseries 326} (1994) 185--189}.

\bibitem{Dobado:1987pj}
A.~Dobado, M.~Herrero and C.~Munoz, \textit{{Production of Leptoquarks From
  Superstring Models in $e p$ Colliders}},
  \href{https://doi.org/10.1016/0370-2693(87)90638-1}{\textit{Phys. Lett. B}
  {\bfseries 191} (1987) 449--455}.

\bibitem{Ilyin:1995jv}
V.~Ilyin, A.~Pukhov, V.~Savrin, A.~Semenov and W.~von Schlippe, \textit{{Single
  leptoquark production associated with hard photon emission in $e p$
  collisions at high-energies}},
  \href{https://doi.org/10.1016/0370-2693(95)00444-P}{\textit{Phys. Lett. B}
  {\bfseries 351} (1995) 504--509},
  [\href{https://arxiv.org/abs/hep-ph/9503401}{{\ttfamily hep-ph/9503401}}].
  [Erratum: Phys.Lett.B 352, 500--500 (1995)].

\bibitem{Gunion:1987ge}
J.~F. Gunion and E.~Ma, \textit{{Production of $E(6)$ Scalar Leptoquarks in $e
  p$ Collisions}},
  \href{https://doi.org/10.1016/0370-2693(87)91205-6}{\textit{Phys. Lett. B}
  {\bfseries 195} (1987) 257--264}.

\bibitem{Bandyopadhyay:2020klr}
P.~Bandyopadhyay, S.~Dutta and A.~Karan, \textit{{Investigating the Production
  of Leptoquarks by Means of Zeros of Amplitude at Photon Electron Collider}},
  \href{https://doi.org/10.1140/epjc/s10052-020-8083-7}{\textit{Eur. Phys. J.
  C} {\bfseries 80} (2020) 573},
  [\href{https://arxiv.org/abs/2003.11751}{{\ttfamily 2003.11751}}].

\bibitem{Bandyopadhyay:2020wfv}
P.~Bandyopadhyay, S.~Dutta and M.~Jakkapu, \textit{{Exploring angular
  distributions at the LHC}},
  \href{https://arxiv.org/abs/2007.12997}{{\ttfamily 2007.12997}}.

\bibitem{Angelescu:2018tyl}
A.~Angelescu, D.~Be\v{c}irevi\'c, D.~Faroughy and O.~Sumensari,
  \textit{{Closing the window on single leptoquark solutions to the $B$-physics
  anomalies}}, \href{https://doi.org/10.1007/JHEP10(2018)183}{\textit{JHEP}
  {\bfseries 10} (2018) 183},
  [\href{https://arxiv.org/abs/1808.08179}{{\ttfamily 1808.08179}}].

\bibitem{19436}
\textit{{HERA - A Proposal for a Large Electron Proton Colliding Beam Facility
  at DESY}}.
\newblock 1981.

\bibitem{Klein:2008di}
M.~Klein and R.~Yoshida, \textit{{Collider Physics at HERA}},
  \href{https://doi.org/10.1016/j.ppnp.2008.05.002}{\textit{Prog. Part. Nucl.
  Phys.} {\bfseries 61} (2008) 343--393},
  [\href{https://arxiv.org/abs/0805.3334}{{\ttfamily 0805.3334}}].

\bibitem{Agostini:2020fmq}
{\scshape LHeC, FCC-he Study Group} collaboration, P.~Agostini et~al.,
  \textit{{The Large Hadron-Electron Collider at the HL-LHC}},
  \href{https://arxiv.org/abs/2007.14491}{{\ttfamily 2007.14491}}.

\bibitem{Bordry:2018gri}
F.~Bordry, M.~Benedikt, O.~Br\"uning, J.~Jowett, L.~Rossi, D.~Schulte et~al.,
  \textit{{Machine Parameters and Projected Luminosity Performance of Proposed
  Future Colliders at CERN}},
  \href{https://arxiv.org/abs/1810.13022}{{\ttfamily 1810.13022}}.

\bibitem{Davidson:1993qk}
S.~Davidson, D.~C. Bailey and B.~A. Campbell, \textit{{Model independent
  constraints on leptoquarks from rare processes}},
  \href{https://doi.org/10.1007/BF01552629}{\textit{Z. Phys. C} {\bfseries 61}
  (1994) 613--644}, [\href{https://arxiv.org/abs/hep-ph/9309310}{{\ttfamily
  hep-ph/9309310}}].

\bibitem{Davies:1990sc}
A.~J. Davies and X.-G. He, \textit{{Tree Level Scalar Fermion Interactions
  Consistent With the Symmetries of the Standard Model}},
  \href{https://doi.org/10.1103/PhysRevD.43.225}{\textit{Phys. Rev. D}
  {\bfseries 43} (1991) 225--235}.

\bibitem{10.1093/ptep/ptaa104}
{Particle Data Group}, P.~A. Zyla et~al., \textit{{Review of Particle
  Physics}}, \href{https://doi.org/10.1093/ptep/ptaa104}{\textit{Prog. Theor.
  Exp. Phys.} {\bfseries 2020} 083C01 (2020)}.

\bibitem{Leurer:1993em}
M.~Leurer, \textit{{A Comprehensive study of leptoquark bounds}},
  \href{https://doi.org/10.1103/PhysRevD.49.333}{\textit{Phys. Rev. D}
  {\bfseries 49} (1994) 333--342},
  [\href{https://arxiv.org/abs/hep-ph/9309266}{{\ttfamily hep-ph/9309266}}].

\bibitem{Leurer:1993qx}
M.~Leurer, \textit{{Bounds on vector leptoquarks}},
  \href{https://doi.org/10.1103/PhysRevD.50.536}{\textit{Phys. Rev. D}
  {\bfseries 50} (1994) 536--541},
  [\href{https://arxiv.org/abs/hep-ph/9312341}{{\ttfamily hep-ph/9312341}}].

\bibitem{Carpentier:2010ue}
M.~Carpentier and S.~Davidson, \textit{{Constraints on two-lepton, two quark
  operators}},
  \href{https://doi.org/10.1140/epjc/s10052-010-1482-4}{\textit{Eur. Phys. J.
  C} {\bfseries 70} (2010) 1071--1090},
  [\href{https://arxiv.org/abs/1008.0280}{{\ttfamily 1008.0280}}].

\bibitem{Mandal:2019gff}
R.~Mandal and A.~Pich, \textit{{Constraints on scalar leptoquarks from lepton
  and kaon physics}},
  \href{https://doi.org/10.1007/JHEP12(2019)089}{\textit{JHEP} {\bfseries 12}
  (2019) 089}, [\href{https://arxiv.org/abs/1908.11155}{{\ttfamily
  1908.11155}}].

\bibitem{Adloff:1997fg}
{\scshape H1} collaboration, C.~Adloff et~al., \textit{{Observation of events
  at very high $Q^{2}$ in $e p$ collisions at HERA}},
  \href{https://doi.org/10.1007/s002880050383}{\textit{Z. Phys. C} {\bfseries
  74} (1997) 191--206}, [\href{https://arxiv.org/abs/hep-ex/9702012}{{\ttfamily
  hep-ex/9702012}}].

\bibitem{Breitweg:1997ff}
{\scshape ZEUS} collaboration, J.~Breitweg et~al., \textit{{Comparison of ZEUS
  data with standard model predictions for $e^{+} p \to e^{+} X$ scattering at
  high x and $Q^{2}$}}, \href{https://doi.org/10.1007/s002880050384}{\textit{Z.
  Phys. C} {\bfseries 74} (1997) 207--220},
  [\href{https://arxiv.org/abs/hep-ex/9702015}{{\ttfamily hep-ex/9702015}}].

\bibitem{Blumlein:1996qp}
J.~Blumlein, E.~Boos and A.~Kryukov, \textit{{Leptoquark pair production in
  hadronic interactions}},
  \href{https://doi.org/10.1007/s002880050538}{\textit{Z. Phys. C} {\bfseries
  76} (1997) 137--153}, [\href{https://arxiv.org/abs/hep-ph/9610408}{{\ttfamily
  hep-ph/9610408}}].

\bibitem{Chekanov:2003af}
{\scshape ZEUS} collaboration, S.~Chekanov et~al., \textit{{A Search for
  resonance decays to lepton + jet at HERA and limits on leptoquarks}},
  \href{https://doi.org/10.1103/PhysRevD.68.052004}{\textit{Phys. Rev. D}
  {\bfseries 68} (2003) 052004},
  [\href{https://arxiv.org/abs/hep-ex/0304008}{{\ttfamily hep-ex/0304008}}].

\bibitem{Abe:1995fj}
{\scshape CDF} collaboration, F.~Abe et~al., \textit{{A Search for second
  generation leptoquarks in $p\bar{p}$ collisions at $\sqrt{s} = 1.8$ TeV}},
  \href{https://doi.org/10.1103/PhysRevLett.75.1012}{\textit{Phys. Rev. Lett.}
  {\bfseries 75} (1995) 1012--1016}.

\bibitem{Abe:1996dn}
{\scshape CDF} collaboration, F.~Abe et~al., \textit{{Search for third
  generation leptoquarks in $\bar{p}p$ collisions at $\sqrt{s} = 1.8$ TeV}},
  \href{https://doi.org/10.1103/PhysRevLett.78.2906}{\textit{Phys. Rev. Lett.}
  {\bfseries 78} (1997) 2906--2911}.

\bibitem{Staub:2013tta}
F.~Staub, \textit{{SARAH 4 : A tool for (not only SUSY) model builders}},
  \href{https://doi.org/10.1016/j.cpc.2014.02.018}{\textit{Comput. Phys.
  Commun.} {\bfseries 185} (2014) 1773--1790},
  [\href{https://arxiv.org/abs/1309.7223}{{\ttfamily 1309.7223}}].

\bibitem{Belyaev:2012qa}
A.~Belyaev, N.~D. Christensen and A.~Pukhov, \textit{{CalcHEP 3.4 for collider
  physics within and beyond the Standard Model}},
  \href{https://doi.org/10.1016/j.cpc.2013.01.014}{\textit{Comput. Phys.
  Commun.} {\bfseries 184} (2013) 1729--1769},
  [\href{https://arxiv.org/abs/1207.6082}{{\ttfamily 1207.6082}}].

\bibitem{Nocera:2014gqa}
{\scshape NNPDF} collaboration, E.~R. Nocera, R.~D. Ball, S.~Forte, G.~Ridolfi
  and J.~Rojo, \textit{{A first unbiased global determination of polarized PDFs
  and their uncertainties}},
  \href{https://doi.org/10.1016/j.nuclphysb.2014.08.008}{\textit{Nucl. Phys. B}
  {\bfseries 887} (2014) 276--308},
  [\href{https://arxiv.org/abs/1406.5539}{{\ttfamily 1406.5539}}].

\bibitem{Sjostrand:2007gs}
T.~Sjostrand, S.~Mrenna and P.~Z. Skands, \textit{{A Brief Introduction to
  PYTHIA 8.1}},
  \href{https://doi.org/10.1016/j.cpc.2008.01.036}{\textit{Comput. Phys.
  Commun.} {\bfseries 178} (2008) 852--867},
  [\href{https://arxiv.org/abs/0710.3820}{{\ttfamily 0710.3820}}].

\bibitem{Sjostrand:2014zea}
T.~Sj\"ostrand, S.~Ask, J.~R. Christiansen, R.~Corke, N.~Desai, P.~Ilten
  et~al., \textit{{An introduction to PYTHIA 8.2}},
  \href{https://doi.org/10.1016/j.cpc.2015.01.024}{\textit{Comput. Phys.
  Commun.} {\bfseries 191} (2015) 159--177},
  [\href{https://arxiv.org/abs/1410.3012}{{\ttfamily 1410.3012}}].

\bibitem{Cacciari:2011ma}
M.~Cacciari, G.~P. Salam and G.~Soyez, \textit{{FastJet User Manual}},
  \href{https://doi.org/10.1140/epjc/s10052-012-1896-2}{\textit{Eur. Phys. J.
  C} {\bfseries 72} (2012) 1896},
  [\href{https://arxiv.org/abs/1111.6097}{{\ttfamily 1111.6097}}].

\bibitem{Krohn:2012fg}
D.~Krohn, M.~D. Schwartz, T.~Lin and W.~J. Waalewijn, \textit{{Jet Charge at
  the LHC}},
  \href{https://doi.org/10.1103/PhysRevLett.110.212001}{\textit{Phys. Rev.
  Lett.} {\bfseries 110} (2013) 212001},
  [\href{https://arxiv.org/abs/1209.2421}{{\ttfamily 1209.2421}}].

\bibitem{Tokar:2017syr}
{\scshape ATLAS, CMS} collaboration, S.~Tokar, \textit{{Jet charge
  determination at the LHC}},  in \textit{{Parton radiation and fragmentation
  from LHC to FCC-ee}}, pp.~79--84, 2, 2017.

\bibitem{Sirunyan:2017tyr}
{\scshape CMS} collaboration, A.~M. Sirunyan et~al., \textit{{Measurements of
  jet charge with dijet events in pp collisions at $\sqrt{s}=8$ TeV}},
  \href{https://doi.org/10.1007/JHEP10(2017)131}{\textit{JHEP} {\bfseries 10}
  (2017) 131}, [\href{https://arxiv.org/abs/1706.05868}{{\ttfamily
  1706.05868}}].

\end{thebibliography}\endgroup
			\bibliographystyle{Ref}
		\end{document}